\documentclass[11pt,a4paper]{article}

\usepackage[paper=a4paper,margin=1in]{geometry}%

\usepackage[utf8]{inputenc}
\usepackage{graphicx}
\graphicspath{ {./images/} }
\usepackage{xcolor}
\usepackage{multirow}

\usepackage{hyperref}
\hypersetup{
    colorlinks=true,
    linkcolor=blue,
    filecolor=magenta,      
    urlcolor=cyan,
    citecolor=blue
}

\usepackage{todonotes}
\usepackage{url} 
\usepackage[acronym]{glossaries}

\usepackage{titlesec} 
\titlespacing\subsection{0pt}{5pt}{0pt} 
\usepackage[framemethod=default]{mdframed}

\newcommand{\artref}[2]{\hyperref[art:#2]{\textsf{\textbf{[#1]}}}}

\newcommand{\ArtA}{\artref{A}{A}}
\newcommand{\ArtB}{\artref{B}{B}}
\newcommand{\ArtC}{\artref{C}{C}}
\newcommand{\ArtD}{\artref{D}{D}}
\newcommand{\ArtE}{\artref{E}{E}}
\newcommand{\ArtF}{\artref{F}{F}}
\newcommand{\ArtG}{\artref{G}{G}}
\newcommand{\ArtH}{\artref{H}{H}}
\newcommand{\ArtI}{\artref{I}{I}}
\newcommand{\ArtJ}{\artref{J}{J}}
\newcommand{\ArtK}{\artref{K}{K}}
\newcommand{\ArtL}{\artref{L}{L}}
\newcommand{\ArtM}{\artref{M}{M}}
\newcommand{\ArtN}{\artref{N}{N}}
\newcommand{\ArtO}{\artref{O}{O}}
\newcommand{\ArtP}{\artref{P}{P}}
\newcommand{\ArtQ}{\artref{Q}{Q}}
\newcommand{\ArtR}{\artref{R}{R}}
\newcommand{\ArtS}{\artref{S}{S}}
\newcommand{\ArtT}{\artref{T}{T}}
\newcommand{\ArtU}{\artref{U}{U}}
\newcommand{\ArtV}{\artref{V}{V}}
\newcommand{\ArtW}{\artref{W}{W}}
\newcommand{\ArtX}{\artref{X}{X}}
\newcommand{\ArtY}{\artref{Y}{Y}}
\newcommand{\ArtZ}{\artref{Z}{Z}}
\newcommand{\ArtAA}{\artref{AA}{AA}}
\newcommand{\ArtAB}{\artref{BB}{BB}}

\newcommand{\ArtAD}{\artref{DD}{DD}}
\newcommand{\ArtAE}{\artref{EE}{EE}}
\newcommand{\ArtAF}{\artref{FF}{FF}}
\newcommand{\ArtAG}{\artref{GG}{GG}}
\newcommand{\ArtAH}{\artref{HH}{HH}}
\newcommand{\ArtAI}{\artref{II}{II}}
\newcommand{\ArtAJ}{\artref{JJ}{JJ}}
\newcommand{\ArtAK}{\artref{KK}{JJ}}
\newcommand{\ArtAL}{\artref{LL}{LL}}
\newcommand{\ArtAM}{\artref{MM}{MM}}
\newcommand{\ArtAN}{\artref{NN}{NN}}
\newcommand{\ArtAO}{\artref{OO}{OO}}
\newcommand{\ArtAP}{\artref{PP}{PP}}
\newcommand{\ArtAQ}{\artref{QQ}{QQ}}
\newcommand{\ArtAR}{\artref{RR}{RR}}
\newcommand{\ArtAS}{\artref{SS}{SS}}
\newcommand{\ArtAT}{\artref{TT}{TT}}
\newcommand{\ArtAU}{\artref{UU}{UU}}
\newcommand{\ArtAV}{\artref{VV}{VV}}
\newcommand{\ArtAW}{\artref{WW}{WW}}
\newcommand{\ArtAX}{\artref{XX}{XX}}
\newcommand{\ArtAY}{\artref{YY}{YY}}
\newcommand{\ArtAZ}{\artref{ZZ}{ZZ}}

\usepackage{tcolorbox}

\newcounter{note}
\definecolor{noteHead}{RGB}{50,62,72}
\definecolor{noteMain}{RGB}{166,217,214}

\newenvironment{note}[1][]{
\begin{tcolorbox}[colback=noteMain, colframe=noteHead, title= \textsf{\large{\textbf{Note~\thenote. #1}}}]
\refstepcounter{note}
   \rmfamily
   \par\medskip}
{  \medskip\end{tcolorbox}}

\usepackage{tcolorbox}

\newcounter{example}
\definecolor{exHead}{RGB}{118,103,174}
\definecolor{exMain}{RGB}{142,209,241}

\newenvironment{example}[1][]{
\begin{tcolorbox}[colback=exMain, colframe=exHead, title= \textsf{\large{\textbf{Example~\theexample. #1}}}]
\refstepcounter{example}
   \rmfamily
   \par\medskip}
{  \medskip\end{tcolorbox}}

\usepackage{tcolorbox}

\newcounter{stage}
\newcommand{\stage}[1]{\section{Stage~\thestage.~#1\label{stg:\thestage}}\refstepcounter{stage}}

\newcommand{\stageref}[1]{\hyperref[stg:#1]{#1}}

\newcounter{ActivityCounter}
\refstepcounter{ActivityCounter}

\newcommand{\Activity}[2]{\subsection{Activity~\theActivityCounter: {#1}}\label{#2}\refstepcounter{ActivityCounter}}

\newcommand{\actref}[2]{\hyperref[#1]{#2}}

\makeglossaries

\title{SACE}
\author{richard.hawkins}
\date{July 2022}

\begin{document}


\begin{titlepage}
	{\centering
 	{\Large{Guidance on the Safety Assurance of Autonomous Systems in Complex Environments (SACE)
 	}}\par\vspace{1cm}
Richard Hawkins, Rob Alexander, Matt Osborne, Mike Parsons, Mark Nicholson, John McDermid and Ibrahim Habli \par \vspace{0.5cm}
Centre for Assuring Autonomy (CfAA)\\
University of York

\texttt{richard.hawkins@york.ac.uk}

\textbf{Version 1.1, July 2026} 
\par\vspace{5mm}
}

\vspace{11cm}
\begin{centering}
\textsf{The material in this document is provided as guidance only. No responsibility for loss occasioned to any person acting or refraining from action as a result of this material or any comments made can be accepted by the authors or The University of York.}
\end{centering}

\end{titlepage}

\tableofcontents
\newpage

\section{Introduction}

Autonomous systems (AS) are systems that have the capability to take decisions free from direct human control. AS are increasingly being considered for adoption for applications where their behaviour may cause harm, such as when used for autonomous driving, medical applications or in domestic environments. For such applications, being able to ensure and demonstrate (assure) the safety of the operation of the AS is crucial for their adoption. This can be particularly challenging where AS operate in complex and changing real-world environments. Establishing justified confidence in the safety of AS requires the creation of a compelling safety case. This document introduces a methodology for the \textbf{S}afety \textbf{A}ssurance of Autonomous Systems in \textbf{C}omplex \textbf{E}nvironments (SACE). SACE comprises a set of safety case patterns and a process for (1) systematically integrating safety assurance into the development of the AS and (2) for generating the evidence base for explicitly justifying the acceptable safety of the AS.
\section{Using this Document}

The aim of this document is to provide guidance on how to systematically integrate safety assurance into
the development of AS. A primary outcome of this integration is an explicit and structured
safety case. More specifically, SACE offers a set of argument patterns, and the underlying assurance
activities, that can be instantiated and specialised in order to develop the AS safety cases.

This document is aimed at
\begin{enumerate}
    \item Safety engineers who are interested in understanding what must be done to provide the required assurance in the safety of an AS operating in a complex environment
    \item System engineers and developers who are interested in understanding the safety assurance considerations when developing an AS
    \item Reviewers and safety assessors who are interested in understanding what should be the focus of review and the criteria by which the sufficiency of the assurance activities should be judged.

\end{enumerate}

When using this document it is recommended that the reader is aware of other sources of complementary guidance on best practice for the safety of autonomous systems such as UL4000 \cite{UL4600} or SCSC-153B \cite{SASWG2020}.

Throughout the document, the use of "shall" indicates a required element of the guidance. Information marked as a ``NOTE'' or ``EXAMPLE'' is only used for clarification of the associated activities. A ``NOTE'' provides additional information, for clarification or advice purposes. An ``EXAMPLE'' is used to illustrate a particular point that is specific to a domain or technology. An example presented in this document is not meant to be exhaustive. Planned case studies and future experiments will provide fuller examples.

\clearpage
\section{Change History}

\begin{table}[h]
\begin{tabular}{|l|p{10cm}|l|}
\hline
\textbf{Version} & \textbf{Changes}                         & \textbf{Date of Issue}          \\ \hline
0.1               & First draft for internal review.                           & \multicolumn{1}{c|}{07/03/2022} \\ \hline
0.2               & Revised draft for external review.                          & \multicolumn{1}{c|}{19/05/2022} \\ \hline
1                 & First issue.                           & \multicolumn{1}{c|}{29/07/2022} \\ \hline
1.1              & Changes to Stage 8, including names and descriptions of artefacts, plus updates to the AS verification argument pattern. A small number of other minor changes elsewhere in the document. & \multicolumn{1}{c|}{27/07/2026} \\ \hline
\end{tabular}
\end{table}
\clearpage
\section{Overview of SACE}\label{sec:overview}

\begin{figure}[h]
    \centering
    \includegraphics[width=1\linewidth]{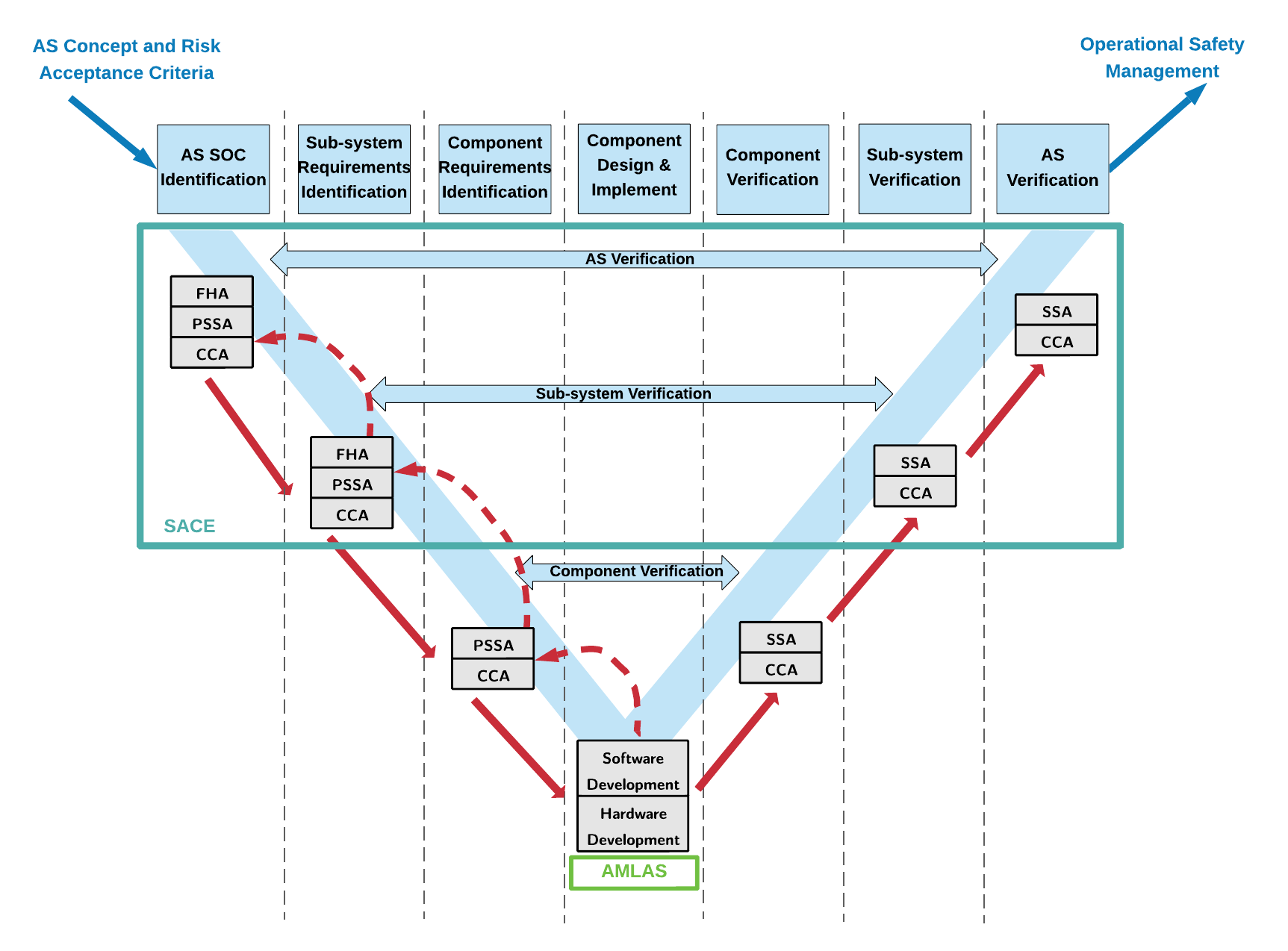}
    \caption{The scope of the SACE process}
    \label{fig:SACEscope}
\end{figure}

The SACE process runs in parallel to and complements the activities undertaken as part of an existing systems engineering and associated system safety assurance process. We assume that such a system safety assurance process is in place and that the activities described as part of the SACE process are modifications, enhancements or additions to that system safety process to specifically deal with the safety assurance challenges of an autonomous system operating in a complex environment. In this guidance we do not describe what that baseline system safety assurance process is, however neither do we assume that any particular system safety process is adopted, instead we define certain characteristics that the safety assurance process should have. Figure \ref{fig:SACEscope} provides a model of this baseline safety process that illustrates the following required features:

\begin{itemize}
    \item Safety requirements are identified at multiple levels of decomposition of the AS design based on analysis of the system.
    \item The safety requirements at each level of decomposition preserve the intent of, and are traceable to the safety requirements at preceding levels.
    \item The safety requirements are allocated to components that implement those requirements.
    \item Verification and assessment at each level of integration provides evidence that the allocated safety requirements are satisfied.
    \item Throughout the process analysis is performed to identify potential common cause failures, which are reflected in the safety requirements.
\end{itemize}

These features are influenced by established good practice system safety processes such as the Aerospace Recommended Practice ARP 4754A \cite{SAE_2010}. 

Note that the system safety assurance process itself runs in parallel to, is informed by, and informs, the system development process. In addition the development of the safety case for the system also runs in parallel to this process. The safety case process is discussed in more detail below. For clarity we do not show either the development process or the safety case development process in Figure \ref{fig:SACEscope}.

This SACE guidance covers only part of this overall system safety assurance process, as indicated by the SACE box in Figure \ref{fig:SACEscope}. SACE starts at the beginning of the development process when a concept for the AS has been determined and continues down to the derivation of requirements for the sub-systems of the AS. SACE also considers verification of the AS at the sub-system and system level. SACE does not include the development of requirements  for the individual system components, or the implementation of those components. These issues are considered as part of other guidance documents. For example safety assurance of components implemented using machine learning is considered as part of the AMLAS (Assurance of Machine Learning for Autonomous Systems) guidance \cite{AAIP2021a}, as indicated in Figure \ref{fig:SACEscope}.

There are other safety assurance aspects that are not within the scope of the SACE guidance. Firstly, the safety assurance activities considered by SACE are only those that are applied during the development phase of the AS lifecycle, that is activities carried out prior to deployment of the AS into operation. Although the SACE activities will have consideration for safety assurance during later lifecycle phases such as operation and maintenance, the safety assurance activities that are actually undertaken during these later phases (often referred to as operational safety management) are not within the scope of this document. Secondly, SACE focuses on safety assurance for an individual AS. Although this involves consideration of the interaction of that AS with other agents (including other AS), SACE does not explicitly consider the additional safety assurance implications of multiple collaborating ASs  (see \cite{gleirscher2020safety} for examples of safety approaches for collaborative robots). Thirdly, SACE does not provide guidance on the legal and ethical considerations surrounding the development and operation of AS. Such issues will be considered as part of separate guidance \cite{porter2022principle}. SACE takes as an input the output from such considerations in the form of defined acceptance criteria. This input to the SACE process is discussed when describing Stage~\stageref{2}.

\subsection{AS Safety Case}

\begin{figure}[p]
    \centering
    \includegraphics[width=0.9\linewidth]{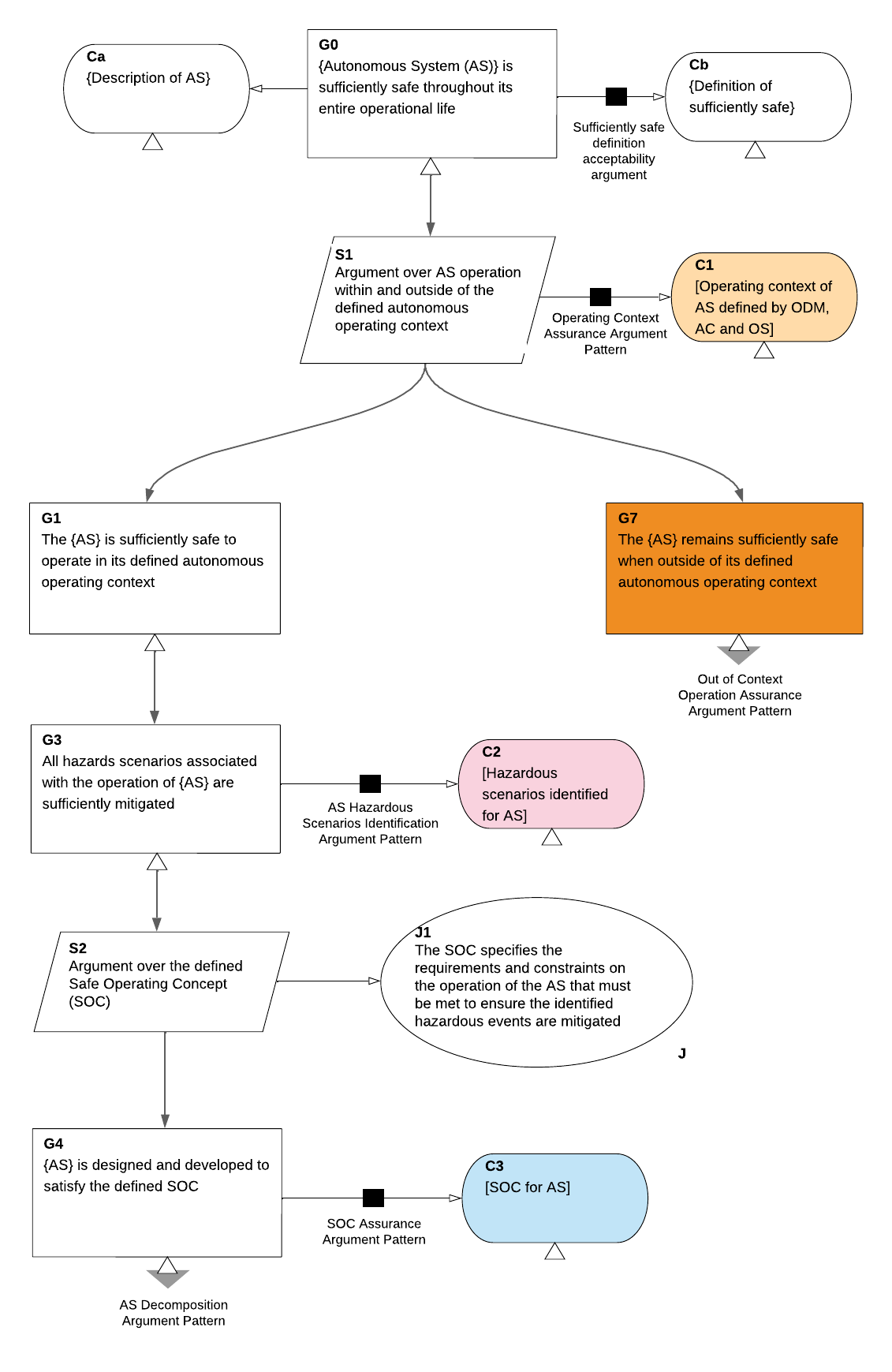}
    \caption{High-level baseline safety argument pattern}
    \label{fig:SACEarg1}
\end{figure}

The key output of the SACE process is a safety case for the AS. We adopt a commonly-used definition of a safety case as a ``structured argument, supported by a body of evidence that provides a compelling, comprehensible and valid case that a system is safe for a given application in a given operating environment.''~\cite{m2017a}. As for the safety assurance process, SACE does not provide general guidance on how to create a compelling safety case for a system, instead SACE focuses on what modifications, enhancements or additions are required with respect to a safety case for a more conventional system in order to address the challenges of autonomy. As the baseline for this we consider a very simple simple argument structure as shown in the safety case patterns in Figures \ref{fig:SACEarg1} and \ref{fig:SACEarg2}. A safety case pattern documents a reusable argument structure and types of evidence that can be instantiated to create a specific safety case instance \cite{kelly1997safety}. 

In these figures, and throughout SACE, safety case patterns are represented using the Goal Structuring Notation (GSN)~\cite{group2021a}. GSN is a graphical notation for explicitly capturing safety arguments that is widely used in many industries for documenting safety cases. For a detailed description of the notation, the reader is advised to consult the publicly available GSN standard~\cite{group2021a}.

Figure \ref{fig:SACEarg1} shows what the top level of a simple safety argument for an AS may look like, documented in the form of a pattern. The intention here is to provide the general form that such an argument should take, rather than to mandate a particular structure. In practice it is expected that a more detailed safety argument would need to be provided, but that it will retain the characteristics defined by the pattern as discussed below. The safety case for the AS also requires further argument relating to the decomposition of the design and requirements throughout the AS development process as discussed earlier. The safety argument pattern for this system decomposition is shown in Figure \ref{fig:SACEarg2} and discussed further later in this section. 

Figure \ref{fig:SACEarg1} identifies in white, the elements of the argument structure that are required in a safety argument for an AS, but that would also need to be addressed in a safety case for any system. These are the elements that are affected much less by the autonomous nature of the system. In contrast, those elements of the argument structure that are coloured in Figure \ref{fig:SACEarg1} represent aspects of the safety argument that are either novel to AS or which are most affected by autonomy. These correspond to the parts of the safety assurance process that are challenged by autonomy as discussed earlier. In this guidance document we provide argument patterns relating to these autonomy-related aspects identified in Figure \ref{fig:SACEarg1}. The details of the patterns are provided as part of the relevant process activity along with descriptions of the relevant artefacts. 

In the argument patterns we make use of assurance claim points (ACPs) \cite{group2021a}, indicated by black squares in the argument pattern, to represent points in the argument at which further argument and evidence demonstrating the confidence in particular elements is required. It can be noted in Figure \ref{fig:SACEarg1} that many of the argument patterns we provide for AS relate to the confidence arguments for ACPs relating to the key artefacts arising from the AS safety process. This reflects the importance that must be placed on managing uncertainty when considering AS. For example in Figure \ref{fig:SACEarg1}, there can be seen to be an ACP relating to the hazards identified for the AS. This shows it is necessary to provide a confidence argument relating to the sufficiency of the identified hazards. The AS Hazard Identification Argument Pattern provides guidance on the structure of that confidence argument.

Below we describe the elements of the argument in Figure \ref{fig:SACEarg1}.

\subsection*{G0}

The top level safety claim that we consider is that the AS is sufficiently safe throughout its operational life. The definition of what is considered to be sufficiently safe for an AS can be extremely challenging since it must take account of complex factors such as:
\begin{itemize}
    \item Risk perception - The tolerability of risk  for an AS may be different from an equivalent non-autonomous system meaning that a direct comparison may not be an effective way to judge sufficiency.
    \item Risk trade-off - Autonomous operation will introduce new risks that must be traded-off against the potential safety benefits that they may bring.
    \item Ethical considerations - Any definition of sufficiently safe must take account of the complex ethical issues that are raised by autonomous operations (see \cite{burton2020mind}, and \cite{zhu2021ai} for example).
\end{itemize}
In addition to this there are many domain-specific and legal factors that inform this definition.

These issues are the subject of extensive on-going research that is outside of the scope of this document. For the purposes of SACE we assume that there exists an acceptable definition of ``sufficiently safe''. Additional guidance in this area can be found at~\cite{AAIP-BoK}.

\subsection*{S1}

The strategy that is adopted is to split the safety argument into a claim regarding the safe operation of the AS when it is operating within the defined autonomous operating context (G1), and also a claim that it remains sufficiently safe when outside of that operating context (G7). This requires that the operating context has been explicitly defined (Stage~\stageref{1} describes how this is done). The operating context description is provided at C1, however it is also crucial for assurance of AS that an argument is provided, supported by evidence, to demonstrate that the operating context that is defined is sufficient to support the safe operation of the AS.

The Operating Context Assurance Argument Pattern (\ArtG) is provided in order to guide the development of this argument, and is discussed in detail in Stage~\stageref{1}. The link to the assurance argument for the operating context is established in Figure \ref{fig:SACEarg1} using an Assurance Claim Point (ACP) \cite{group2021a} (indicated by the black square). ACPs represent points in the argument at which further assurance is provided. 

\subsection*{G1} 

This safety claim relates to the safe operation of the AS when it is operating within the defined autonomous operating context.

\subsection*{G3}

The safe operation of the AS in the defined autonomous operating context is demonstrated through an argument that all of the hazardous scenarios associated with the operation of the AS have been sufficiently mitigated. The hazardous scenarios that are identified are provided at C2. The completeness and correctness of the identified hazardous scenarios for the AS must be demonstrated. The AS Hazardous Scenarios Identification Argument Pattern (\ArtI) is provided in order to guide the development of this argument, and the activities and artefacts are discussed in detail in Stage~\stageref{2}.

\subsection*{G4}

The identified hazards for the AS are mitigated through ensuring that the AS operates in such a manner that the defined Safe Operating Concept (SOC) is satisfied. The SOC provides a specification for safe operation of the AS taking account of the operating context and the identified hazards. The SOC is discussed in detail in Stage~\stageref{3}. It must be demonstrated that the operation specified by the SOC sufficiently mitigates the identified hazards. The SOC Assurance Argument Pattern (\ArtN) is provided in order to guide the development of this argument.

This safety claim is supported by an argument that considers the control and mitigation of hazards and the associated safety assurance of the AS throughout decomposition of the development lifecycle (as presented in Figure \ref{fig:SACEprocess}). This AS Decomposition Argument Pattern is presented in Figure \ref{fig:SACEarg2} and discussed below.

\subsection*{G7}

This safety claim relates to the safety of the AS when operating outside of the defined operating context. Although the AS is only expected to operate autonomously within the defined operating context, there may be occasions when it is operating outside of that. This may be planned non-autonomous operation where a human operator takes control of the system, or it may be unplanned excursions from the defined operating context. In all such cases it is important to be able to demonstrate that the system remains sufficiently safe. The Out of Context Operation Argument Pattern (\ArtAP) is provided in order to guide the development of this argument, and the activities and artefacts are discussed in detail in Stage~\stageref{7}.

\begin{figure}[h]
    \centering
    \includegraphics[width=1\linewidth]{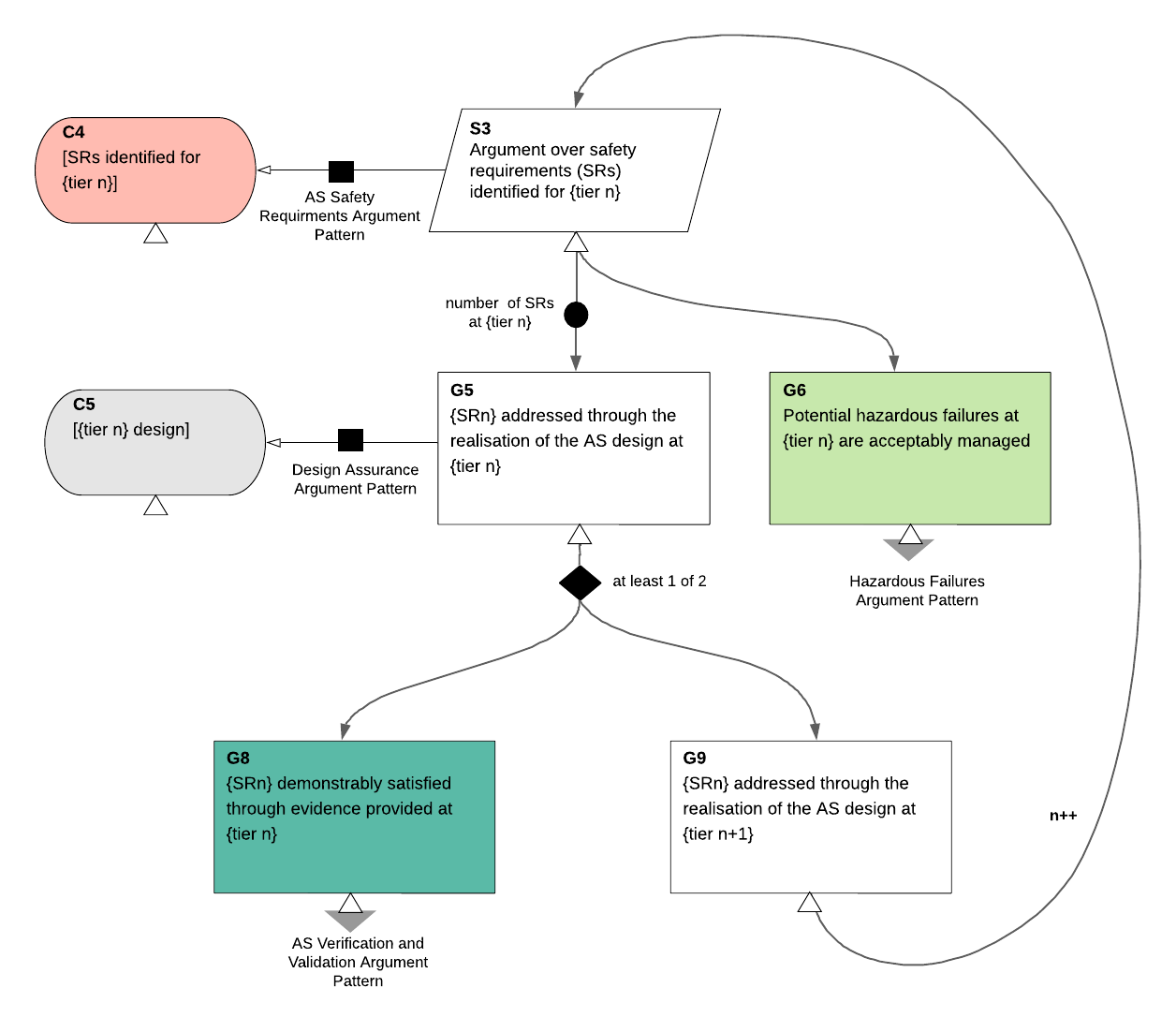}
    \caption{System decomposition baseline safety argument pattern}
    \label{fig:SACEarg2}
\end{figure}

Figure \ref{fig:SACEarg2} shows the further development of the high-level safety argument in Figure \ref{fig:SACEarg1}. This provides a pattern of argument that captures the general form that the safety argument should take for each level of decomposition in the system development process (described as a ``tier" in the argument pattern). As discussed earlier, in this guidance we make no assumptions about the number of levels of decomposition in the AS development process, instead we require that a safety argument is developed that considers safety at each of those levels, irrespective of how many levels there may be. The argument pattern shown in Figure \ref{fig:SACEarg2} is based upon patterns originally developed for the software aspects of systems \cite{hawkins2013software}, but which also apply more generally. The characteristics of this pattern are discussed below.

\subsection*{S3}

In order to demonstrate that the development of the AS satisfies the SOC, the strategy adopted in the safety case is to provide an argument and evidence that considers the safety requirements identified at each tier (level of decomposition). If we refer back to Figure \ref{fig:SACEscope}, this will include requirements at the sub-system and component level, but as discussed, could include requirements for additional tiers as appropriate. 

The safety requirements that have been identified for the tier under consideration are provided as context at C4. It must be demonstrated that these safety requirements have been correctly decomposed, allocated and interpreted from the requirements of the previous tier of development. The AS Safety Requirements Argument Pattern (\ArtS) is provided in order to guide the development of this argument, and the activities and artefacts are discussed in detail in Stage~\stageref{4}.

Two claims are made as part of this strategy. Firstly that the defined safety requirements have been addressed by the implementation of the AS design (G5) and secondly that any potential hazardous failures that may be introduced through the design approach adopted have been managed (G6).

\subsection*{G5}

This claim focuses on demonstrating that each of the safety requirements defined for the tier have been addressed in the design of the AS at that tier. As indicated by the multiplicity in the argument pattern, a claim of this nature must be created for each of the defined safety requirements (this means that a specific argument is provided as to the satisfaction of each safety requirement).

These claims that the safety requirements are addressed can be supported through providing evidence to demonstrate their satisfaction (G8) as well as through further decomposition of the requirement to further tiers of the development lifecycle (G9). 

The design of the AS at the current tier is provided as context at C5. It is necessary to demonstrate that the design decisions that have been made at each tier are appropriate to enable the satisfaction of the safety requirements. The AS Design Assurance Argument Pattern (\ArtU) is provided for this, and the activities and artefacts for design assurance of the AS are discussed in detail in Stage~\stageref{5}.

\subsection*{G6}

There will always exist the potential for hazardous failures to be introduced during the design of an AS as a side effect of design decisions. The nature of those potential failures must be correctly understood so that they can be acceptably managed. Once the design for a tier is known, the potential failures that could result from that design solution can be understood so that appropriate mitigations can be identified (such as design changes or the derivation of further requirements). This claim focuses on the management of those potential hazardous failures. The Hazardous Failures Argument Pattern (\ArtAD) is provided for this, and the activities and artefacts are discussed in detail in Stage~\stageref{6}.

\subsection*{G8}

This claim states that evidence is provided that demonstrates that the safety requirement is satisfied. Such evidence can be provided at the different tiers of the development arising from the verification activities performed, as illustrated on the right hand side of Figure \ref{fig:SACEprocess}. The argument must demonstrate the appropriateness and trustworthiness of the evidence provided. The AS Verification Argument Pattern (\ArtAU) is provided in order to guide the development of this argument, and the activities and artefacts are discussed in detail in Stage~\stageref{8}.

\subsection*{G9}

Where the development of the AS continues to more detailed tiers of the design, a claim is provided that the safety requirement is also addressed through the next decomposition of the AS design. Here the safety argument for the next tier will follow the same form as has been described above (the same argument pattern from Figure \ref{fig:SACEarg2} is applied again for the next tier). This is indicated by the loop from G9 back to S2 which indicates that the argument is repeated.

\clearpage

\subsection{SACE Process}

\begin{figure}[h]
    \centering
    \includegraphics[width=1\linewidth]{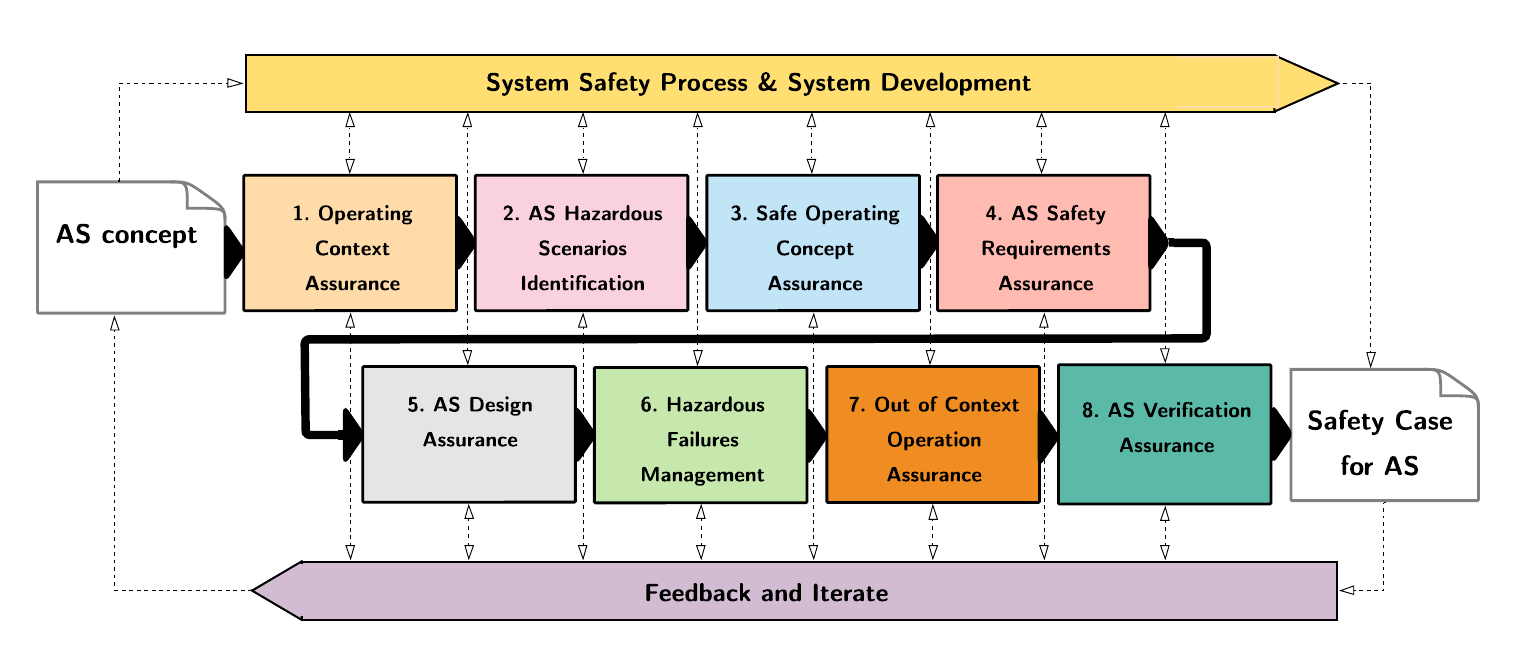}
    \caption{Overview of the SACE process}
    \label{fig:SACEprocess}
\end{figure}

Creating a compelling safety case for an AS of the form discussed above requires that certain activities are undertaken and that certain artefacts are generated. The SACE process describes the required activities and artefacts in order to achieve this. SACE is split into a number of stages, each of which defines a set of activities and artefacts. Each stage corresponds to one of the autonomy related aspects of the system safety case as discussed above and indicated in Figures \ref{fig:SACEarg1} and \ref{fig:SACEarg2} as coloured elements. Figure \ref{fig:SACEprocess} provides an overview of the SACE process indicating each of these stages.

As discussed, an established system safety process should run in parallel to SACE as well as the system development itself.  Like these processes, the SACE process is iterative, as indicated by the feedback loop in Figure~\ref{fig:SACEprocess}. Each stage of the SACE process is linked to the ‘Feedback and Iterate’ thread and could trigger the need to reconsider information generated or consumed by other stages. This is also necessary because of the interdependencies between the different stages, e.g. an activity in one stage might use artefacts produced by another activity in a previous stage. 

The stages of SACE may therefore be performed multiple times throughout the development of the AS, reflecting the iterative nature of AS development processes. For example, the design assurance activities may identify hazards that were not apparent during the initial hazard identification stage. This would require that the hazard identification stage be revisited to update and assure the hazard identification. This would then require that later stages such as the safety requirements assurance are revisited.

In this document, each SACE stage is structured as follows:
\begin{itemize}
    \item Objectives of the stage. 
    \item Inputs to, and outputs from, the stage.
    \item Description of the stage, including development and assurance activities and associated assurance artefacts and safety argument pattern.
\end{itemize}

The description of each stage details the activities to be undertaken and the artefacts produced or required by the activities. Note that detailed guidance is provided for some artefacts, whereas others are simply specified as the documented outputs of particular activities. The description also discusses common issues and misunderstandings relating to each activity; these are generally provided as notes or examples. Importantly, each stage concludes with an activity for instantiating a safety argument pattern based on the artefacts and evidence generated in the stage.

\clearpage
\clearpage
\stage{Operating Context Assurance}

\subsection*{Objectives}
\begin{enumerate}
\item Define the Autonomous Capabilities of the AS.
\item Define and validate the Operational Domain Model (ODM) for the AS.
\item Define and validate the Operating Scenarios within the defined ODM.
\item Create the Operating Context Assurance Argument.
\end{enumerate}

\subsection*{Inputs to the Stage}
\begin{itemize}
\item[\ArtA]: AS Concept definition
\item[\ArtG]: AS Operating Context Assurance Argument Pattern
\end{itemize}

\subsection*{Outputs of the Stage}
\begin{itemize}
\item[\ArtB]: Operational Domain Model
\item[\ArtC]: ODM Validation Report
\item[\ArtD]: Autonomous Capabilities Definition
\item[\ArtE]: Operating Scenarios Definition
\item[\ArtF]: Operating Scenarios Validation Report
\item[\ArtH]: AS Operating Context Assurance Argument
\end{itemize}

\subsection*{Description of the Stage}

\begin{figure}[h]
    \centering
    \includegraphics[width=1\linewidth]{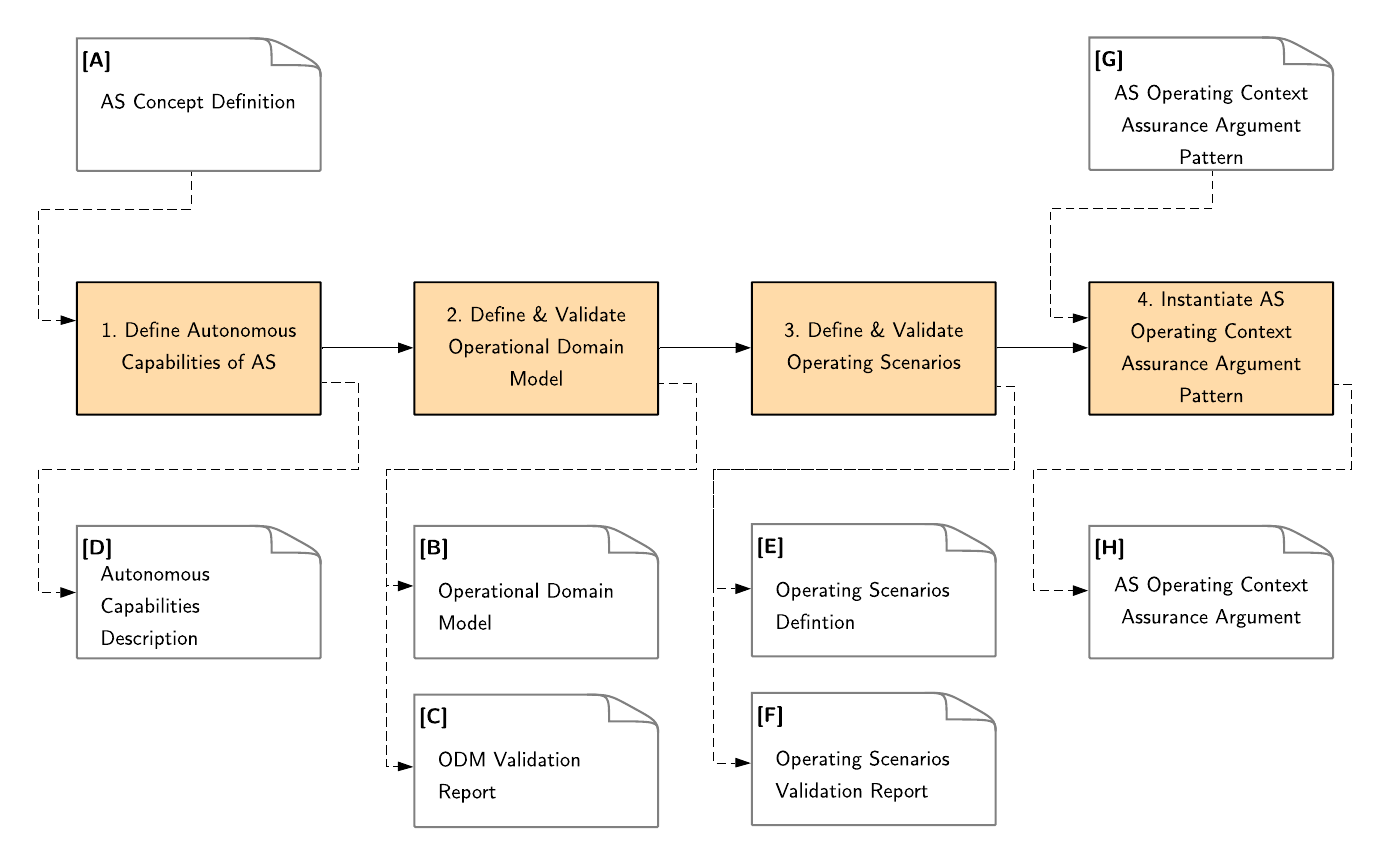}
    \caption{Operating Context Assurance Process}
    \label{fig:ScopingProcess}
\end{figure}

As shown in Figure~\ref{fig:ScopingProcess}\footnote{In the SACE process diagrams, rectangles represent activities. Document symbols represent input or output artefacts. Each document symbol has a unique ID (top left) that is used to refer to the artefact in the guidance text or the argument pattern, e.g. \ArtA~ is a reference to artefact A.}, this stage consists of four activities that are performed to define and validate the operating context for an AS. The artefacts generated from this stage are used to instantiate the AS operating context assurance argument pattern as part of Activity~\ref{act:contextPattern}.  

Additional guidance on assuring the safe operating context for autonomous systems can be found at~\cite{AAIP-BoK}.

\Activity{Define Autonomous Capabilities of AS \label{art:D} \ArtD}{act:capability}

In defining the context of operation of an AS it is important to first define the scope of the system's autonomous capabilities. It should be noted that this will in many cases be a sub-set of the full capability of a system where only some of the functionality is provided autonomously, or the system is only capable of performing tasks autonomously some of the time.

\begin{note}
Autonomous capabilities of an AS may be provided entirely by the system itself or with shared responsibility between a human, or humans, and the AS. This includes systems where a human is required to monitor the autonomous operation of the AS and intervene if required. Again the limitations on such capabilities should be clearly defined along with the separation of responsibilities.
\end{note}


\begin{example}
In defining the autonomous capability of an autonomous robot deployed for delivering small packages around an office environment it is specified that the robot will not be capable of autonomously loading/unloading the packages. A human operator is relied on for the safe loading of packages.
\end{example}

\begin{example}
{The autonomous capabilities of a self-driving car vary for different levels of driving automation. For example an SAE level 3 vehicle \cite{standard2018j3016} may provide autonomous traffic jam pilot capabilities under defined conditions, whereas a level 4 vehicle may provide fully autonomous driving capability under defined conditions.}
\end{example}

\begin{note}
The definition of autonomous capabilities of the AS will often be an iterative process with activity \ref{act:odm} since the capabilities must be chosen in order to match the requirements of a given ODM.
\end{note}

\subsection*{Artefact \ArtA: AS Concept Definition\label{art:A}}

The required autonomous capabilities shall be informed by an understanding of the high-level user requirements and objectives of the AS operation. This AS concept definition should be explicitly documented and agreed with the relevant stakeholders. This could, for example, include a set of use case descriptions for the AS.

\Activity{Define and Validate the Operational Domain Model (ODM) \label{art:B} \ArtB \label{art:C} \ArtC}{act:odm}

\begin{figure}[h]
    \centering
    \includegraphics[width=0.7\linewidth]{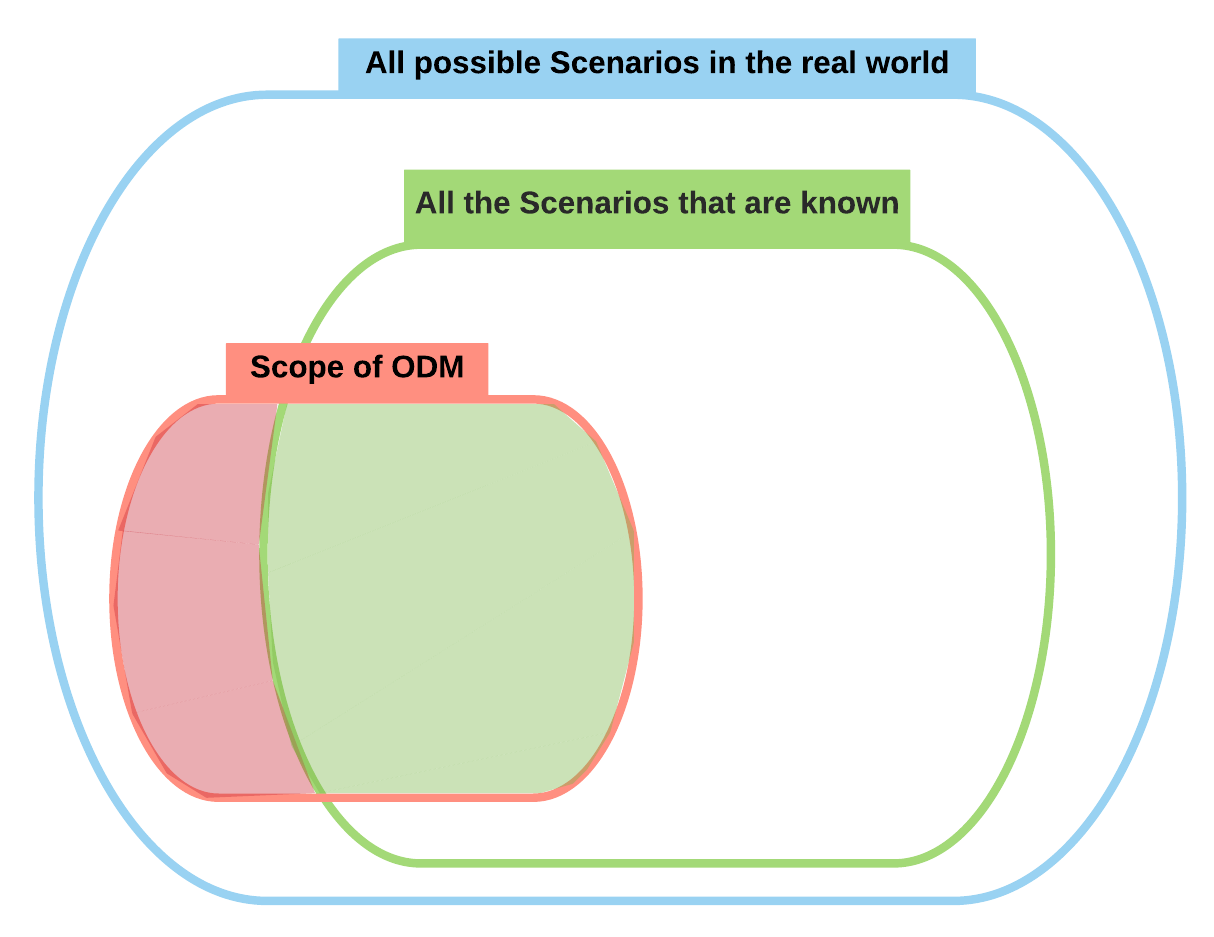}
    \caption{Defining an Operational Domain Model}
    \label{fig:odm}
\end{figure}

The Operational Domain Model (ODM) defines the scope of the operation of an AS within which the AS can be demonstrated to be acceptably safe when carrying out its autonomous capability. When AS are developed and put into operation it is based on assumptions made about when, where and under what conditions that AS will need to operate. The operational scope defined by those assumptions determines the scenarios that may be encountered by the AS as it interacts with its environment. It is illustrated in Figure \ref{fig:odm} how the ODM effectively reduces the number of possible scenarios that the AS may need to deal with when performing its autonomous capability.

The assumptions must be made explicit in the form of a defined ODM that characterises the operating domain. If the ODM is insufficiently defined then the AS may encounter scenarios during its operation that were not considered during the development of the system, and which could therefore be unsafe and for which no assurance is provided. It is crucial therefore that all relevant elements of the operating domain, including those with with the AS may interact are included within the ODM. In identifying relevant elements of the ODM it is important that ``non-mission interactions'' are also considered \cite{harper2021towards}.

\begin{figure}[h]
    \centering
    \includegraphics[width=1\linewidth]{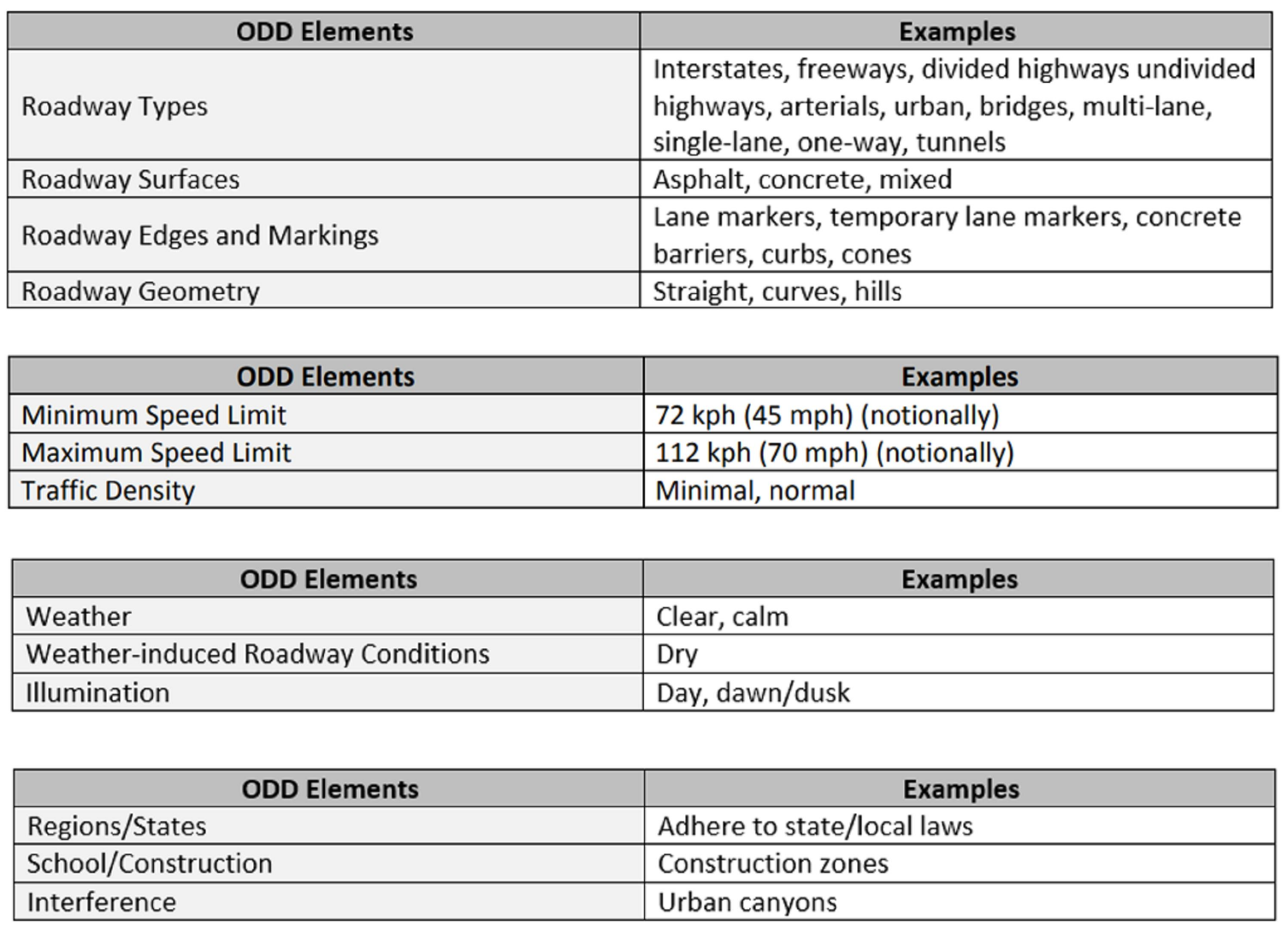}
    \caption{Example baseline ODD for HWD function (from \cite{thorn2018framework})}
    \label{fig:oddEx}
\end{figure}

\begin{example}
For autonomous cars, the term Operational Design Domain (ODD) \cite{koopman2019many} is often used to refer to the ODM. A number of approaches to defining the ODD for an autonomous car have been proposed. NHTSA \cite{thorn2018framework} defined and categorised an ODD taxonomy based on a review of over 50 sources of literature in automotive and other domains. The taxonomy is intended to be descriptive, recognising that other organisations of the elements are possible. The report provides a set of sample baseline ODDs for different automated driving features or capabilities, such as those shown in Figure \ref{fig:oddEx} for an Automated Highway Drive (HWD) function.
\vspace{0.5em}

Note that the examples in Figure \ref{fig:oddEx} are used to define what is asserted as being within the ODM for that particular capability (function). If the operation of the vehicle is outside of this definition then the safety of that capability is no longer assured.\footnote{We return to the issue of operation outside of the defined operating context at Stage~\stageref{2}.}
\end{example}

\begin{note}
One of the major challenges when defining an ODM can be identifying an appropriate level of detail at which to specify elements of the ODM. If elements are not specified with sufficient granularity then important differences in the features that may impact on safety may be missed. On the other hand, too much granularity can make any analysis based on the ODM intractable due to the high number of element combinations.

It may be necessary to revisit the granularity of the ODM as a result of information learned from other activities; for example it may be informed by the nature of the relevant operating scenarios identified at activity \ref{act:os}.
\end{note}

\begin{example}
In the ODM for an autonomous robot operating in an office environment, the ODM may include `walls' or `doors' as part of the model. However, relying only on `wall' or `door' as an element of an ODM will miss crucial attributes of walls or doors that will affect the detection capability of an AS. A wall or door made of translucent  material would not be capable of detection by some technologies (e.g. LiDAR). In the ODM, the material of the walls and doors should also be specified.
\end{example}

\begin{example}
In the ODM for an autonomous shuttle bus operating in pedestrianised areas, the ODM must include people as part of the model. However defining ``people'' as a feature of the ODM may miss crucial differences between children and adults that affect the safety of the AS operation. ``Children'' and ``adults'' should therefore be defined as separate elements of that particular ODM. The appropriate elements to choose and their granularity would be expected to differ for each ODM depending upon the particular operational context of the system.
\end{example}


Once created the ODM shall be validated to check that both the scope of the defined model and its level of detail is appropriate. The results of the validation activity shall be explicitly documented (\ArtC).

The process of defining and validating the ODM is an iterative process, and any initial ODM specification may need to be expanded or reduced based upon the outputs of later process activities. This may include new information that is obtained about the design, operation, capabilities, limitations and/or potential failure conditions of the AS.

\Activity{Define and Validate the Operating Scenarios \label{art:E} \ArtE \label{art:F} \ArtF}{act:os}

An operating scenario of an AS represents a particular set of actions and events that the AS may undertake in a defined environment situation. We adopt the characterisation used in \cite{de2020ontology} as a description of the AS, its activities and/or goals, its static environment, and its dynamic environment.

\begin{note}
The following distinction is often made between ``scenes'' and ``scenarios'' \cite{ulbrich2015defining}:
\begin{itemize}
    \item A scene describes a snapshot of the environment including the scenery and dynamic elements, as well as all actors’ and observers’ self-representations, and the relationships among those entities. Scenes are sometimes also called ``situations''.
    \item A scenario describes the temporal development between several scenes in a sequence. Every scenario starts with an initial scene, and the temporal development is characterised by a set of actions and events.  
\end{itemize}
\end{note}

\begin{example}
An example of a scene could be a typical office environment, with office furniture, artificial lighting, and an autonomous robot in a corridor that is also used by human staff members. A scenario could develop where an autonomous robot and a human are approaching a right-angle in a corridor from opposite directions.
\end{example}

\begin{figure}[h]
    \centering
    \includegraphics[width=0.5\linewidth]{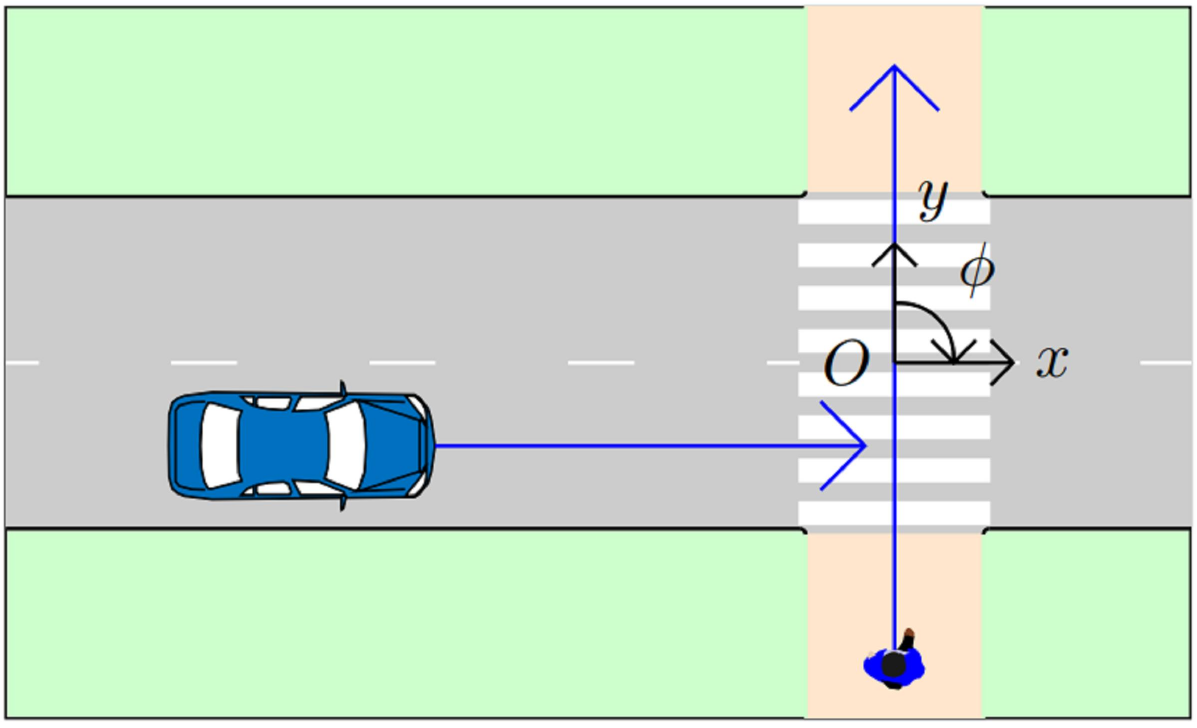}
    \caption{Schematic overview of the scenario of both an autonomous vehicle and a pedestrian approaching a non-signalized pedestrian crossing (from \cite{de2020ontology}).}
    \label{fig:scenExIm}
\end{figure}

\begin{figure}[h]
    \centering
    \includegraphics[width=1\linewidth]{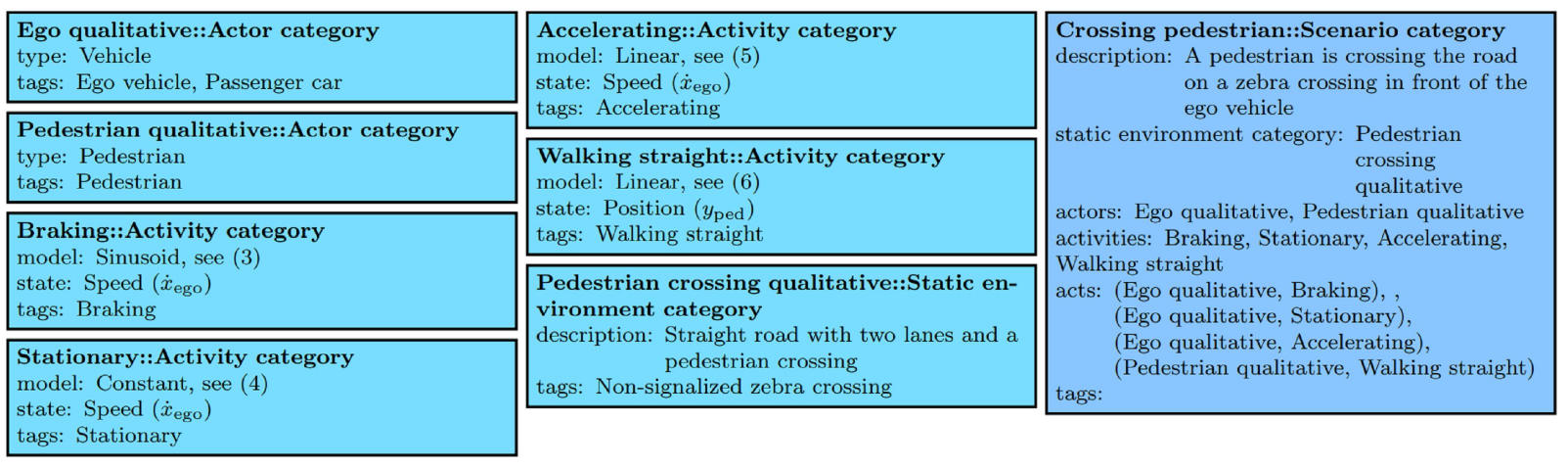}
    \caption{A qualitative description of the same scenario (from \cite{de2020ontology}).}
    \label{fig:scenExDes}
\end{figure}

\begin{example}
Figures \ref{fig:scenExIm} and \ref{fig:scenExDes}, taken from \cite{de2020ontology} provide an example of a scenario description for an autonomous vehicle.
\end{example}

It is crucial for safety assurance of the AS that the operating scenarios that are relevant to the AS within its defined ODM are identified as completely and correctly as possible. If we refer to Figure \ref{fig:odm}, we must identify the operating scenarios present in the area inside the orange boundary defining the scope of the ODM. The green area in the diagram represents all of the identified scenarios within this area. It is not possible for a complex, open environment, such as an urban street, to provide an exhaustive detailed specification of the operating scenarios, since the set of scenarios is simply too large. 

Where operating scenarios exist within the scope of the ODM, but are not correctly identified, these could pose a safety risk to the AS since the scenario could be hazardous, but would not be assessed as part of the safety assurance process and therefore no mitigations put in place. This is represented by the red area in Figure \ref{fig:odm}. It is very important therefore for safety assurance to have confidence that this area is as small as possible by ensuring the operating scenarios are sufficiently well defined.

\begin{note}
It is partly as a result of the possible existence of the red region in Figure \ref{fig:odm} that resilience is such an important requirement when designing ASs; these unanticipated operating scenarios require that the system is able to adapt in a safe manner to unplanned events. This is discussed in more detail at Stage~\stageref{6}.  
\end{note}

The operating scenario descriptions shall be validated to provide evidence that they are sufficiently complete and correct. Evidence can be provided based upon review of the scenario specification. A rigorous specification to a defined format makes the scenarios more amenable to review and reduces ambiguity. It is important that review is carried out by a range of experienced stakeholders. Providing simulations of the defined scenarios may help to ensure the reviewers have a correct understanding of the scenarios. The defined scenarios can also be checked against collected field data from AS in operation to ensure that all encountered scenarios are captured in the operating scenario specification. The results of the data validation activity shall be explicitly documented (\ArtF).

The definition of scenarios should be an iterative process, where the scenarios are refined based on increased understanding of the relevant and important aspects of the scenario space. This refinement may be informed by the outcome of the validation activities described above, particularly the results of simulation and field-based validation tests. 

\Activity{Instantiate AS Operating Context Assurance Argument Pattern \label{art:H} \ArtH}{act:contextPattern}

This activity requires as input the AS operating context assurance argument pattern (\ArtG), as well as the artefacts from Activities~\actref{act:odm}{1}, ~\actref{act:capability}{2} and ~\actref{act:os}{3} (\ArtB, \ArtC, \ArtD, \ArtE and \ArtF). The activity uses these artefacts to create an instantiated AS operating context assurance argument (\ArtH) which demonstrates that the defined ODM is sufficient to support the safe operation of the AS.

\subsection*{Artefact \ArtG: AS Operating Context Assurance Argument Pattern}\label{art:G}

\begin{figure}[h]
    \centering
    \includegraphics[width=1.1\linewidth]{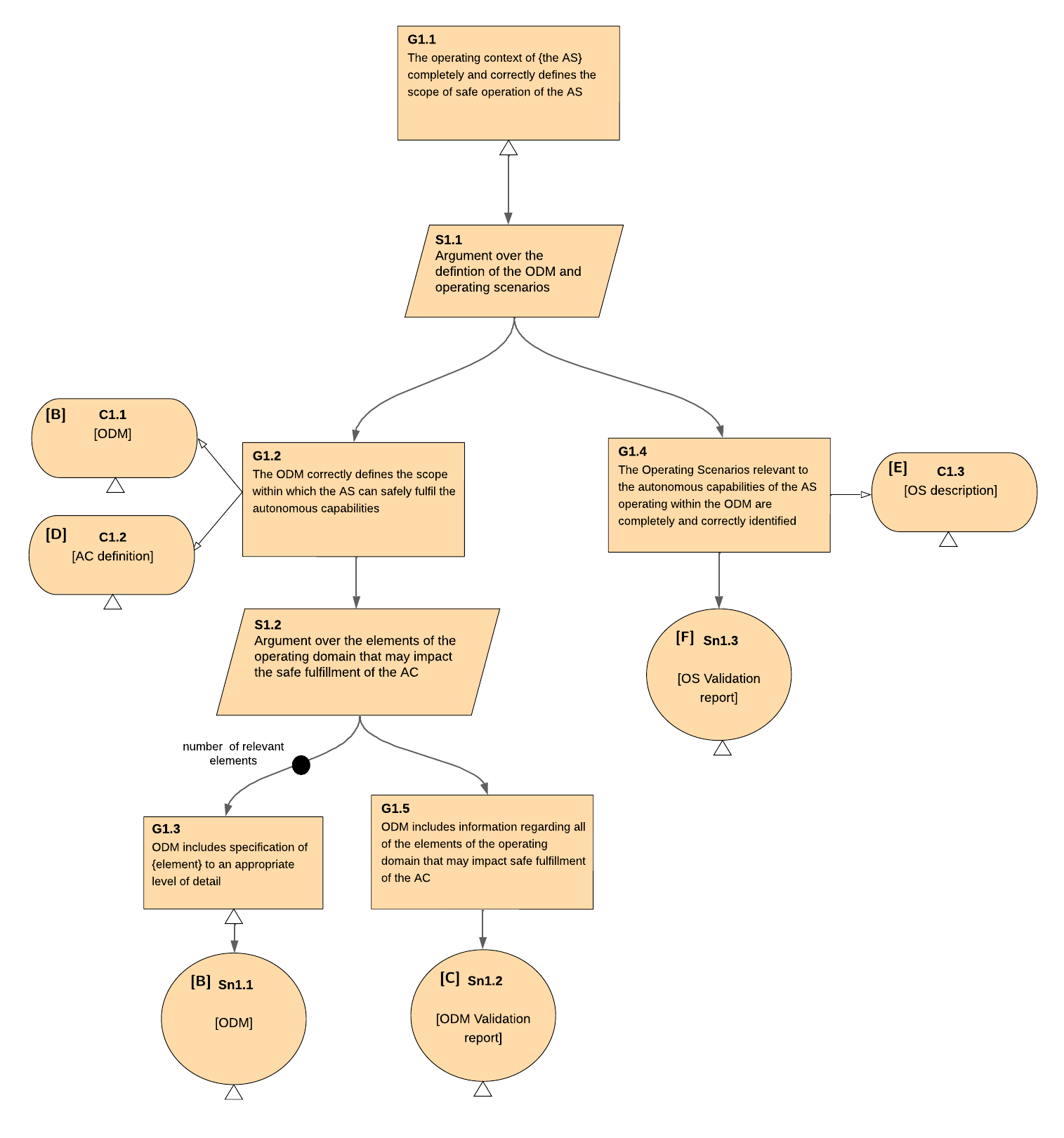}
    \caption{\ArtG~:~Argument Pattern for AS Operating Context Assurance} 
    \label{fig:ArgumentScoping}
\end{figure}

The argument pattern relating to this stage is shown in Figure~\ref{fig:ArgumentScoping} and key elements from the pattern are described in the following sections.

\subsection*{G1.1}
The top claim in this argument pattern is that the defined operating context represents a complete and correct definition of the scope within which safe operation of the AS is assured. The argument to support this must consider both the ODM (G1.2) as well as the defined operating scenarios (G1.4).

\subsection*{G1.2}
Regarding the ODM, it must be demonstrated that, as defined, the ODM supports the AS fulfilling its autonomous capabilities in a safe manner. That is to say that those capabilities can be safely carried out within the entirety of the defined ODM. The ODM (\ArtB) and the Autonomous Capabilities Definition (\ArtD) provide the context to this claim. The claim is supported by arguing over the features that are included as part of the ODM. It is demonstrated that all of the features that could impact on the ability of the AS to safely perform its required capabilities have been identified (G1.5). The ODM Validation Report (\ArtC) is used as evidence to support this claim. Then for each of these features it is demonstrated that this has been included in the ODM specification with an appropriate level of detail (G1.3). 

\subsection*{G1.4}

It must be demonstrated that all of the operating scenarios that are relevant to performing the autonomous capabilities in the ODM (provided as context through artefact \ArtE) have been identified. The OS Validation report (\ArtF) provides evidence for this.
\clearpage
\stage{AS Hazardous Scenarios Identification}

\subsection*{Objectives}
\begin{enumerate}
\item Identify and define potentially hazardous scenarios for the AS.
\item Validate the AS Hazardous Scenarios.
\item Create the AS Hazardous Scenarios Assurance Argument.
\end{enumerate}

\subsection*{Inputs to the Stage}
\begin{itemize}
\item[\ArtB]: Operational Domain Model
\item[\ArtE]: Operating Scenarios Definition
\item[\ArtI]: AS Hazardous Scenarios Assurance Argument Pattern
\end{itemize}

\subsection*{Outputs of the Stage}
\begin{itemize}
\item[\ArtAW]: AS Decision Analysis Report
\item[\ArtAX]: AS Hazardous Scenarios Definition
\item[\ArtAY]: AS Hazardous Scenarios Validation Report
\item[\ArtJ]: AS Hazardous Scenarios Assurance Argument
\end{itemize}

\subsection*{Description of the Stage}
Hazardous scenarios are those scenarios that the AS may encounter during its operation that could, under certain conditions, lead to an unsafe outcome. For AS we focus in particular on the interactions between the AS and elements of the operating environment, and on the decisions that are made by the AS as part of its autonomous capability. The hazardous scenario for the AS should therefore be described using the general form:

\textbf{\emph{<AS operating scenario><relevant environment state(s)> AND <decision>}}, where:

\begin{itemize}
    \item An AS Operating Scenario describes what the AS is undertaking (identified from \ArtE)
    \item A Relevant Environment State is one or more states occurring within or emanating from the operating environment. Environment states may be identified by considering elements of the ODM (\ArtB).
    \item The decision is the selected course of action as a result of the Operating Scenario and relevant environment state(s).
\end{itemize}

\begin{figure}[h]
    \centering
    \includegraphics[width=\linewidth]{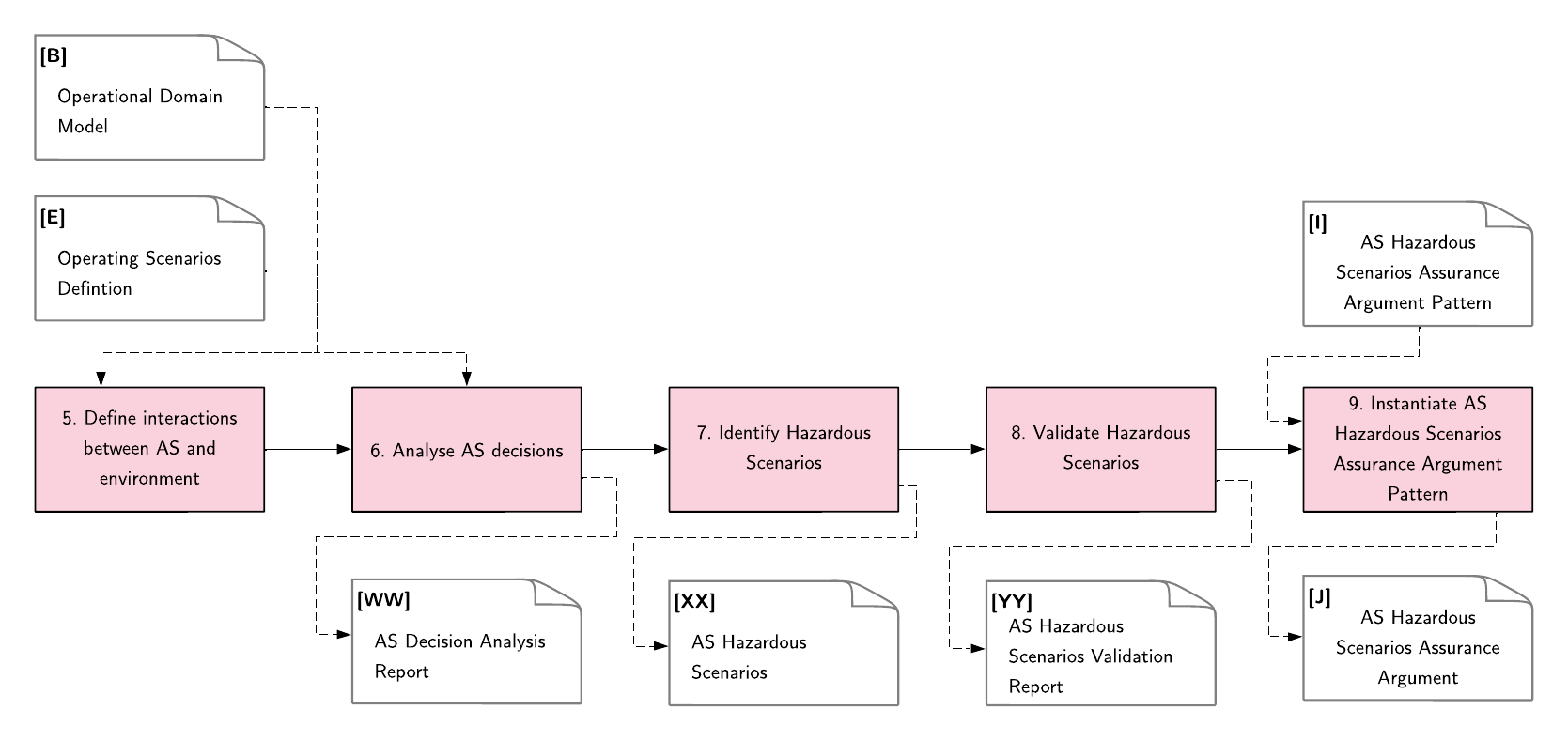}
    \caption{AS Hazardous Scenarios Assurance Process}
    \label{fig:ASHazScen}
\end{figure}

As shown in Figure \ref{fig:ASHazScen}, this stage consists of  activities that are performed to identify and validate the potentially hazardous scenarios associated with the operation of the AS.  The artefacts generated from this stage are used to instantiate the AS hazardous scenarios assurance argument pattern as part of Activity~\ref{act:ASHazpattern}.

\begin{note}
This guidance focuses specifically on hazardous scenarios related to the deployment of autonomous technology. It is assumed that consideration of the hazardous scenarios associated with the more conventional (non-autonomous) aspects of a system are considered concurrently and additionally to this.
\end{note}



\begin{figure}[h]
    \centering
    \includegraphics[width=\linewidth]{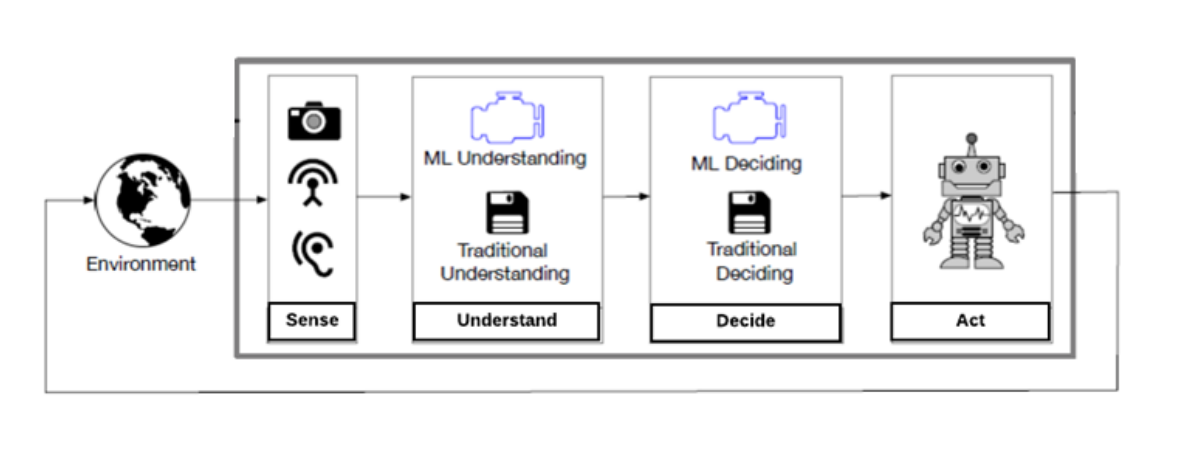}
    \caption{SUDA agent model of an AS}
    \label{fig:SUDA}
\end{figure}

\begin{note}
It is clear that understanding the decisions that may need to be taken by an AS during its operation is crucial to identifying the potential hazardous scenarios. It is important therefore to understand what is meant by `decision-making' for an AS. The decision determines which action the AS should take in any given situation, and an incorrect decision can lead directly to an unsafe action. In order to make a decision, an AS must understand the state of the environment and the system, as shown in figure \ref{fig:SUDA}.

It can sometimes be difficult to differentiate between `Understanding' and `Deciding'. For example, in the case of a mobile robot, the act of detecting a static object represents `understanding', whereas the decision is on whether moderation of speed or course alteration is required (we would not therefore characterise the identification of the object as a `decision' in this case). In the case of an autonomous medical device the act of classifying patient vital signs represents `understanding', the decision is whether to increase medication or not (again, the classification of vital signs would not be characterised as a `decision' in this case). Note here that a decision NOT to increase medication could in itself lead to harm due to delayed treatment. 

In both of these cases, although incorrect understanding may be a causal factor, it is the decision that is made that ultimately determines if the outcome of a scenario is safe or not.
\end{note}

\Activity{Define interactions between the AS and the environment}{act:ASHazInt}  
  
Correctly understanding the interactions that the AS has with elements of the operating environment is a crucial part of identifying the hazardous scenarios.  This activity focuses on defining those interactions.

A complex operating environment can have a large impact on any hazardous behaviour of an AS, since it increases the probability of the system encountering unusual and unanticipated scenarios. Although complex environments are not unique to AS, traditional systems often rely on the involvement of a human as an operator of the system to deal with any resulting unanticipated or unusual scenarios. For autonomous operation, a human operator is not available to deal with these scenarios, so the AS itself must be able to deal safely with such situations so that they do not become hazardous (through inappropriate decision making). This requires therefore that analysis of hazardous scenarios for an AS incorporates consideration of the operating environment in a more systematic and explicit manner than is currently the case for human-controlled systems.

The ODM (\ArtB) is a key input used to identify the interactions that the AS may encounter during operation .

\begin{note}
The consideration of interactions is not limited solely to the interactions between an AS and other agents explicitly required for the purpose of carrying out a task (sometimes referred to as `mission interactions'). Consideration must also be given to unexpected interactions with features such as the type of terrain encountered, as well as with unexpected agents that are not ordinarily involved in the task (`non-mission interactions') \cite{harper2021towards}.
\end{note}

\begin{example}
\begin{itemize}
\item The ODM for an Autonomous Robot operating in an office environment may identify possible interactions with office workers, visitors, other robots etc.
\item The ODM for an Autonomous Car may identify possible interactions with different types of road surfaces and weather conditions etc.
\item The ODM for an Autonomous Insulin Infusion Pump may identify possible interactions with different healthcare professionals (e.g. doctors, nurses), as well as the patient, other medical devices etc.
\end{itemize}
\end{example}

Having identified the elements of the ODM with which the AS may potentially interact, the defined operating scenarios (\ArtE) can then be used to identify the particular interactions that may occur with those elements. This requires each scenario to be broken down into a set of steps that are undertaken by the AS as described below:

\begin{enumerate}
    \item \textbf{Define start and end points of the scenario}: identify the logical start/end points (the former may often be the event that triggers the scenario, and the latter asserted as the point at which the scenario is successfully completed).
    \item \textbf{Define the steps}: establish the steps involved in undertaking the scenario. Some of these steps may represent an `understanding point' in the scenario where the AS requires information about a particular feature of the operating environment. These understanding points should be captured as questions with a binary (yes/no) output, such as ``is an obstacle present in the planned path of the robot''. Other steps may represent a `decision point' where the AS must make a choice, such as to reduce speed.
    \item \textbf{Identify the Interactions}: Determine which of the defined steps may involve interaction with elements of the operating environment defined in the ODM.

\end{enumerate}

\Activity{Analyse AS Decisions \label{art:WW} \ArtAW}{act:ASHazDec} 

Having modelled the steps and interactions of the operating scenario, each of the identified decision points can then be analysed in order to determine the nature of any hazardous scenarios that may arise. 

Each of the decision points identified in Activity \ref{act:ASHazInt} should be selected. For each, the possible environmental states that may arise within or emanate from the operating environment are identified before enumerating the decisions that the AS could select at that decision point.
    
\begin{example}
A decision point for an autonomous robot moving in a building may correspond to the detection of an object in the robot's path.  The options available to the AS at this point are:
\begin{enumerate}
    \item Continue on the current path at the current speed, OR
    \item Continue on the current path at a reduced speed, OR
    \item Change path to avoid the object, OR
    \item Stop and wait.
\end{enumerate}

\end{example}
    
The different possible scenarios and environmental states that relate to the decision can then be identified by considering the real world state and the belief state of the AS at the point at which the decision is made, along with each of the possible options. The real world state represents the actual state of the operating environment, whilst the belief state represents the understanding that the AS has of the state of the operating environment.
    
\begin{example}
    An example of some enumerated situations for an autonomous robot encountering an object in its path are shown in Table \ref{SitRobPath}. The real world and belief states are represented as Boolean states where `True' represents the presence of the object and `False' represents that the object is not present. So situation 5 in Table \ref{SitRobPath} represents the situation where there is an object in the path of the robot, but the robot is unaware of the object and continues on its current path at its current speed.
\end{example}

\begin{table}[]
\centering
\begin{tabular}{|cccll|}
\hline
\multicolumn{5}{|l|}{\textbf{Operating Scenario: Autonomous Robot Following Planned Path}} \\ \hline
\multicolumn{5}{|l|}{\textbf{Decision: How to respond when an object is present on the planned path}} \\ \hline
\multicolumn{1}{|c|}{\textbf{\begin{tabular}[c]{@{}c@{}}Potentially\\ Hazardous\\ Scenario\end{tabular}}} & \multicolumn{1}{c|}{\textbf{\begin{tabular}[c]{@{}c@{}}Real \\ World\\ Environment\\ State\end{tabular}}} & \multicolumn{1}{c|}{\textbf{\begin{tabular}[c]{@{}c@{}}System\\ Environment\\ Belief\\ State\end{tabular}}} & \multicolumn{1}{l|}{\textbf{Option}} & \textbf{Outcome} \\ \hline
\multicolumn{1}{|c|}{1} & \multicolumn{1}{c|}{\multirow{4}{*}{T}} & \multicolumn{1}{c|}{\multirow{4}{*}{T}} & \multicolumn{1}{l|}{\begin{tabular}[c]{@{}l@{}}1. Continue on Current\\ Path at Current Speed\end{tabular}} & \begin{tabular}[c]{@{}l@{}}Hazardous - \\ Collision\end{tabular} \\ \cline{1-1} \cline{4-5} 
\multicolumn{1}{|c|}{2} & \multicolumn{1}{c|}{} & \multicolumn{1}{c|}{} & \multicolumn{1}{l|}{\begin{tabular}[c]{@{}l@{}}2. Continue on Current\\ Path at Reduced Speed\end{tabular}} & \begin{tabular}[c]{@{}l@{}}Hazardous - \\ Collision\\ Reduced Severity\end{tabular} \\ \cline{1-1} \cline{4-5} 
\multicolumn{1}{|c|}{3} & \multicolumn{1}{c|}{} & \multicolumn{1}{c|}{} & \multicolumn{1}{l|}{\begin{tabular}[c]{@{}l@{}}3. Change Path to \\ Avoid Breach of \\ Safe Separation \\ Minima\end{tabular}} & Safe \\ \cline{1-1} \cline{4-5} 
\multicolumn{1}{|c|}{4} & \multicolumn{1}{c|}{} & \multicolumn{1}{c|}{} & \multicolumn{1}{l|}{4. Stop and Wait} & \begin{tabular}[c]{@{}l@{}}Possible\\ Obstruction\\ Hazard\end{tabular} \\ \hline
\multicolumn{1}{|c|}{5} & \multicolumn{1}{c|}{\multirow{4}{*}{T}} & \multicolumn{1}{c|}{\multirow{4}{*}{F}} & \multicolumn{1}{l|}{\begin{tabular}[c]{@{}l@{}}1. Continue on Current\\ Path at Current Speed\end{tabular}} & \begin{tabular}[c]{@{}l@{}}Hazardous - \\ Collision\end{tabular} \\ \cline{1-1} \cline{4-5} 
\multicolumn{1}{|c|}{6} & \multicolumn{1}{c|}{} & \multicolumn{1}{c|}{} & \multicolumn{1}{l|}{\begin{tabular}[c]{@{}l@{}}2. Continue on Current\\ Path at Reduced Speed\end{tabular}} & \begin{tabular}[c]{@{}l@{}}Hazardous - \\ Collision\\ Reduced Severity\end{tabular} \\ \cline{1-1} \cline{4-5} 
\multicolumn{1}{|c|}{7} & \multicolumn{1}{c|}{} & \multicolumn{1}{c|}{} & \multicolumn{1}{l|}{\begin{tabular}[c]{@{}l@{}}3. Change Path to \\ Avoid Breach of \\ Safe Separation \\ Minima\end{tabular}} & \begin{tabular}[c]{@{}l@{}}Not possible\\ N/A\end{tabular} \\ \cline{1-1} \cline{4-5} 
\multicolumn{1}{|c|}{8} & \multicolumn{1}{c|}{} & \multicolumn{1}{c|}{} & \multicolumn{1}{l|}{4. Stop and Wait} & \begin{tabular}[c]{@{}l@{}}Possible\\ Obstruction\\ Hazard\end{tabular} \\ \hline
\multicolumn{1}{|c|}{9} & \multicolumn{1}{c|}{F} & \multicolumn{1}{c|}{F} & \multicolumn{1}{l|}{\begin{tabular}[c]{@{}l@{}}1. Continue on Current\\ Path at Current Speed\end{tabular}} & Safe
\end{tabular}
\caption{Extract from a Table showing the analysis of an AS decision, for the purpose of identifying potentially hazardous scenarios}
\label{SitRobPath}
\end{table}

For each of the identified scenarios, the outcome can be defined in order to identify those that may be potentially hazardous.

\begin{example}
For an autonomous car turning right at a roundabout,one decision point is the car entering the roundabout. The options for the car at this decision point are:
\begin{enumerate}
    \item Enter the roundabout
    \item Stop and wait.
\end{enumerate}
A possible interaction at this point is with another road user, in this example a cyclist. Relevant scenarios can be identified by considering the real world presence of the cyclist in combination with the car's belief that a cyclist is present, and each of the options identified above. For example, one situation is that a cyclist is not present, but the car believes that there is a cyclist and decides to stop and wait. This scenario could lead to a hazardous outcome as an unnecessary and unexpected stop could potentially cause a rear-end collision. \footnote{Although the severity of this situation would be less than a situation that resulted in the potential for impact with the cyclist.}
\end{example}

\begin{example}
For an autonomous insulin pump that is monitoring a patient's blood sugar level, one decision point is whether to alter the infusion rate to the patient. The options for the pump at this decision point are:
\begin{itemize}
    \item Increase the insulin infusion
    \item Decrease the insulin infusion
    \item Maintain the current rate of insulin infusion
    \item Stop the insulin pump
\end{itemize}
Relevant scenarios can be identified by considering the real world change in the patient's blood sugar level in combination with the pump's belief in the current sugar level, and each of the options identified above. For example, one scenario is that the patient's blood sugar level rises, the pump also has a belief state that the blood sugar level has risen and yet decides to maintain the current rate of insulin infusion. This scenario could lead to a hazardous outcome as the patients blood sugar level could increase to unsafe levels.
\end{example}

Considering the possibly hazardous scenarios for an autonomous robot encountering an object in its path in Table \ref{SitRobPath}, it is possible to also assign a severity of outcome, and consider whether minor or serious injuries (or even fatalities) are likely.

It is also possible to further consider how this severity could be impacted by factors pertaining to the operating enviornment.

\begin{example}
An autonomous robot moving in a building that fails to detect an object in its path, and therefore continues on the current path at the current speed, could impact a static object.  If the static object is an adult human, the severity of the impact could be minor (bruise or laceration).  Should the static object be a small child, then the severity of the impact could be major (broken bones).

The robot may have detected the static object, but decided to maintain its current path, but reduce speed. Wet floor surfaces may reduce the effectiveness in braking (through wheel slippage), and this higher then expected collision speed could impact the severity of the collision.
\end{example}

\begin{example}
For an autonomous car turning right at a roundabout, the autonomous car may fail to detect a dynamic object that would intersect it should it enter the roundabout.  If this dynamic object is a car, and was collided into by the autonomous vehicle, the occupants may suffer only minor injuries (if any at all).  Should the dynamic object be a cyclist, then the collision may result in a fatality.
Should the road surface be icy, the force of impact may displace an impacted car onto a fixed object in the environment, thereby increasing the severity of the collision for the car's occupants.
\end{example}

\begin{example}

For an autonomous insulin pump that is monitoring a patient’s blood sugar level, the autonomous infusion pump could decrease the insulin infusion without medical need, should the patient be experiencing a spike in blood sugar levels at the time the infusion was reduced, then the patient may become hyperglycaemic.  Should the patient have additional health complications, or the reduction in insulin infusion go unobserved by a clinician, then the severity could be impacted and result in a fatality. 
\end{example}

The analysis undertaken in this activity shall be documented (\ArtAW).

\Activity{Identify AS Hazardous Scenarios \label{art:XX} \ArtAX}{act:ASHazScen} 

Based upon the analysis performed in Activity \ref{act:ASHazDec} it is possible to define the hazardous scenarios associated with the operation of the AS in the defined operating context. These may be specified using the general form presented earlier:

\textbf{\emph{<AS operating scenario><relevant environment state(s)> AND <decision>}}

\begin{example}
\begin{itemize}
    \item An AS Hazardous Scenario for an autonomous robot is: \textbf{<}The autonomous robot is following a planned path\textbf{><} with a static object present in the path\textbf{>} \textbf{AND <} the robot maintains its current speed and direction\textbf{>}
    \item An AS Hazardous Scenario for an autonomous car is: \textbf{<}the car is approaching a roundabout\textbf{><}with a cyclist on the roundabout to the vehicle's right\textbf{> AND <}the car enters the roundabout\textbf{>}
    \item An AS Hazardous Scenario for an autonomous insulin infusion pump is: \textbf{<}the insulin pump is monitoring the patient's blood sugar level \textbf{><} when the sugar level rises\textbf{>} \textbf{AND <} the pump maintains the current insulin infusion rate\textbf{>}
\end{itemize}
\end{example}

\begin{note}
The System Environmental Belief State is NOT an element of what constitutes a Hazardous Scenario, as belief states are merely hypothesised causes of a failure.
\end{note}

These AS Hazardous Scenarios shall be explicitly documented (\ArtAX)

\Activity {Validate the AS Hazardous Scenarios \label{art:YY} \ArtAY}{act:ASHazVal} 

The AS Hazardous Scenarios documented in \ArtAX~shall be validated in order to check that:

\begin{itemize}
    \item All AS Hazardous Scenarios have been identified
    \item The analysis of hazardous scenarios included consideration of all the relevant:
        \subitem Agents
        \subitem ODM Elements
        \subitem Decision points and their associated outcomes
    \item Each AS Hazardous Scenario is unambiguously specified in sufficient detail to support the elicitation of mitigating safety requirements (see Stage~\stageref{3}).
    \end{itemize}

There are a number of approaches that can be used to validate the AS Hazardous Scenarios. One approach is for an independent review of the hazardous scenarios defined (\ArtAX) and their rationale (\ArtAW) to be undertaken by a suitably qualified and experienced person. Another approach is to perform additional analysis to corroborate or identify problems in the existing analysis. This could for example include simulation of operating scenarios in order to explore different possible situations, decisions and outcomes.

The results of the validation activities shall be documented (\ArtAY).

\Activity{Instantiate AS Hazardous Scenarios Assurance Case Pattern \label{art:J} \ArtJ}{act:ASHazpattern}

This activity requires as input the AS hazardous scenarios assurance argument pattern (\ArtI), as well as the artefacts from Activities~\actref{act:ASHazInt}{5} to~\actref{act:ASHazVal}{8} (\ArtAW, \ArtAX, \ArtAY). The activity uses these artefacts to create an instantiated AS hazardous scenarios assurance argument (\ArtJ) which demonstrates that the hazardous scenarios have been correctly identified.

\subsection*{Artefact \ArtI: AS Hazardous Scenarios Assurance Case Pattern \label{art:I}}

\begin{figure}[h]
    \centering
    \includegraphics[width=1\linewidth]{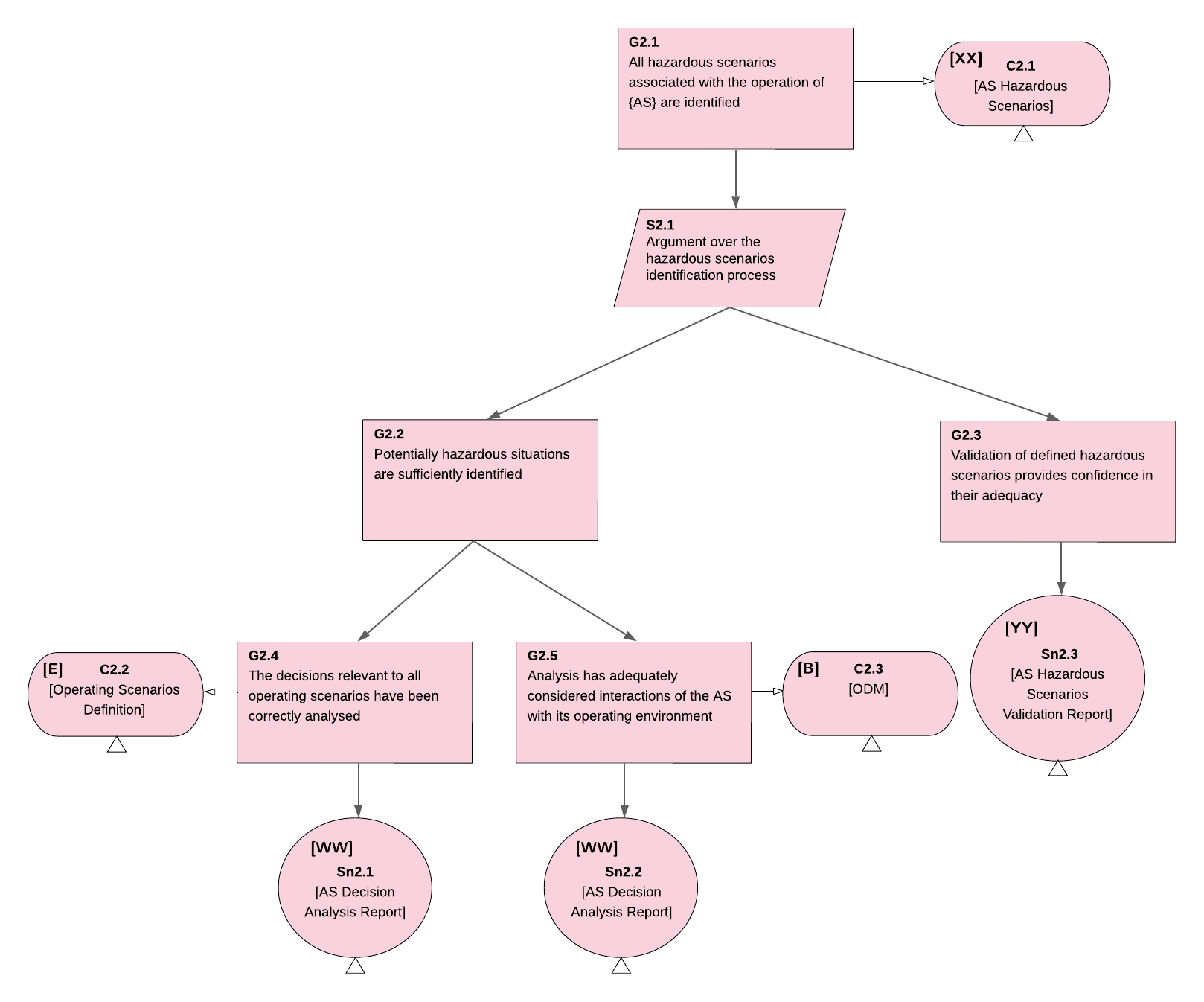}
    \caption{\ArtI~:~Argument Pattern for AS Hazardous Scenarios Assurance} 
    \label{fig:ArgumentHazards}
\end{figure}

The argument pattern relating to this stage is shown in Figure~\ref{fig:ArgumentHazards} and key elements from the pattern are described in the following sections.

\subsection*{G2.1} This argument demonstrates that the hazardous scenarios relating to the operation of the AS have been sufficiently identified. This is done by considering the way in which the hazardous scenarios were identified (G2.2), and by validating those identified scenarios (G2.3).

\subsection*{G2.2} The hazardous scenarios have been identified using a process that identifies and analyses the decisions made by the AS and the interactions between the AS and its operating environment. The argument therefore demonstrates that all the relevant decisions have been correctly analysed (G2.4), and that the interactions with the environment have been adequately considered as part of that analysis (2.5). The evidence for both of these claims should be provided by the analysis report (\ArtAW). These claims are made in the context of the defined operating scenarios (\ArtE) and ODM (\ArtB).

\subsection*{G2.3} Confidence in the identified scenarios is provided through the validation activities performed at Activity \ref{act:ASHazVal}. The validation report (\ArtAY) is used as evidence of the adequacy of the hazardous situations identified.
\clearpage
\stage{Safe Operating Concept Assurance}

\subsection*{Objectives}
\begin{enumerate}
\item Define the Safe Operating Concept for the AS.
\item Validate the Safe Operating Concept.
\item Create the Safe Operating Concept Assurance Argument.
\end{enumerate}

\subsection*{Inputs to the Stage}
\begin{itemize}
\item[\ArtB]: Operational Domain Model
\item[\ArtD]: Autonomous Capabilities Definition
\item[\ArtE]: Operating Scenarios Definition
\item[\ArtAX]: AS Hazardous Scenarios
\item[\ArtK]: Definition of sufficiently safe
\item[\ArtN]: SOC Assurance Argument Pattern
\end{itemize}

\subsection*{Outputs of the Stage}
\begin{itemize}
\item[\ArtL]: Safe Operating Concept Definition
\item[\ArtM]: SOC Justification Report
\item[\ArtO]: SOC Assurance Argument
\end{itemize}

\subsection*{Description of the Stage}

\begin{figure}[h]
    \centering
    \includegraphics[width=1\linewidth]{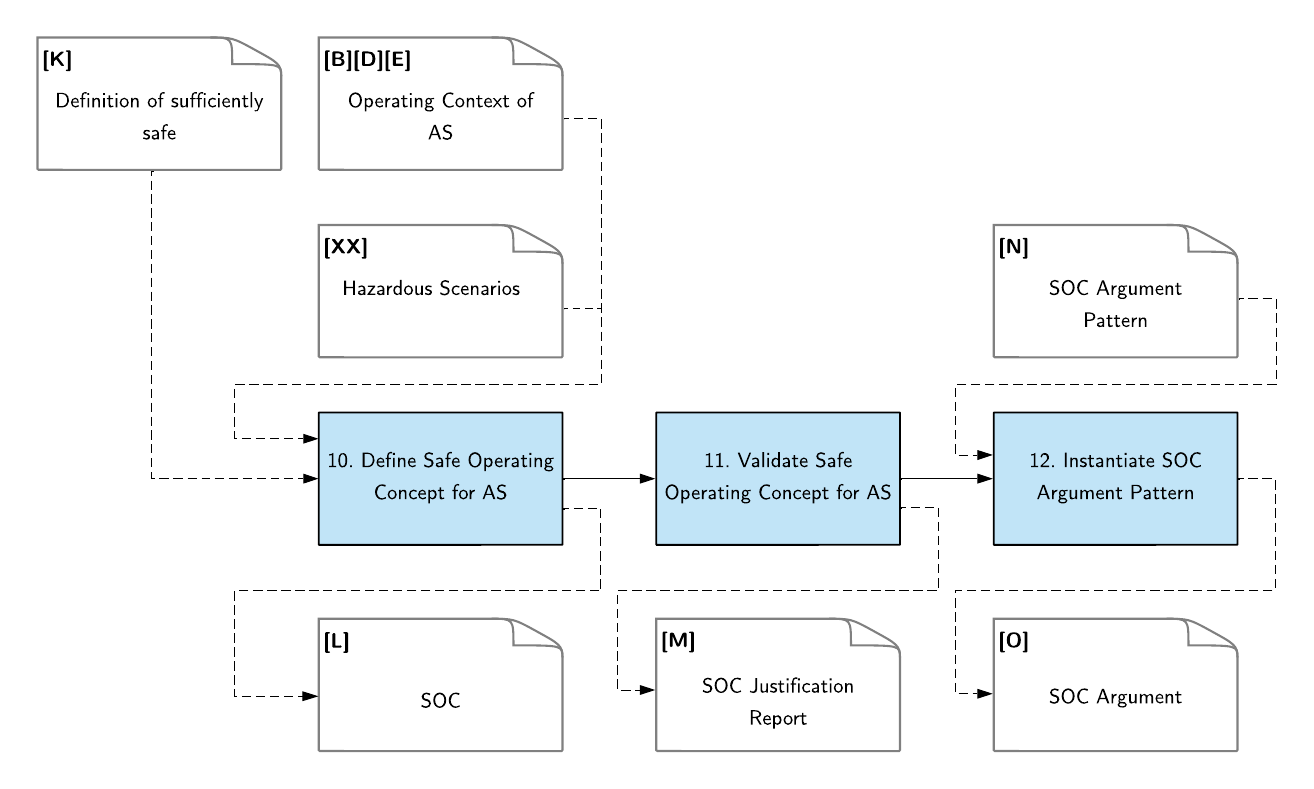}
    \caption{Safe Operating Concept Assurance Process}
    \label{fig:SOCProcess}
\end{figure}

As shown in Figure~\ref{fig:SOCProcess}, this stage consists of three activities that are performed to define and validate the safe operating concept for an AS. The artefacts generated from this stage are used to instantiate the SOC assurance argument pattern as part of Activity~\ref{act:SOCpattern}.

\Activity{Define Safe Operating Concept of AS \label{art:L} \ArtL}{act:socDefine}

The SOC specifies sufficiently safe operation of the AS within the operating context defined by \ArtB, \ArtD \hspace{0.08cm} and \ArtE \hspace{0.08cm} taking account of the identified hazardous scenarios from \ArtAX \hspace{0.08cm} at Stage~\stageref{2}. The SOC consists of:
\begin{enumerate}
    \item A definition of “sufficiently safe”
    \item A set of system level safety requirements that specify how the AS must behave in order to provide a sufficient mitigation for the hazardous scenarios. 
    \item (If necessary) A description of one or more reduced operating domains (RODs) and the conditions under which the ROD applies.
\end{enumerate}

The system safety requirements may be specified for the AS using either natural language or more formalised representations. Whatever representation is chosen, of most importance is that the requirements are clear and unambiguous.

\clearpage

\begin{note}
There is existing good practice for requirements specification that can be used to guide the specification of system safety requirements as part of the SOC, such as \cite{Mavin} which provides a generic requirements syntax:

\vspace{5mm}
\textbf{\emph{<optional preconditions> <optional trigger> the <system name> shall <system response>}}

\vspace{5mm}
In addition to five templates for different requirement types:

\vspace{5mm}
\textbf{Ubiquitous requirement} 

These are requirements that always hold and therefore have no preconditions or trigger (sometimes referred to as invariants). These requirements should have the form:

\textbf{\emph{The <system name> shall <system response>}}

\vspace{5mm}
\textbf{Event-driven requirement}

These are requirements that are initiated when, and only when, a triggering event (such as a change in the operating environment) is detected at the system boundary. These requirements should have the form:

\textbf{\emph{When <optional preconditions> <trigger> the <system name> shall <system response>}}

\vspace{5mm}
\textbf{Requirements to handle unwanted behaviour}

These requirements are a variant of the event-driven requirements that specifically deal with unwanted events (such as failures). These requirements should have the form:

\textbf{\emph{If <optional preconditions> <trigger>, then the <system name> shall <system response>}}

\vspace{5mm}
\textbf{State-driven requirement}

These are requirements that are valid while the system is in a defined state. These requirements should have the form:

\textbf{\emph{While <in a specific state> the <system name> shall <system response>}}

\vspace{5mm}
\textbf{Optional feature requirement }
These are requirements that are valid only in systems that include a given feature or capability. These requirements should have the form:

\textbf{\emph{Where <feature is included> the <system name> shall <system response>}}

\vspace{5mm}
\textbf{Requirements with complex conditions}

Combinations of keywords \emph{When, While} and \emph{Where} can be use to build complex expressions to specify more complex As behaviours. For example:

\textbf{\emph{While the robot is moving, when a person is present, the robot shall issue an audible warning}}

\vspace{5mm}
A structured syntax approach such as that described above provides a method for mitigating against \emph{syntactic ambiguity}.

Mitigating against \emph{semantic ambiguity} requires ways to ensure that the meaning of words and phrases is clearly defined within the relevant context and that only that meaning is used. This can be achieved through the use of domain or project specific dictionaries or through developing context-specific ontologies. These should include definitions of concepts and elements included in the ODM.
\end{note}

\begin{example} \label{ex:sepEx}
The SOC for an autonomous car includes the following safety requirement: ``The AV shall maintain sufficient distance between itself and any vehicle in front in order to provide enough time to react if the car in front suddenly brakes.'' This requirement has been specified in order to mitigate the hazardous event of the autonomous car colliding into the rear of another vehicle.  For such a requirement, the distance can be calculated for different driving speeds based upon a number of assumed properties such as the minimum and maximum braking performances of the cars and the frictional coefficient of the road surface (see \cite{shalev2017formal} for details of such calculations). These assumptions must be validated and explicitly documented as part of the assurance case.
\end{example}

\begin{example}
The SOC for an autonomous excavator operating on a construction site includes the following safety requirement: ``The excavator shall ensure maximum tilting angle is never exceeded.'' This requirement has been specified in order to mitigate the hazardous event of the excavator toppling whilst digging \cite{seward2007safe}. This could occur due to an incorrect configuration of the excavator whilst digging on uneven or unstable ground or as a result of resistance to the bucket.
\end{example}

\begin{example}
The SOC for an Autonomous Insulin Infusion Pump used on a busy Intensive Care Unit (ICU) treating multiple patients concurrently may contain the following safety requirement: ``The alarms and warnings function of the autonomous insulin infusion pump shall not unnecessarily distract or disturb ICU nurses from their other tasks''.
\end{example}

\begin{note}
The safety requirements defined as part of the SOC should be defined with reference to the system-level behaviour of the AS, that is behaviour that is visible on the boundary of the AS, rather than referring to any particular sub-system or component. For instance, the safety requirement relating to rear collision in Example 17 should not be written as a requirement on the braking system, but, as is shown, on required behaviour of the vehicle as a whole. This requirement and its assumptions may then lead to safety requirements being derived upon the braking system through a consideration of the system design later in the lifecycle (see Stage 5).
\end{note}

\begin{figure}[h]
    \centering
    \includegraphics[width=0.8\linewidth]{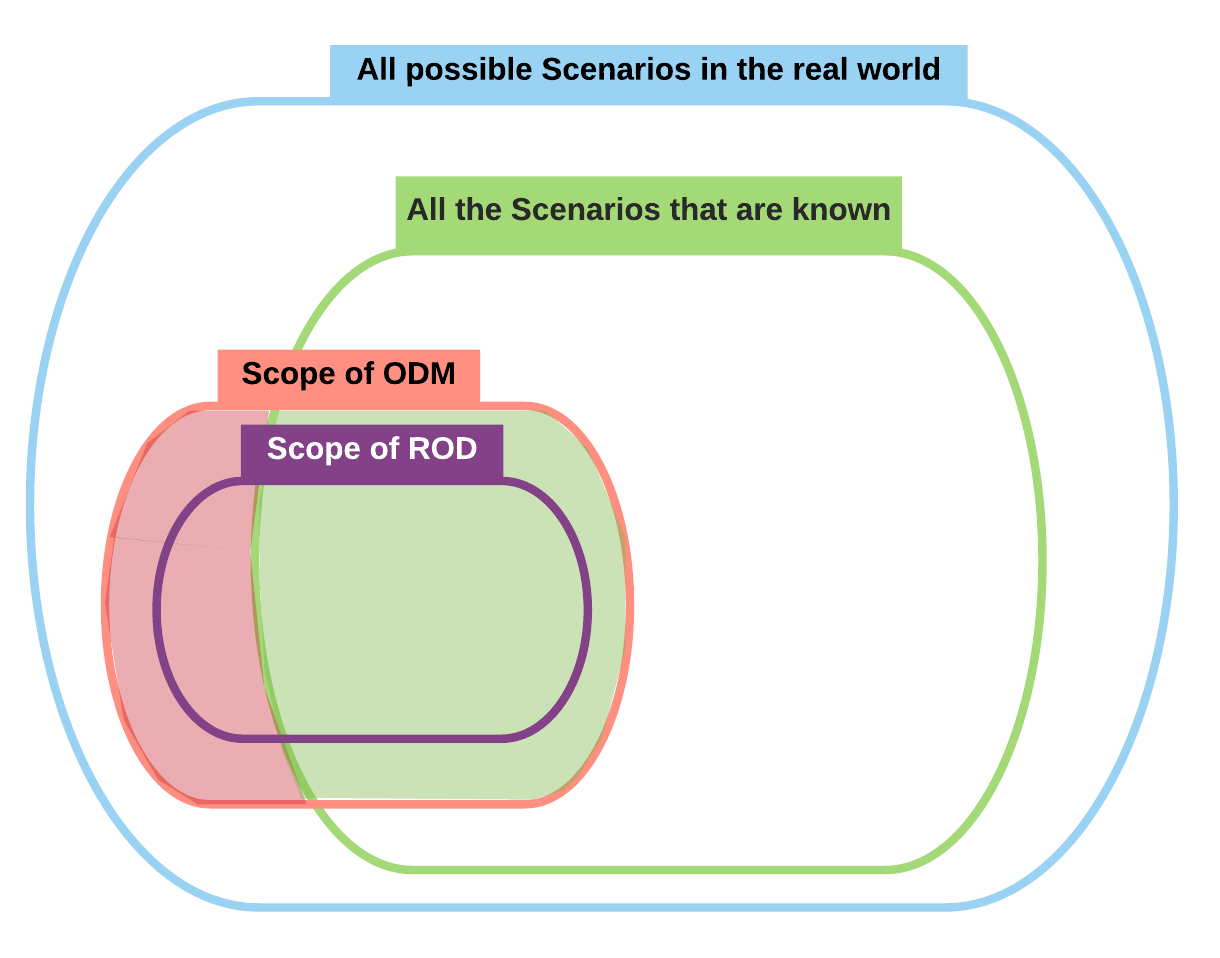}
    \caption{Defining a Reduced Operating Domain}
    \label{fig:rod}
\end{figure}

In addition to specifying system safety requirements, as part of the SOC it may also be necessary to define a reduced operating domain (ROD). As shown in Figure \ref{fig:rod}, a ROD is a definition of additional constraints on the ODM in order to reduce the operational scope of an autonomous capability under certain conditions. By operating in a ROD, the intention is that the AS is not exposed to scenarios that may lead to hazardous events under those defined conditions. 

There are a number of reasons why a ROD may need to be defined. Often a ROD will be defined with respect to a particular system state where it would be unsafe to operate within the entire scope of the ODM. This may often be in response to a particular system or component failure mode.

\begin{example}
If there is a failure of the long range object detection sensors on an autonomous passenger shuttle, the vehicle enters a reduced operating domain that requires a low maximum operating speed for the vehicle. This enables the vehicle to transport the passengers to a safe location, but to do so without increasing the risk of collision whilst relying only on short range sensing capability.
\end{example}

\begin{example}
If there is a data communication failure between an Autonomous Insulin Infusion Pump, and the centralised electronic patient healthcare records system, the autonomous pump may be prevented from increasing or decreasing the rates of insulin infusion until either communications are restored, or action is taken by a healthcare professional.
\end{example}

Although RODs may be identified early in the lifecycle, it is likely that RODs will often be formulated or revised later in the lifecycle in response to understanding gained through later activities such as hazardous failures identified in \ArtAB  at Stage~\stageref{6}.

\begin{note}
In order to define an SOC that provides sufficient mitigation for the hazardous scenarios under the specified conditions (including failures modes), there is generally a choice as to the safety strategy that is adopted:
\begin{enumerate}
    \item The AS continues to provide all of the Autonomous Capability (AC) under those specified conditions but a ROD is defined that limits the scope of operation under which that AC is provided.
    \item The AS continues to operate under the same ODM but the AC is limited.
    \item A combination of 1) and 2) is adopted.
\end{enumerate}
\end{note}

\begin{example}
An autonomous car driving on a highway enters a zone with emergency roadworks in place. There are different safety strategies that could be adopted under these conditions. The car could continue to operate fully autonomously, but the maximum allowable speed could be reduced (Strategy 1). The car could continue to operate up to the normal speed but only provide lane keeping capability, handing other control to the driver (Strategy 2). Or the car could have both its speed constrained as well as restricting its capability to lane keeping only (Strategy 3).
\end{example}

\subsection*{Artefact \ArtK: Definition of sufficiently safe}\label{art:K}

Defining the SOC for an AS requires a judgement of what is considered ``sufficiently safe'' operation. This is a complex judgement that will require consideration of legal and ethical factors \cite{burton2020mind}, \cite{zhu2021ai} as well as the risk tolerance of the identified stakeholders for the AS operation. Determining and justifying the judgement of sufficiently safe is therefore a very broad issue that is outside of the scope of this document, detailed guidance on this will be provided in a separate document.

\begin{example}
As an example of the challenges of providing a definition of sufficiently safe for an AS we can consider autonomous cars, where there has been much debate and disagreement on the best approach. One approach is to judge sufficiency in comparison to a human driver. A report from the UK Law Commission \cite{law2022} sought the views of the general public on what an acceptable comparison may be. The following three options were provided:
\begin{enumerate}
    \item as safe as a competent and careful driver
    \item as safe as a human driver who does not cause at fault accidents
    \item safer than the average human driver
\end{enumerate}

The results of this consultation were inconclusive, with none of these options receiving a majority response. An additional complication with all of these options is the huge range of driving capabilities that exist in human drivers which can vary enormously based on factors such as age, health, experience, level of distractions etc. 

An alternative approach that has been proposed for judging what is sufficiently for an autonomous car is to judge the performance of the vehicle in specific scenarios, rather than assessing an overall average performance. The attraction of this approach is that it should help to ensure that the risk exposure of humans to the operation of the autonomous car is fairly distributed.

\end{example}

\Activity{Validate Safe Operating Concept for AS \label{art:M} \ArtM}{act:socVal}

The SOC defined in \ArtL shall be validated in order to check that:

\begin{itemize}
    \item The AS safety requirements defined as part of the SOC specify sufficient mitigation for all of the  identified hazardous scenarios (\ArtAX).
    \item The AS safety requirements are clear and unambiguous.
    \item The strategies defined by the RODs and reduced ACs provide a sufficient mitigation for the identified hazardous scenarios under the specified conditions.
\end{itemize}

Validation of the SOC will often require the involvement of multiple stakeholders, particularly the system developers and operators who have the necessary domain knowledge and understanding of the system operation. The stakeholders should provide an independent view on whether each of the points above is satisfied by the SOC definition. 

\begin{note}
One option for demonstrating the SOC for the purposes of validation is to use simulation. Simulation allows for earlier validation of the system, as well as more more rapid exploration of the operational space, than would be possible with the real system. Since simulating the entire AS can be prohibitive, a hardware in the loop approach may be used where pre-captured real-world sensor data is input to the simulation which simulates the AS response.
\end{note}

\Activity{Instantiate SOC Assurance Argument Pattern \label{art:O} \ArtO}{act:SOCpattern}

This activity requires as input the SOC assurance argument pattern (\ArtN), as well as the artefacts from Activities~\actref{act:socDefine}{10} and ~\actref{act:socVal}{11} (\ArtL and \ArtM). The activity uses these artefacts to create an instantiated AS operating context assurance argument (\ArtO) which demonstrates that the defined SOC sufficiently mitigates the hazardous scenarios identified for the AS operation.

\subsection*{Artefact \ArtN: SOC Argument Pattern}\label{art:N}

\begin{figure}[h]
    \centering
    \includegraphics[width=1\linewidth]{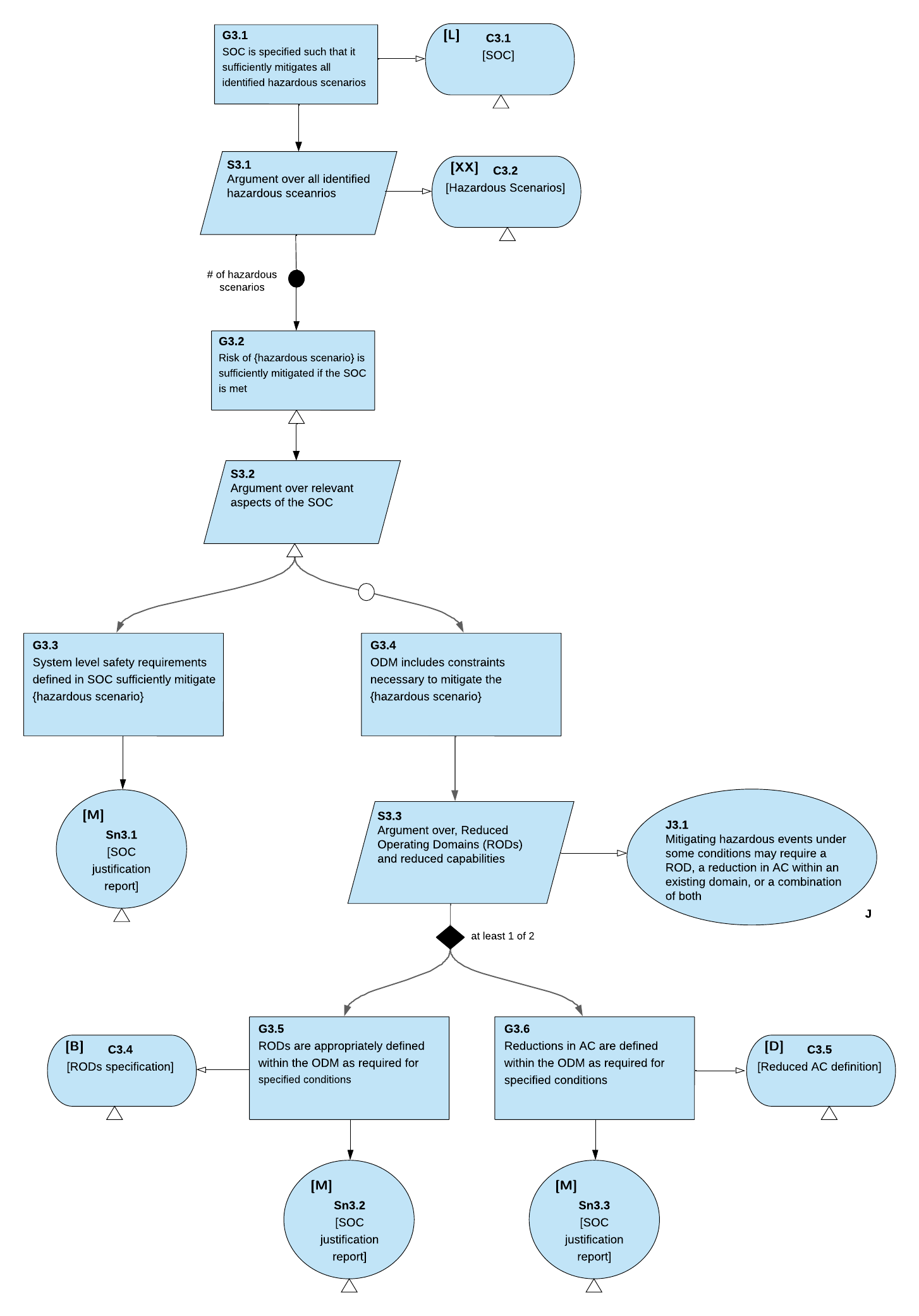}
    \caption{\ArtN~:~Argument Pattern for SOC Assurance} 
    \label{fig:SOCarg}
\end{figure}

The argument pattern relating to this stage is shown in Figure~\ref{fig:SOCarg} and key elements from the pattern are described in the following sections.

\subsection*{G3.1}
The top claim in this argument pattern is that the SOC that has been specified defines a sufficient mitigation for all of the hazardous scenarios identified at Stage~\stageref{3} and provided as context to the argument. This is demonstrated through considering each of the identified hazardous scenarios in turn and providing a claim (G3.2) about the sufficiency of the SOC in mitigating each hazardous scenario.

\subsection*{G3.2}
For each of the identified hazardous scenarios, it must be demonstrated that if the SOC is met by the system during operation, the risk associated with that hazardous scenario is sufficiently mitigated. A claim of this nature must be supported for each hazardous scenario. The strategy to demonstrate this is to consider the aspects of the SOC that are relevant to the hazardous scenario under consideration. This will include the relevant safety requirements, and may also include any relevant ROD specifications. 

\subsection*{G3.3}
This claim considers the sufficiency of the defined safety requirements in providing mitigation for the hazardous scenario. The SOC justification report (\ArtM) should demonstrate that this is the case. It is therefore important that the SOC justification report is systematic in its consideration of each identified hazardous scenario and provides explicit justification for each.

\subsection*{G3.4}
Where it has been identified that additional constraints are required as part of the mitigation for the hazardous scenario, it must be demonstrated that those constraints are sufficient\footnote{Since additional constraints may not always be required as part of the SOC, G3.4 is an optional element of the argument. This is indicated in GSN by the use of an open circle in the argument structure.}. As discussed in the guidance, the constraints may take the form of RODs, and/or reductions in autonomous capability. In each case a claim must be made (G3.5 and G3.6 respectively) that justifies the nature of those constraints with respect to the hazardous scenario under consideration. The SOC justification report (\ArtM) is used as evidence to support these claims.

\clearpage
\stage{AS Safety Requirements Assurance}
\label{Stage 5}

\subsection*{Objectives}
\begin{enumerate}
\item Define safety requirement for each tier of decomposition in the development of the AS.
\item Validate the defined safety requirements.
\item Create the Safe Requirements Argument.
\end{enumerate}

\subsection*{Inputs to the Stage}
\begin{itemize}
\item[\ArtP]: Safety Requirements from tier n-1
\item[\ArtW]: tier n Design
\item[\ArtS]: Safety Requirements Argument Pattern
\end{itemize}

\subsection*{Outputs of the Stage}
\begin{itemize}
\item[\ArtQ]: Safety Requirements for tier n 
\item[\ArtR]: Safety Requirements justification report
\item[\ArtT]: Safety Requirements Argument
\end{itemize}

\subsection*{Description of the Stage}
\begin{figure}[h]
    \centering
    \includegraphics[width=1\linewidth]{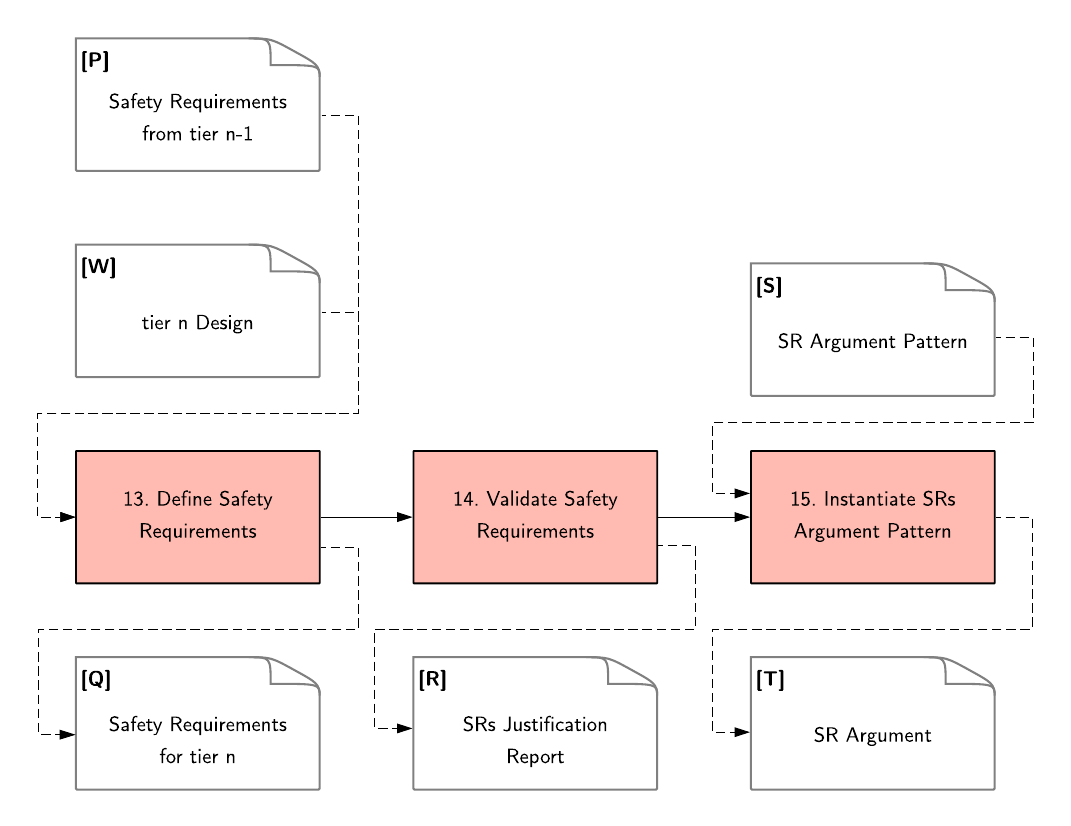}
    \caption{AS Safety Requirements Assurance Process}
    \label{fig:SRprocess}
\end{figure}

As shown in Figure~\ref{fig:SRprocess}, this stage consists of three activities that are performed to define and validate the safety requirements. This stage will need to be repeated multiple times as the development of the AS proceeds through decomposition of each tier of development. This stage considers how the safety requirements are allocated, decomposed and interpreted at each tier. The artefacts generated from this stage are used to instantiate the safety requirements assurance argument pattern as part of Activity~\ref{act:SRpattern}.

\Activity{Define Safety Requirements \label{art:Q} \label{art:P} \ArtQ, \ArtP}{act:reqDef}

This activity considers how the safety requirements are defined at each tier of decomposition in the system development process such that they adequately capture the intent of the safety requirements that were established at the previous tier of decomposition (\ArtP). This requires consideration of the design that is proposed for this tier of decomposition (\ArtW) such that safety requirements may be adequately decomposed and allocated to the relevant system components. The safety requirements must also be correctly interpreted for the allocated component to reflect the component design.

\begin{figure}[h]
    \centering
    \includegraphics[width=1\linewidth]{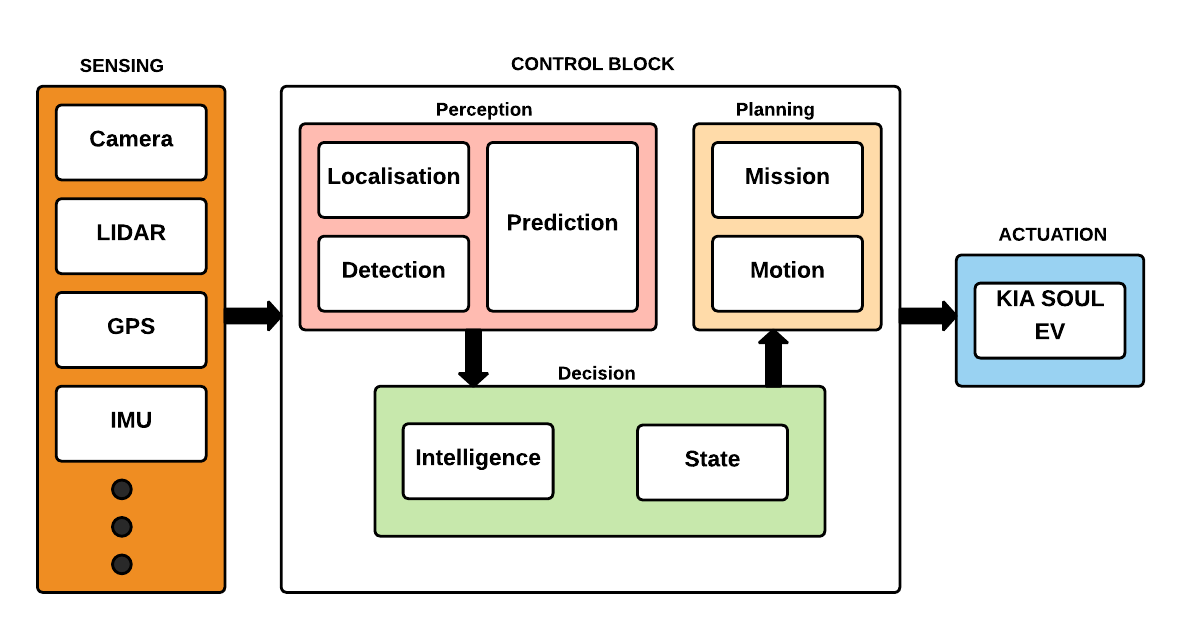}
    \caption{A simple example of the logical architecture for the controller of an autonomous vehicle (from \cite{munir2018autonomous})}
    \label{fig:arch}
\end{figure}

\begin{example}
We can identify the different tiers for a typical AS by considering the example in Figure \ref{fig:arch}, taken from \cite{munir2018autonomous}, which shows a typical logical architecture for the controller of an autonomous vehicle. At the highest level we could consider the AS itself as a tier of decomposition. At this level the interactions between the AS and other entities would be considered. The next tier in Figure \ref{fig:arch} would decompose this to the three main subsystems (Sensing, Control Block and Actuation) and their interactions. Each of these subsystems can then be seen to decompose at the next tier to components such as camera, LIDAR and GPS for the sensing subsystem, or Perception, Decision and Planning for the Control Block. Whereas the components may be considered the lowest level tier for the Sensing subsystem (since these are provided as developed units to the AS developer), further tiers of decomposition may be required for other components such as the Perception component, which is broken down to Localisation, Detection and Prediction. As illustrated in this example, the number of tiers required as part of the development process may not be uniform across all elements of an AS.

\end{example}

\begin{note}
In theory, all of the information required to demonstrate the adequacy of the decomposed safety requirements would be captured in the initial high-level requirement. In practice however, some of the information will always remain implicit. For AS, this could particularly be due to the complexity of the operating environment, but also to the fact that design decisions will always be made later in the development lifecycle that require greater detail in the requirements. This detail cannot be properly known until the design decisions have been made. 
For this reason it is important to ensure that the \emph{intent} of the safety requirements is maintained between levels of abstraction, with consideration being given to information that may not be fully specified in the requirement. 
\end{note}

\begin{figure}[h]
    \centering
    \includegraphics[width=1\linewidth]{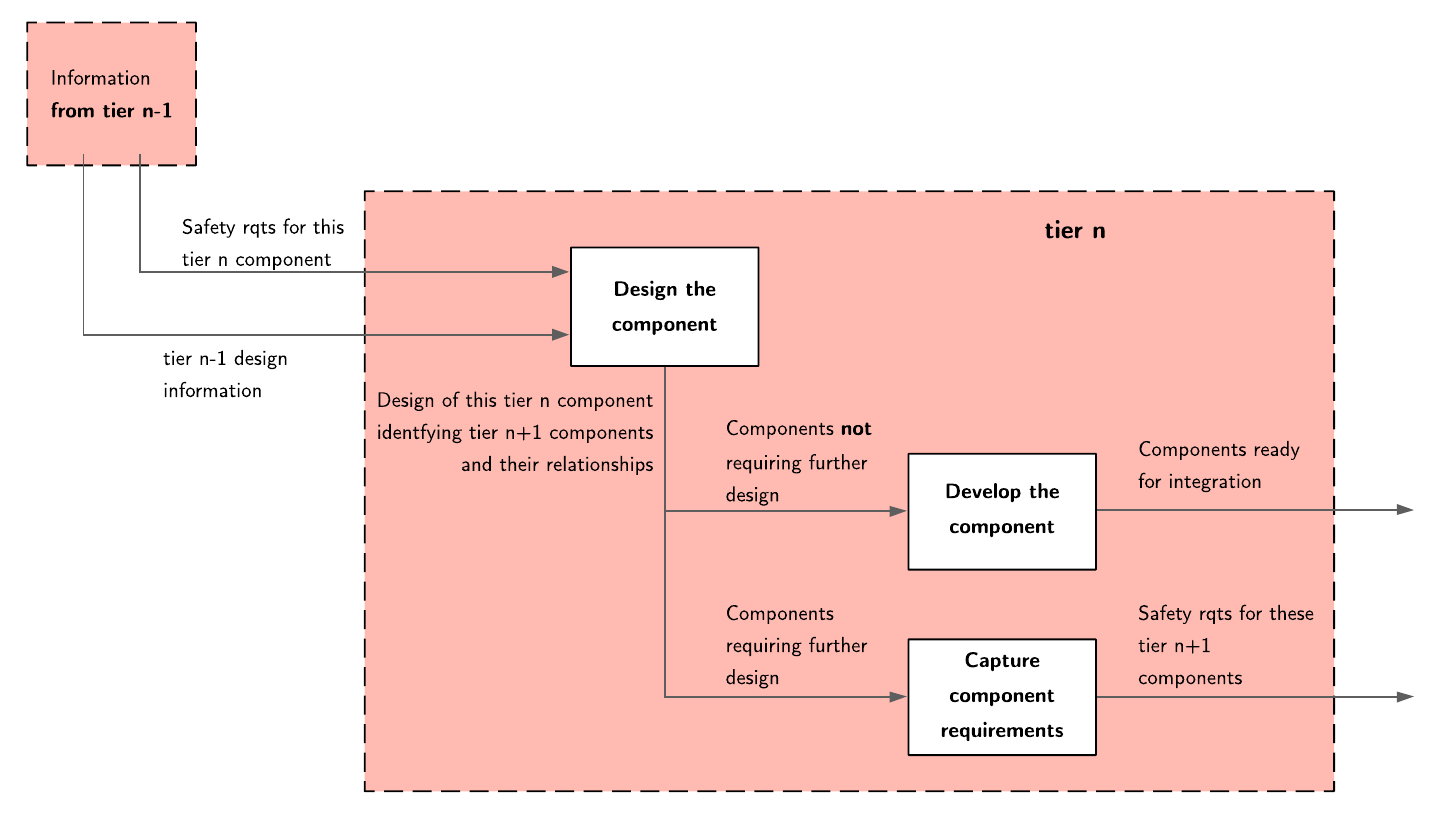}
    \caption{Defining safety requirements for components of a tier of design decomposition.}
    \label{fig:tierRqts}
\end{figure}

Figure \ref{fig:tierRqts}, adapted from \cite{jaffe2008progress}, illustrates how safety requirements are derived for a component of a single tier of decomposition in the development of an AS. This takes as input information from the previous tier of decomposition relating to the safety requirements placed on the components at this tier, as well as the design information. This information is used to create a design for the component (as discussed in Stage~\stageref{5}). This design will specify the sub-components that are required and the relationships between those sub-components.

\begin{note}
At the highest level of abstraction, which is the overall AS system level (``tier 0'') the safety requirements will be those defined by the SOC (see Stage~\stageref{4}). These will then be decomposed to more detailed safety requirements at the next tier of design decomposition (``tier 1''). Each time the safety requirements are further decomposed to more detailed tiers it is expected that additional safety requirements will be needed. Thus the total number of safety requirements specified for the AS will increase as more detail is added to the design of the system.

It may also be the case when safety requirements are derived for a component, that additional safety requirements relevant to a previous tier are also identified. These safety requirements should be fed back up, in order to be correctly decomposed to the relevant system components. 
\end{note}

\begin{figure}[h]
    \centering
    \includegraphics[width=1\linewidth]{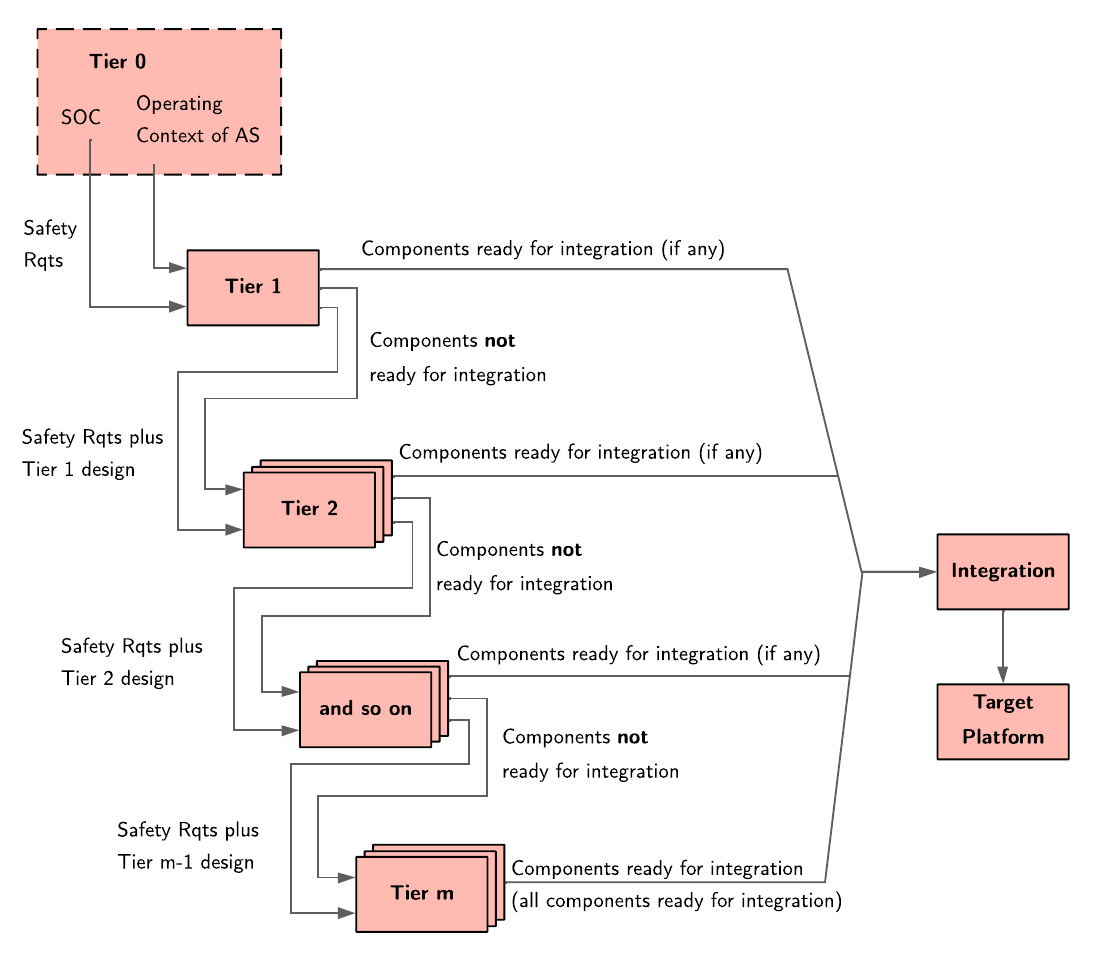}
    \caption{The decomposition of AS safety requirements}
    \label{fig:tierDecomp}
\end{figure}

Some of the sub-components defined for a tier may be ready for implementation. For these components no further design activity is required; these components are at the stage where they can be implemented according to the requirements that are in place. These components are ready for verification and integration (see Stage~\stageref{8}). For other sub-components further design may be required before they are ready for implementation. For these components a further tier of requirements decomposition will be required (along with further design activity). As shown in Figure \ref{fig:tierDecomp}, adapted from \cite{jaffe2008progress}, different components may be ready to be integrated to the target AS platform at different tiers. Irrespective of the number of tiers in the development lifecycle of the system, the safety assurance considerations for the requirements at each tier remain the same: to derive safety requirements that adequately capture the intent of the the safety requirements specified at the previous tier, and to justify the sufficiency of those safety requirements.

\begin{note}
The definition of safety requirements is intimately linked to the design process of the AS. It is important to note therefore that safety requirements may need to be re-defined throughout the lifecycle based upon the architectural and design decisions that are taken. For example, many AS make extensive use of third-party components to implement parts of the design (such as sensors). If a third-party component is selected to be used, it may be found that the safety requirements that were originally derived for that component cannot be fully met by that third-party component (the same could be true for legacy or re-used components). This would therefore necessitate changes to the system design to mitigate those limitations (such as introducing additional components). This would then require a re-definition of the safety requirements allocated to the components to reflect the changes to the design. It may then be possible to demonstrate that the third party component can meet these redefined safety requirements.

Essentially it may be necessary to revisit the safety requirements at any tier multiple times as changes are made to the proposed design solution. What is crucial is that throughout this process, as changes are made to the design and the safety requirements, traceability is maintained to the safety requirements of the higher tier, and ultimately to the SOC for the AS operation.
\end{note}

\begin{note}
Where safety requirements are allocated to a component that uses machine learning, the safety assurance of that component may be undertaken in accordance with the AMLAS guidance \cite{AAIP2021a}.
\end{note}

\Activity{Validate Safety Requirements \label{art:R} \ArtR}{act:reqDef}

The safety requirements documented for each tier in \ArtQ~shall be validated in order to check that they adequately capture the intent of the more abstract safety requirements defined at the previous tier. This will require that it is checked that each of the higher level requirements can be satisfied if the safety requirements for this tier are correctly implemented. It is therefore important that the validation activities focus on the semantic equivalence of the safety requirements at different tiers.

\begin{note}
Demonstrating that the intent of the safety requirements is captured requires more than simply stating that a relationship exists between safety requirements at different tiers. Some explanation and justification for the sufficiency of that relationship must be provided. Concepts such as ``Rich Traceability'' \cite{dick2002rich} could help in this regard.

Ensuring the intent of the safety requirements is maintained throughout decomposition may often be more challenging for AS than it is for traditional systems due to the sometimes large ''semantic gaps`` that can exist. The nature of these gaps and the challenges of addressing them for AS is discussed in more detail in \cite{burton2020a}.
\end{note}

The output from the safety requirements validation activity shall be documented in the Safety Requirement Justification Report (\ArtR).

\Activity{Instantiate Safety Requirements Argument Pattern \label{art:T} \ArtT}{act:SRpattern}

This activity requires as input the safety requirements argument pattern (\ArtS), as well as the artefacts from the previous activities of this stage (\ArtP, \ArtW, \ArtQ and \ArtR). The activity uses these artefacts to create an instantiated safety requirements assurance argument for the AS (\ArtT) which demonstrates that the defined safety requirements sufficiently capture the intent of the SOC for the AS.

\subsection*{Artefact \ArtS: Safety Requirements Argument Pattern}\label{art:S}

\begin{figure}[h]
    \centering
    \includegraphics[width=1\linewidth]{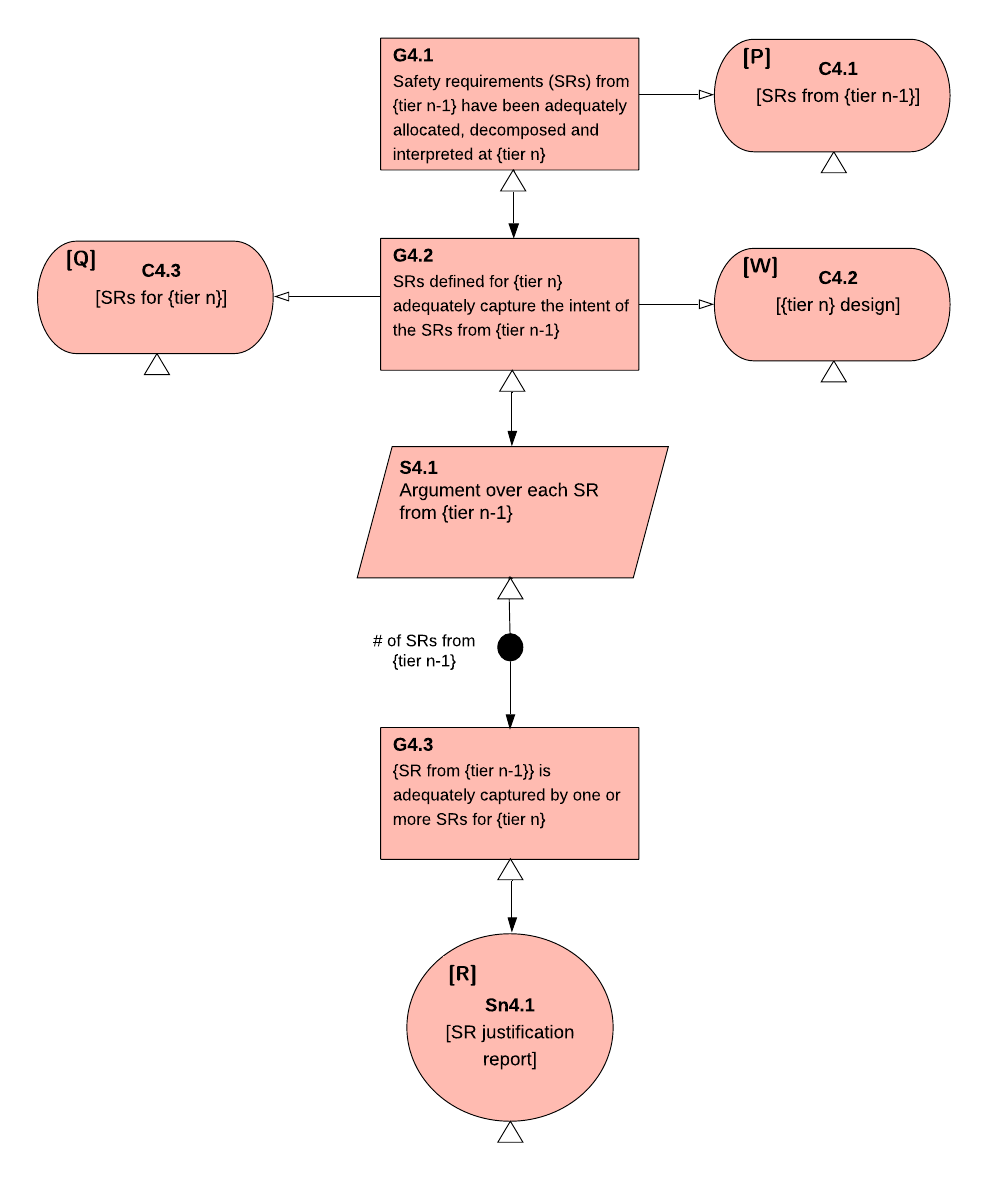}
    \caption{\ArtS~:~Argument Pattern for AS Safety Requirements Assurance} 
    \label{fig:SRarg}
\end{figure}

The argument pattern relating to this stage is shown in Figure~\ref{fig:SRarg} and key elements from the pattern are described in the following sections.

\subsection*{G4.1}
This claim, which is made for each tier of the AS development, demonstrates that the safety requirements that were defined at the previous tier have been adequately allocated, decomposed and interpreted through the safety requirements that have been defined at the current tier.

\subsection*{G4.2}
It must be shown that the safety requirements at the current tier are sufficient to capture the intent of the inherited safety requirements. The definition of safety requirements must take account of the context of the design that is in place for this tier (\ArtW). To ensure that all the safety requirements from the previous tier are considered, the argument explicitly considers each of those safety requirements in turn though the creation of a claim (G4.3) for each requirement.

\subsection*{G4.3}
A claim is made for each of the safety requirements from the previous tier, demonstrating that one or more of the safety requirements defined at the current tier adequately capture the intent of that requirement. This is demonstrated using the evidence documented in the SR justification report (\ArtR).
\clearpage
\stage{AS Design Assurance}

\subsection*{Objectives}
\begin{enumerate}
\item Create a design at tier n that ensures the safety requirements for tier n can be satisfied.
\item Justify the sufficiency of the design created at tier n with respect to the defined safety requirements.
\item Review the sufficiency of the design created at tier n with respect to the defined safety requirements.
\item Create the AS Design Assurance Argument.
\end{enumerate}

\subsection*{Inputs to the Stage}
\begin{itemize}
\item[\ArtQ]: Safety Requirements for tier n
\item[\ArtX]: Design Process for tier n
\item[\ArtU]: AS Design Assurance Argument Pattern
\end{itemize}

\subsection*{Outputs of the Stage}
\begin{itemize}
\item[\ArtW]:  tier n Design
\item[\ArtV]:  AS Development Log
\item[\ArtY]:  AS Design Justification
\item[\ArtZ]:  AS Design Review
\item[\ArtAA]:  AS Design Assurance Argument
\end{itemize}

\subsection*{Description of the Stage}

This stage aims to provide assurance regarding the AS design. This stage is iterative as it considers the assurance of the design of the AS across multiple levels of design decomposition. This stage is also highly integrated with stages \stageref{4} and \stageref{6} in that it involves creating design proposals to meet the safety requirements (defined at Stage~\stageref{4}), doing analysis of that design proposal (Stage~\stageref{6}), then perhaps changing the design in response, updating the safety requirements, doing some more analysis, and so on across multiple tiers in order to create and assure a sufficient design.

\begin{figure}[h]
    \centering
    \includegraphics[width=1\linewidth]{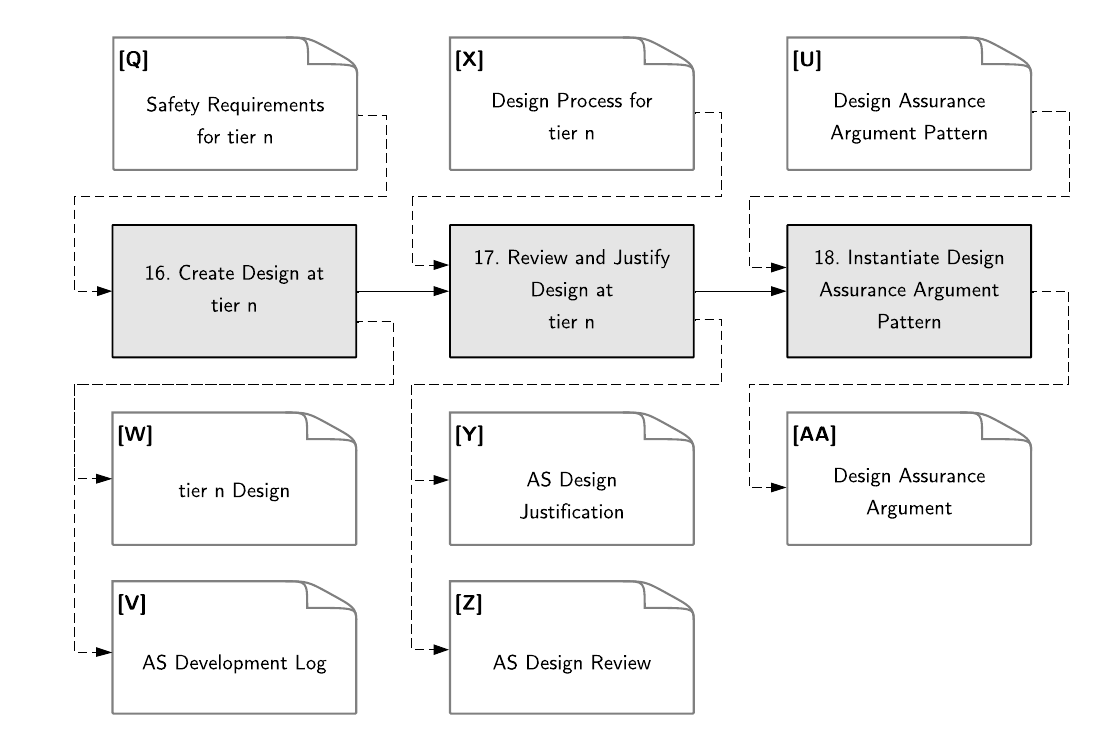}
    \caption{AS Design Assurance Process}
    \label{fig:DesignProcess}
\end{figure}

\Activity{Create Design at tier n \label{art:V} \label{art:W} \ArtV \ArtW }{act:DesCreate}

Given a set of safety requirements defined at a tier of decomposition (\ArtQ) of the AS design, a design for that tier shall be produced that ensures those safety requirements can be met. This will involve making design decisions that are appropriate given the overall context of the safety requirements, the operating context and the known failure modes. In particular decisions relating to the system architecture are of particular importance when considering the satisfaction of safety requirements for an AS.

\begin{note}
In the context of AS, we use the following definition of system architecture, adapted from \cite{alves2018considerations}:

``An Architecture comprises the rules and constraints for the design of a system, including
its structure and behavior, that would meet all of its established requirements and restrictions under the anticipated environmental conditions and component failure scenarios.''
\end{note}

\begin{note}
A system architecture may capture either the logical or physical structure of the system. Decomposition of an AS design may often involved a transition from a logical architecture to a physical one. During such a transition there is often an increased likelihood that errors may be introduced to the design.
\end{note}

The design of the system architecture at each tier shall consider the need for robustness, fault tolerance and runtime monitoring in order to satisfy the safety requirements defined for that tier.

\textbf{\textit{Robustness}} can be defined as the delivery of a correct service in implicitly-defined adverse situations arising due to an uncertain system environment \cite{lussier2004fault}. Robustness is therefore a mechanism that is particularly important for AS since it enables the AS to mitigate hazardous system failures associated with hard to predict, and thus unexpected, changes in a complex operational environment. This may for example include objects in the environment that were not included as part of the ODM which the system fails to detect, or unexpected effects of particular lighting conditions that lead to ``phantom objects'' being detected. Since these events have not been anticipated, specific hazardous failures will not have been identified as part of the assurance process. However, it is expected that hazardous failures of this type would be identified, leading to a requirement for the AS to be designed such that it is robust to those types of failures.

Robustness in AS systems is typically achieved through redundancy in the architecture. The purpose of the redundancy is to provide compensation in the design of the system for potential limitations in system components that could result in unexpected system failures. Such redundancy may be required for both hardware and software elements of the architecture, and may be used at multiple architectural levels.

\begin{note}
Redundancy in the architecture alone will not necessarily provide the robustness for the AS that is required. Identical components will contain the same limitations and failure modes. This would mean that the components, when exposed to the same set of inputs, would suffer the same failures. To be effective, there must therefore be diversity in the redundant architectural elements. This requires some level of independence between the components, for example by using components developed by different teams of people in an attempt to ensure the same mistakes are not replicated in each component, or by providing conceptual diversity through providing different specifications for each of the redundant components.

Ensuring diversity for software components particularly challenging \cite{littlewood2001modeling}, \cite{avizienis1984fault} and therefore requires particular attention for AS (with their high reliance on software). The use of artificial intelligence techniques such as machine learning (ML) for developing components also provides unique considerations for diversity.
\end{note}


\textbf{\textit{Fault tolerance}} can be defined as the delivery of a correct service despite
faults arising from the AS itself. The focus of fault tolerance is therefore on the detection and recovery from anticipated failures, which should be identified as part of the hazard assessment of the AS (see Stage~\stageref{6}). Since the hazardous failures are known during AS development, checks can be included in the system design in order to detect the faults. This could include, for example detecting inconsistencies between the system state as characterised by the sensors and the system state predicted by a model \cite{lussier2004fault}. Once a fault is detected, fault recovery strategies enable the continued provision of the service. Standard software fault tolerance strategies can be applied effectively to AS \cite{powell2011tolerance}.

\begin{note}

There are three basic fault tolerance approaches that can be used: Recovery blocks, N-version programming and N-self-checking software \cite{laprie1990definition}. 

A recovery block approach provides fault tolerance for a component by using alternate variants of the component plus an adjudicator. The adjudicator tests the outputs from the components. If a component fails then the next alternate component is used and so on. This mechanism can be used to ensure that the behaviour of the component continues to be provided to the AS once a failure occurs in the component. The tests performed by the adjudicator must be sufficient to identify the hazardous failure based on consideration of the safety analysis results \ArtAB obtained from Stage~\stageref{6}.

In contrast, with an N-version programming approach, all variants provide outputs to the adjudicator which then decides by considering all the outputs together which result to use. All variants of the component must be functionally-equivalent, but diversely-designed. An advantage of this approach is that it does not require tests to be defined for the adjudicator, so can be used for situations where it may be challenging to specify specific test criteria for a hazardous failure.

N-self-checking software is a hybrid approach that involves the use of self-checking software components (SCSC). Each SCSC could use a recovery block or an N-version program approach. At least two SCSC are required, whose outputs are then checked. Such an approach provides fault tolerance for components or subsystems which are themselves fault tolerant.

\end{note}

\textbf{\textit{Runtime monitoring}}

The provision of a runtime monitor as part of the AS design enables the behaviour of the AS during operation to be checked against defined constraints or behavioural predictions. The runtime monitor should be independent from the components it is monitoring, but may take the same inputs. One of the biggest challenges with using runtime monitors is being able to correctly define the constraints or bounds on behaviour that the monitor will check. This definition should make reference to the SOC (\ArtL) and the hazardous scenarios (\ArtG) identified from previous stages.

Runtime monitoring may, in addition to monitoring for unsafe behaviour of the AS, also monitor behavioural trends of the AS over periods of time. This would enable the identification of, for example, the deterioration in performance of the AS which may act as a leading indicator of future unsafe behaviour. In this case it is important to determine which information to monitor and how that can be interpreted as representing a threat to the safety of the AS.

The key design decisions that are taken at each tier shall be documented in the AS Development Log (\ArtV).

\subsection*{Artefact \ArtX: Design Process for tier n \label{art:X}}

When designing the AS, a rigorous design process shall be followed to help ensure that potentially hazardous errors are not introduced into the design. The design process that is followed shall be documented (\ArtX), ensuring that where different approaches are required for different tiers or for different types of component (such as for software or hardware components) these are made explicit. Justification for the appropriateness of the design process used should be included in (\ArtX).

\Activity{Review and Justify Design at tier n \label{art:Y} \label{art:Z} \ArtY \ArtZ}{act:DesJustify}

Justification shall be provided for how each of the key design decisions that have been made at tier n (\ArtV) help to ensure that the safety requirements can be met by the AS. In particular this should consider how the design decisions relate to the robustness and fault tolerance of the AS, and the way in which this supports the safety of the system. It is also important that the design decisions are reviewed to check that no inappropriate decisions are taken that mean that the safety requirements cannot be satisfied by the proposed design. The justification for the design at each tier shall be documented in the AS design justification report (\ArtY). 

Even if appropriate design decisions have been taken, the design should also be reviewed to check that errors have not been made in the design. It is particularly important for AS, where substantial use is made of software components, that this includes review of the proposed software design. In particular, the review should check for potentially hazardous errors. The review of the design shall be undertaken by a suitable person who is independent from the design activity itself. The results of the design review at each tier shall be documented in the AS design review report (\ArtZ). 

The sufficiency of the design process used (as documented in \ArtX) shall also be reviewed to check it is sufficiently rigorous. It should also be checked that the defined process has been correctly followed when developing the design.

\begin{note}
The level of independence that is required of the people responsible for the review may vary depending upon the level of risk associated with the AS. For high risk systems it may be required for the review to be undertaken by personnel from a different organisation to the design organisation. For lower risk systems it may be sufficient for the review to be undertaken by personnel from the same organisation who are capable of providing an independent view.
\end{note}

\Activity{Instantiate Design Assurance Argument Pattern \label{art:AA} \ArtAA}{act:DesPattern}

This activity requires as input the design assurance argument pattern (\ArtU), as well as the artefacts from the previous activities of this stage (\ArtV, \ArtW, \ArtX, \ArtY and \ArtZ). The activity uses these artefacts to create an instantiated design assurance argument for the AS (\ArtAA) which demonstrates that all tiers of the AS design are sufficient to ensure that the defined safety requirements can be met.

\subsection*{Artefact \ArtU: AS Design Assurance Argument Pattern}\label{art:U}

\begin{figure}[h]
    \centering
    \includegraphics[width=1\linewidth]{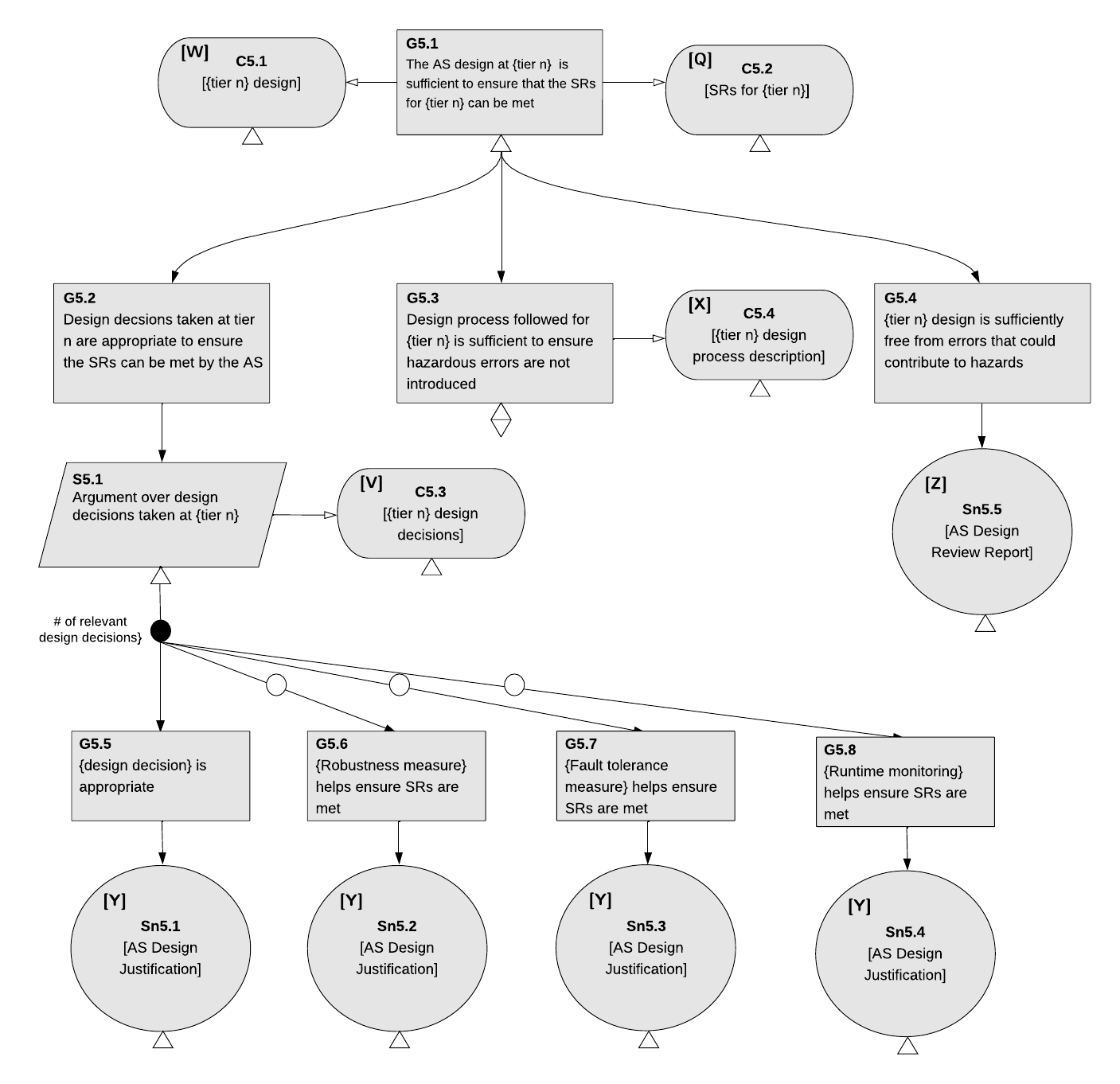}
    \caption{\ArtU~:~Argument Pattern for AS Design Assurance} 
    \label{fig:DesignArg}
\end{figure}

The argument pattern relating to this stage is shown in Figure~\ref{fig:DesignArg} and key elements from the pattern are described in the following sections.

\subsection*{G5.1}
This claim, which is made for each tier of the AS development, demonstrates that the design of the AS is sufficient to ensure that the safety requirements that have been defined at the current tier can be satisfied. This claim is demonstrated by considering the design decisions that have been taken (G5.2), the design process that has been followed (G5.3), and checking for hazardous errors that may have been introduced to the design (G5.4).

\subsection*{G5.2}

A justification must be provided that the key design decisions that have been taken are appropriate to help ensure that the safety requirements can be met by the AS. An argument is presented that considers each of the relevant design decisions in turn. The justifications documented in the AS design justification (\ArtY) are used as evidence that the decisions are appropriate. In many cases, where required by the safety requirements, the design decisions taken will include measures to achieve robustness, fault tolerance and runtime monitoring. The argument pattern indicates how claims regarding these measures could be included as design decisions in the argument (G5.6, G5.7 and G5.8). These claims are indicated as optional elements, since they may not always be required depending upon the nature of the system.

\subsection*{G5.4}

It must be demonstrated that the design at each tier has been checked for errors that may have been made in the design that could result in unsafe outcomes. The design review report (\ArtZ) can provide evidence to support this claim.
\clearpage
\stage{Hazardous Failures Management}

\subsection*{Objectives}
\begin{enumerate}
\item Identify the AS Hazardous Failures that the proposed design could realise
\item Elicit mitigations for the identified AS Hazardous Failures
\item Instantiate the Hazardous Failures Assurance Argument Pattern

\end{enumerate}

\subsection*{Inputs to the Stage}
\begin{itemize}
\item[\ArtB]: Operational Domain Model
\item[\ArtW]: Design at Tier n
\item[\ArtAD]: Hazardous Failures Argument Pattern
\end{itemize}

\subsection*{Outputs of the Stage}
\begin{itemize}
\item[\ArtAB]: AS Safety Analysis Report
\item[\ArtY]: AS Design Justification
\item[\ArtQ]: Safety Requirements for Tier n
\item[\ArtAE]: Hazardous Failures Argument
\end{itemize}

\subsection*{Description of the Stage}

This stage considers the identification and mitigation of hazardous failures of the AS. This considers the design of the AS at each tier to determine how hazardous failures could arise as a result of that design. This is a crucial activity since, even where the design has implemented completely all of the identified safety requirements, it still may be the case that the AS may be capable of doing something else, under certain conditions, that may be hazardous. It is therefore crucial that the potential hazardous failures are identified, and sufficient mitigations put in place.

\begin{figure}[h]
    \centering
    \includegraphics[width=0.8\linewidth]{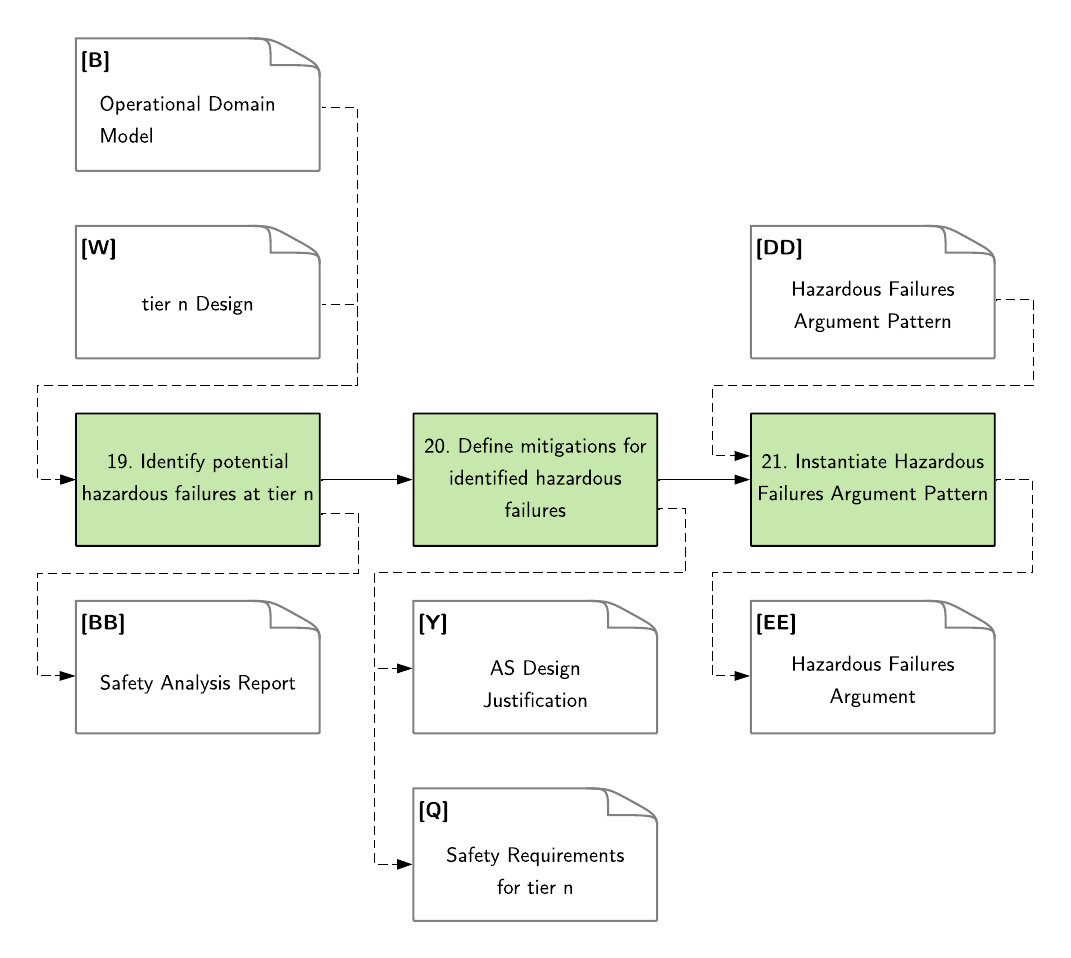}
    \caption{AS Hazardous Failures Process}
    \label{fig:DesignProcess}
\end{figure}

\Activity{Identify Potential AS Hazardous Failures at Tier n \label{art:BB}  \ArtAB}{act:ASHazFail}

The proposed design of the AS at each tier shall be analysed to identify the potential hazardous failures that could arise as a result of that design. Although an AS has been development to satisfy the identified safety requirements, it still may be the case that the AS may be capable of doing something else, under certain conditions, that may be hazardous. Such hazardous failures can often not be identified until detail of the design solution, and in particular the characteristics of the system components is understood.

Potentially hazardous failures are identified by considering possible deviations from intended behaviour that may arise for the AS. Analysis shall be undertaken based upon the information regarding the design at the current tier under consideration (\ArtW). The analysis shall consider information about the particular components proposed as part of that design. In particular the known limitations of the components, their known failure modes, and the conditions under which the components may fail, or under which their performance may deteriorate, shall be considered in the analysis.

\begin{example}
The cameras chosen to be used on an Autonomous Passenger Shuttle to determine its distance from the footpaths adjacent to the road are found to not function well in conditions that present sunlight that is both of a high intensity, and at an acute angle to the road. This is a potentially hazardous failure since it may mean that the shuttle drives too close to the footpath.
\end{example}

Analysing the AS design for possible deviations must be undertaken predictively at higher levels of abstraction where specific component solutions have not been chosen. For this analysis, a technique such as HAZOP \cite{guiochet2016hazard} that uses a set of defined guidewords applied to elements of the design in order to prompt the identification of possible deviations, can be applied. For more detailed levels of design, once the properties of the actual components used are known a more specific deviation analysis techique such as failure modes and effects analysis (FMEA) \cite{BS60812} can be used, perhaps in combination with Fault Tree Analysis \cite{5222114} in order to understand the causal path. Safety analysis such as this forms part of traditional safety engineering processes. For AS it is important that deviations are specifically identified relating to the autonomy of the system (particularly understanding and decision making deviations).

\begin{note}
Although a fairly traditional HAZOP analysis may be applicable to AS, particular attention must be paid to subtle deviations when interpreting the guidewords. For example, when analysing a proposed perception component, careful consideration must be given to how the following guidewords may be defined and interpreted:
\begin{itemize}
    \item More (more than one object detected when only one is present)
     \item Less (less objects are detected than are actually present)
     \item As well as (an extra area is classified as navigable as well as the intended route)
     \item Part of (only part of an object is detected)
     \item Other than (an object is classified other than what it is)
\end{itemize}

It may also be the case that additional guidewords are required when considering AS such as:
\begin{itemize}
    \item \textbf{Intermittent} (considers intermittent detection and/or classification)
    \item \textbf{Erroneous but Credible} (elicits failure conditions relating to information that is incorrect but `believable')
\end{itemize}

\end{note}

Having identified the possible deviations, the hazardous failures are determined by considering which of the deviations, if they occurred in the AS, could result in a hazardous outcome. 

The results of the analyses, including the identified hazardous failures shall be documented in the safety analysis report (\ArtAB). The safety analysis report shall also provide a justification for the sufficiency of the analysis approach used, including the suitability of the approach for the particular tier under consideration and the appropriateness of any modifications made to existing techniques to consider autonomy related deviations.

\Activity{Define Mitigations for Identified Hazardous Failures \ArtQ \ArtY}{act:ASHazMit}

Mitigations shall be defined for all the identified potential hazardous AS failures (\ArtAB). The mitigations could take various forms such as:
\begin{itemize}
    \item defining required design changes
    \item limitations to the operating concept 
    \item deriving additional safety requirements 
\end{itemize}

Where changes are made to the AS design in order to mitigate identified hazardous failures then Activity~19 shall be repeated to ensure additional hazardous failures have not been introduced by those changes. The suitability of the design changes shall be justified as part of the design justification report (\ArtY).

Limitations on the operating concept may include changes to the ROD for the AS to provide additional constraints. The changes to the ROD shall be reflected in the SOC definition (\ArtL).

Any additional safety requirements that are derived shall be added to the existing safety requirements definition (\ArtQ) for implementation. For some of the identified potential hazardous failures it may be determined that the existing design is already sufficient to mitigate those failures (such as through redundancy in the architecture). Where this is the case, this justification shall be documented as part of the design justification report (\ArtY).

\begin{example}
For an autonomous robot operating in an office building, a potential hazardous failure identified from analysis of an object detection component is it may under certain conditions fail to detect walls made of translucent material. In mitigation to this a design change is proposed to add an additional sensor of a different type.
\end{example}

\Activity{Instantiate Hazardous Failures Argument Pattern \label{art:EE} \ArtAE}{act: HazFailAP}

This activity requires as input the hazardous failures argument pattern (\ArtAD), as well as the relevant artefacts from previous activities (\ArtAB, \ArtQ and \ArtY). The activity uses these artefacts to create an instantiated hazardous failures argument for the AS (\ArtAE) which demonstrates that the potentially hazardous failures identified at each tier are acceptably managed.

\subsection*{Artefact \ArtAD: AS Hazardous Failures Argument Pattern}\label{art:DD}

\begin{figure}[h]
    \centering
    \includegraphics[width=1\linewidth]{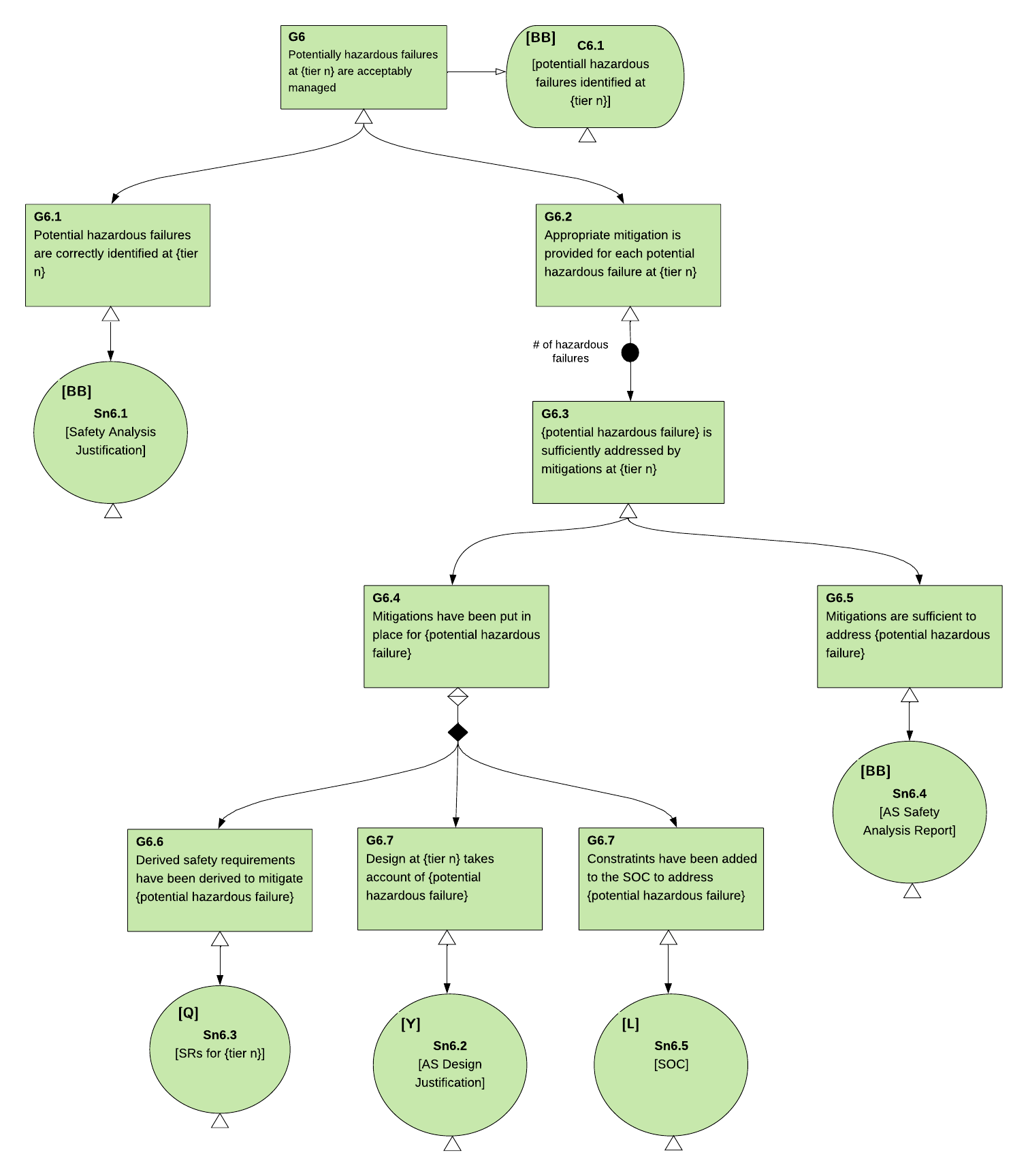}
    \caption{\ArtF~:~Argument Pattern for AS Hazardous Failures} 
    \label{fig:FailuresArg}
\end{figure}

The argument pattern relating to this stage is shown in Figure~\ref{fig:FailuresArg} and key elements from the pattern are described in the following sections.

\subsection*{G6}

The top claim in this argument pattern is that all of the potentially hazardous failures that are identified for tier n (as documented in \ArtAB) are acceptably managed. To demonstrate this it must be shown that the potential hazardous failures have been identified correctly (G6.1) and also that appropriate mitigations have been put in place for each (G6.2).

\subsection*{G6.1}

The details of the safety analysis performed at tier n, as documented in the safety analysis justification report (\ArtAB) is used as evidence that the potentially hazardous failures have been completely and correctly identified.

\subsection*{G6.2}

For each of the identified potential hazardous failures it must be demonstrated that they are sufficiently addressed by the mitigations that are put in place. A separate claim (G6.3) is therefore made for each of the potential hazardous failures.

\subsection*{G6.3}

To demonstrate that each potential hazardous failure is sufficiently addressed, it must first be shown that mitigations for that failure have been put in place (G6.4). As discussed in Activity \ref{act:ASHazMit}, the mitigations could be provided in a number of ways. There is therefore a choice as to how this is demonstrated. In the pattern three choices are provided based upon the use of evidence from safety requirements (Sn6.3), design mitigations (Sn6.2) or constraints placed upon the operating concept of the AS (Sn6.5). Other forms of mitigation may be provided where necessary, and for any hazardous failure multiple forms of mitigation may be used as appropriate. The sufficiency of the chosen mitigations must be justified (G6.5). 
\clearpage
\stage{Out of Context Operation Assurance}

\subsection*{Objectives}
\begin{enumerate}
\item  Demonstrate that the AS will be aware if it is leaving the defined autonomous operating context
\item Implement a strategy that ensures the AS remains sufficiently safe even if it leaves the defined autonomous operating context
\item  Instantiate the Out of Context Operation Assurance Argument Pattern
\end{enumerate}

\subsection*{Inputs to the Stage}
\begin{itemize}
\item[\ArtB]: Operational Domain Model
\item[\ArtAF]: Key features of Environment Outside ODM
\item[\ArtAJ]: ODM Transition Model
\item[\ArtAL]: Stakeholder Risk Acceptance Definition
\item[\ArtAP]: Out of Context Operation Assurance Argument Pattern
\end{itemize}

\subsection*{Outputs of the Stage}
\begin{itemize}
\item[\ArtAG]: Out of Context Analysis Report
\item[\ArtAH]: Interpretation of ODM Boundary
\item[\ArtAI]: ODM Boundary Assessment Report
\item[\ArtAK]: Transition Assessment Report
\item[\ArtAM]: Outside ODM Minimum Risk Strategy
\item[\ArtAN]: Outside ODM Strategy Justification Report
\item[\ArtAO]: Outside ODM Verification Report
\item[\ArtAQ]: Out of Context Operation Assurance Argument
\end{itemize}

\subsection*{Description of the Stage}

\begin{figure}[h]
    \centering
    \includegraphics[width=\linewidth]{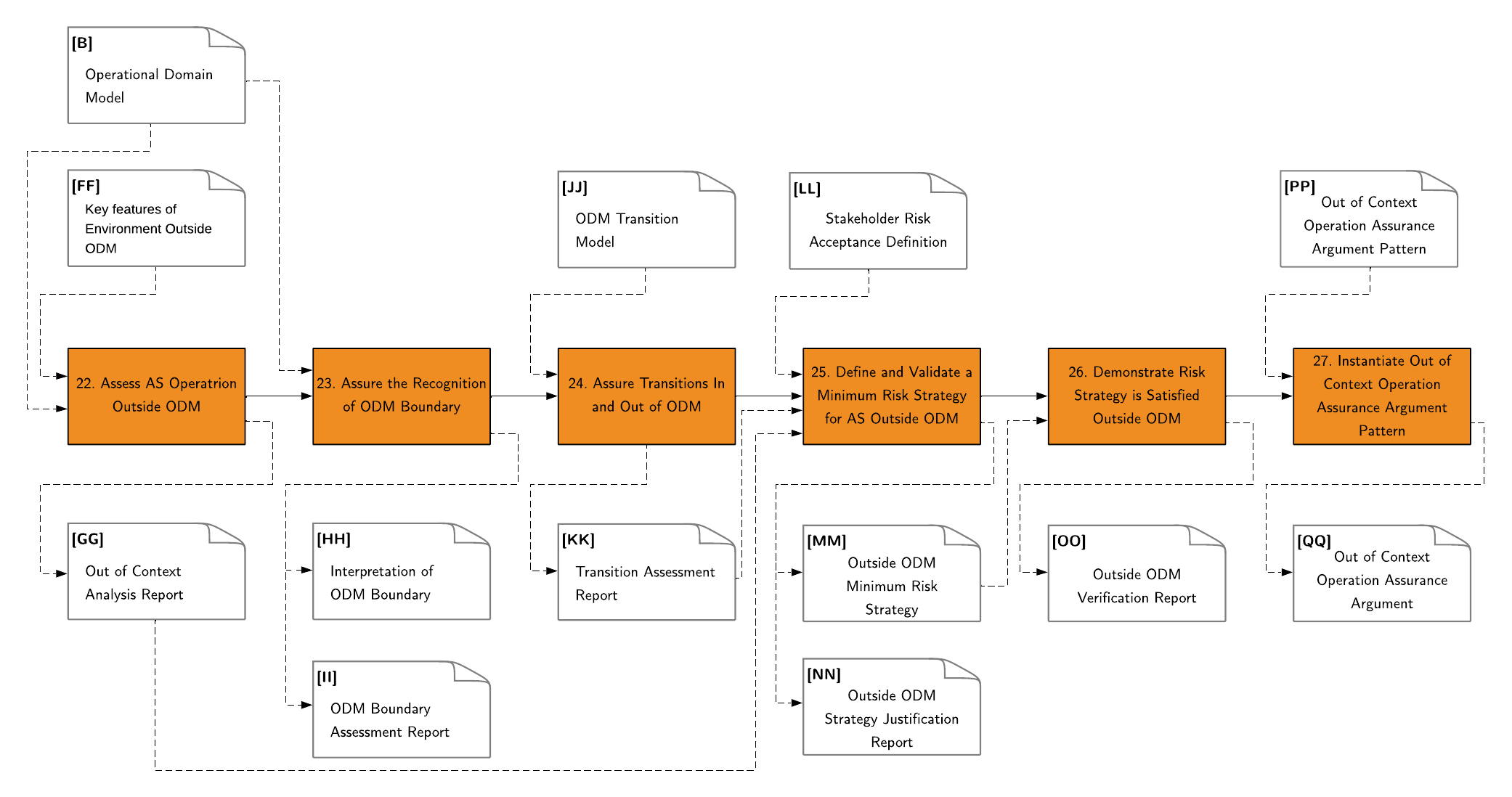}
    \caption{Out of Context Operation Assurance Process}
    \label{fig:outProc}
\end{figure}

As shown in Figure \ref{fig:outProc}, this stage consists of six activities that are performed to define and validate the safe out of context operation for an AS. The artefacts generated from this stage are used to instantiate the Out of Context Operation Assurance Argument Pattern as part of Activity \ref{act:ASoutPattern}.

An AS may spend some time operating outside the defined ODM (\ArtB) whilst still be operating autonomously. This could be unsafe, since autonomous operation is only assured for safety within the defined ODM. There are several situations where operation outside the ODM may occur: 

\begin{enumerate}
    \item The environment or context of the AS suddenly changes without warning (see example \ref{floodEx}).
    \item The AS fails to recognise the boundary of the ODM (see example \ref{sleetEx}).
    \item The boundary of the ODM is poorly defined, ambiguous or has dynamically changed. As in example \ref{sleetEx}, the transition between classes of weather conditions may for example be ambiguously defined.
    \item The AS does not recognise the boundary of the ODM within an acceptable period of time.
    \item The AS recognises the boundary but is unable to hand over to another function or an operator (either because none are available or the transition itself fails) and therefore continuation in autonomous mode is the safest option.
    \item The AS fails to transition out of autonomous mode quickly enough (this could take seconds or even minutes).
\end{enumerate}

\begin{example} \label{floodEx}
An autonomous road vehicle encounters flash flooding on the road. Such flood conditions are not within the ODM as they  cannot be handled safely by the autonomous driving function. Since the flash flood conditions arise suddenly and unexpectedly it is not possible for the vehicle to anticipate and avoid these conditions.
\end{example}

\begin{example} \label{sleetEx}
The ODM for an agricultural robot includes rain but not snow as the robot cannot operate safely in snowy conditions. During operation in heavy rain, the low temperature causes the rain to become sleet, followed by a transition to snow. The AS is unclear as to when sleet becomes snow and hence when it has moved outside of the ODM.
\end{example}

\begin{note}
It is expected that time spent in autonomous mode outside the ODM should be limited, or indeed transient. Note that one option for dealing with outside ODM operating is to get back within the ODM as soon as possible.
\end{note}

\Activity{Assess AS Operation Outside ODM \label{art:GG} \ArtAG}{act:assOut2}

In assessing the operation of an AS outside its ODM, it is important to understand the characteristics of the outside ODM environment relevant to the AS and its behaviours. Therefore this activity requires a description of the relevant key features that are anticipated in the environment outside of the ODM (\ArtAF). This is used to establish the scenarios that may arise due to excursions outside of the ODM.

\begin{example}
A drone may be blown out to sea due to severe weather conditions and lose contact with its base station. It must be aware that it is now outside of its defined ODM, and that a landing on water may not be appropriate. 
\end{example}

Those scenarios shall then be analysed to determine those which may be hazardous. There are a number of techniques that can be applied for this analysis such as Hazop \cite{guiochet2016hazard}, STAMP/STPA \cite{leveson2018stpa} or FRAM  \cite{erik2017fram}. The analysis should involve personnel who understand the operating environments and can establish the most likely hazards for the AS if outside of the ODM. 
The identified hazardous scenarios due to excursions outside the ODM shall be documented in the Out of Context Analysis Report (\ArtAG) along with details of the analysis performed and any limitations on the analysis, e.g. assumptions regarding the environment outside the ODM.

\subsection*{Artefact \ArtAF: Description of Key Features of Environment Outside ODM}\label{art:FF}

This arteafact should describe the pertinent features of the outside ODM environment and outlines how this may need to impact on the AS and its behaviour if the AS exits the ODM while in autonomous operation. A tabular approach may be appropriate where the key aspects inside and outside the ODM with their differences are detailed. It is not expected that this approach is exhaustive; only those aspects of the outside ODM environment which can be imagined can be analysed. Generalisations and approximations may be appropriate in some cases. In many cases the outside ODM environment may not be particularly different to the ODM environment.	 

\begin{note}
It may be extremely difficult to ascertain all aspects of the ODM environment and therefore approximations, models or heuristics may need to be to be applied. Sensing and communications may be key, as the existing sensor set must be able to make enough sense of the outside ODM environment to be able to manage the situation.
\end{note}

\Activity{Assure the Recognition of the ODM Boundary \label{art:HH} \label{art:II}  \ArtAH, \ArtAI}{act:boundRecog}

Safe operation of an AS requires that the ODM boundary is correctly recognised. If the AS is unaware that its operation has moved outside of the ODM as defined in \ref{art:B} then its safety may not be assured. 

\begin{example}
A car is only capable of operating autonomously on a motorway, but during operation there are lane closures and move to a minor road running due to an accident. In this case the car must recognise that this situation represents an ODM boundary, as minor roads are not included in the ODM.
\end{example}

The approach that the AS will use during operation to determine and interpret the ODM boundary shall be determined based upon a consideration of the capability of the AS to sense and understand the ODM. Since perfect recognition of the ODM boundary will not be possible, it will always be necessary to make approximations and assumptions to reflect the AS sensing capabilities. In some cases it may not be possible to directly detect the ODM features using the sensors available to the AS. In such cases it may be that ``proxy'' measurements are required to be used to recognise the ODM boundary.

\begin{example}
The ODM for an autonomous car specifies a maximum permitted intensity of rainfall. The car is fitted with a rainfall sensor as part of the automatic windscreen wiper system. The vehicle makes use of this sensor for recognising the ODM boundary. As such the rapid wiper threshold is used as a proxy for the maximum rainfall intensity permitted by the ODM. Since the rapid wiper threshold is less than the intensity defined for the ODM this is determined to be acceptable to use for ODM recognition.
\end{example}

\begin{note}
Recognition of the ODM boundary is often challenging for a number of reasons including:
\begin{itemize}
    \item Many parameters may interact to form the ODM boundary, such as weather conditions, speed etc. For instance, an autonomous vehicle may only be able to progress through fog when visibility is at least 10 metres, during daytime, when speed is less than 60mph.
    \item There may be a complex boundary shape/envelope/volume with ‘holes’ or difficult geometry. For example a medical image recognition system which can be used for detection of tumours in radiological images only for patient age ranges 40-45 and 65-85 (due to the extent of the available training data).
    \item The AS itself may have to use an interpretation of the boundary which may be a different, or simplified approximation of the actual ODM boundary, to give appropriate margins. For example instead of a more complex boundary condition where a drone is able to fly in wind speeds less than 15 km/h and gusts of up to 20 km/h, the interpretation of the boundary the drone uses is wind speeds less than 12 km/h.
    \item It may take some time for the processing and analysis of the sensor data to establish whether the AS is near or has crossed the ODM boundary.
    \item The sensor data used to determine the ODM boundary may be approximate, noisy or subject to infrequent updates. All sensors have an accuracy, resolution and a reading lag time; some also have a polling interval. Sensors may age and deteriorate and readings drift or become subject to bias or noise over time. Individual sensor data may be subject to errors or variations and may have to be averaged with others or over time. This is particularly a problem in harsh environments such as marine, automotive or aviation. In this case the sensing of the ODM may be delayed, or incorrect for some time. Margins therefore have to be implemented for operation, i.e. working to a smaller ODM, so that in all reasonable scenarios the actual ODM boundary can be sensed.
    \item `Flip-flopping' between boundary states may occur, i.e rapid switching between in and out detections. The maximum rate at which changes with respect to the ODM boundary are detected by the AS must be considered. It may be better to indicate the AS as outside of ODM for a minimum period of time if there are likely to be many boundary detection events in a short period of time. Hysteresis biased towards early recognition, i.e. conservative detection, may be needed.
\end{itemize}

\end{note}

The approach for determining the ODM boundary during operation shall be documented (\ArtAH) together with any  assumptions and approximations made.

It shall be demonstrated that the AS is able to recognise the ODM boundary as interpreted during operation. This will involve consideration of at least four recognition cases:

\begin{enumerate}
    \item AS is approaching the ODM boundary from within the ODM.
    \item AS is crossing the ODM boundary.
    \item AS is approaching the ODM boundary from outside the ODM.
    \item AS is re-entering the ODM from outside.
\end{enumerate}

For each of the four cases above, it shall be determined how these may be unsafe through consideration of:

\begin{enumerate}
    \item Timeliness - ODM boundary recognised too early or too late.
    \item Accuracy - false positive recognition (ODM boundary recognised mistakenly) or false negative recognition (ODM boundary not recognised when it should be).
    \item Hysteresis - AS holds on to a ODM boundary recognition state for too short or too long a period.
\end{enumerate}

For any of the cases that are determined to be potentially hazardous it shall be demonstrated that those cases are sufficiently mitigated by the AS. There are a number of approaches for this including testing of scenarios relating to each of these cases. Testing alone however may not be able to provide sufficient evidence where the ODM boundary is complex (as testing can only sample the boundary space). Simulations and analysis of the ODM boundary recognition may therefore also be needed (see Stage~\stageref{8} for further guidance on this). 

The results of the ODM boundary recognition assessment shall be documented (\ArtAI).

\Activity{Assure Transitions In and Out of ODM \label{art:KK} \ArtAK}{act:boundTrans}

\begin{figure}[p]
    \centering
    \includegraphics[width=1\linewidth]{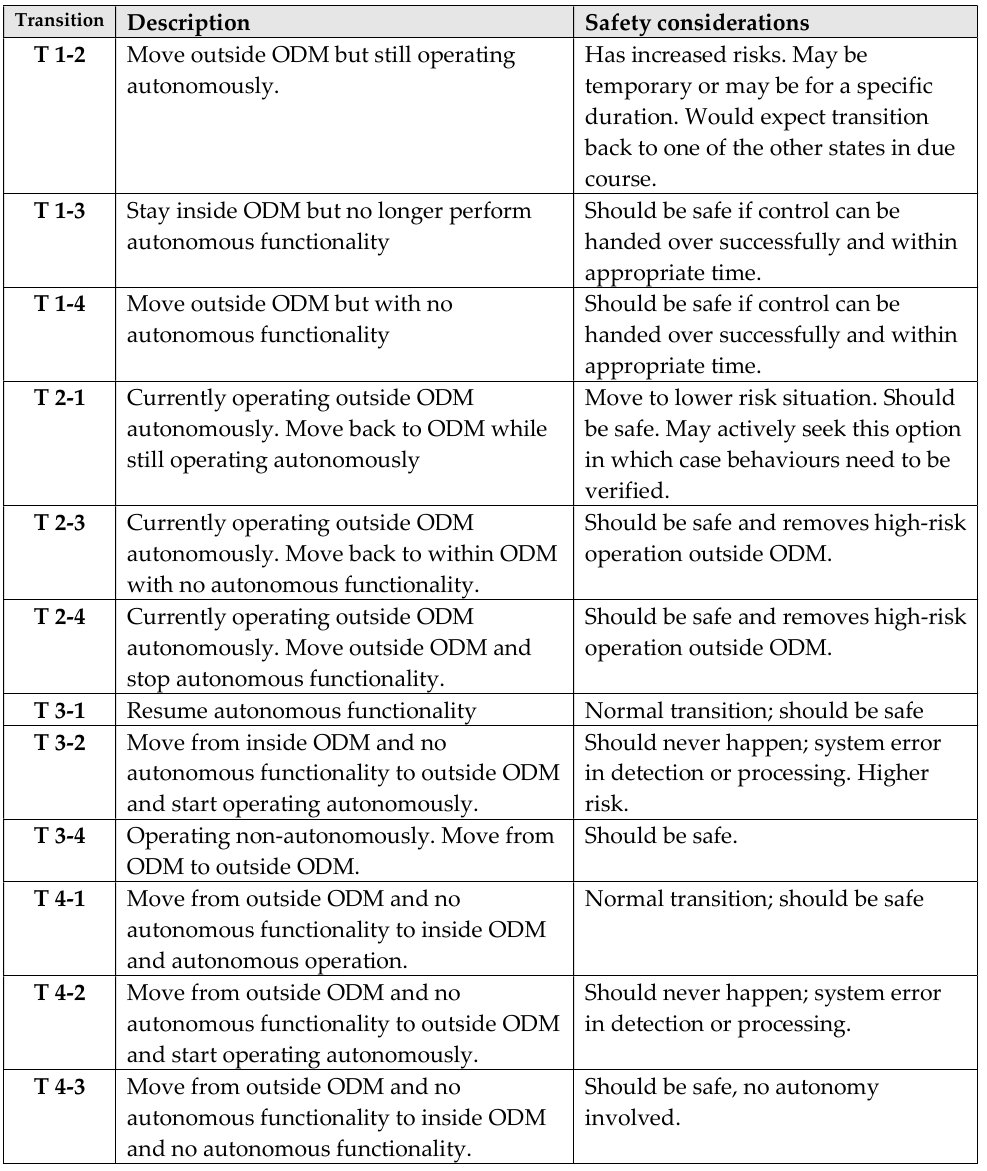}
    \caption{Typical Unsafe Transitions for an AS Across the ODM Boundary} 
    \label{fig:transitionTable}
\end{figure}

This activity shall consider the transitions of the AS from autonomous to non-autonomous operation when it crosses the ODM boundary. This seeks to identify ways in which those transitions could be unsafe such that suitable mitigations can be provided. This requires as input an ODM Transition Model (\ArtAJ) for the AS. Based on the transition model, the ways in which the possible AS transitions may be unsafe shall be determined. This will allow measures to be defined to minimise the risk of the unsafe transitions for the AS operation. The table in Figure \ref{fig:transitionTable} provides examples of the nature of typical unsafe transitions. The transitions in the table refer to those from the ODM Transition Model (\ArtAJ) shown in Figure \ref{fig:transModel}.

\begin{example}
An autonomous ship can self-navigate without intervention as long as it has sufficient channel width margins. If it progresses down a seaway of varying width, the autonomous navigation function may need to engage and disengage, repeatedly handing control to the ship’s crew. In this scenario there are a number of different unsafe transitions that could occur. This includes remaining in autonomous control when the channel width becomes too narrow (which may be beyond the safe capability of the autonomy), as well as relinquishing control when in a wide channel (when the crew may not be in a position to safely take control). In certain situations the ship may attempt to mitigate such deviations by getting back inside the defined ODM, for example by taking a route that involves wider channels.
\end{example}

The results of assessment of the transitions shall be documented along with any assumptions made (\ArtAL).

\subsection*{Artefact \ArtAJ: ODM Transition Model}\label{art:JJ}

\begin{figure}[h]
    \centering
    \includegraphics[width=0.8\linewidth]{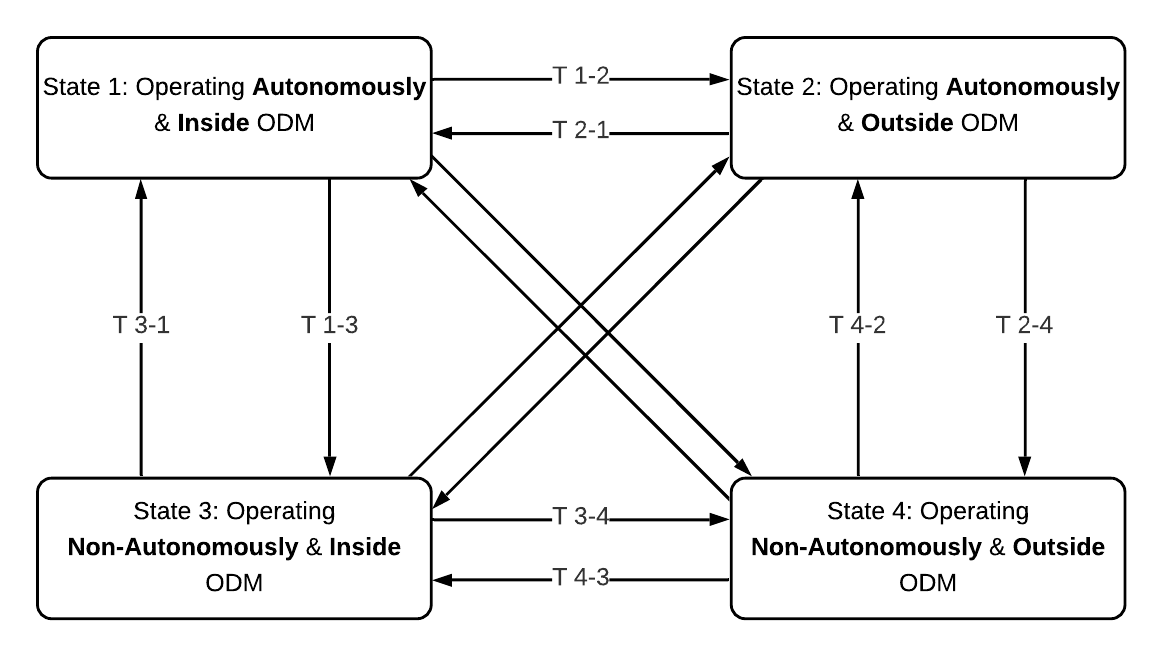}
    \caption{Generic ODM Transition Model (adapted from \cite{birch2020structured})} 
    \label{fig:transModel}
\end{figure}

An ODM transition model should define the states of the AS with respect to the defined ODM boundary, and the transitions that may occur between those states. Figure \ref{fig:transModel}, adapted from \cite{birch2020structured} shows a generic model of ODM transitions. For the particular AS under consideration a transition model of this form should be created to document the transitions that may occur during operations and the conditions under which this may come about.

\Activity{Define and Validate a Minimum Risk Strategy for AS Outside ODM \label{art:MM} \label{art:NN} \ArtAM \ArtAN}{act:riskStratOut}

Based upon the assessments of the hazardous scenarios outside the ODM (\ArtAG) and the transitions across the ODM boudary (\ArtAK) undertaken in activities 22 and \ref{act:boundTrans}, a minimum risk strategy for the operation of the AS outside the ODM shall be defined. A minimum risk strategy determines how the AS should respond when moving outside the defined ODM in order to minimise safety risk. The definition of the minimum risk strategies shall take account of the risk acceptance of the relevant stakeholders, as documented in (\ArtAL).

The minimum risk strategy will vary depending upon the particular AS and its operation. 
For example in many cases the safest strategy may simply be for the AS to return within the ODM as quickly as possible (for instance if an autonomous submersible vehicle sinks to a depth outside its ODM it should rise back up to a level that is within the ODM). In many cases however, this may be difficult or even impossible for the AS to achieve (for instance if an autonomous car encounters extreme whether conditions that are outside of the ODM, the car cannot change these conditions and an alternative way of minimising the risk under those conditions must be defined).
The strategies employed will be dependent upon the state of the AS and the state of the environment and it is possible that a set of principles or heuristics may need to be employed to cover all situations. In some cases the minimum risk strategy will be a set of behaviours rather than a specific action (akin to self-preservation behaviours, called ‘Safe Modes’ in spacecraft). Where it is possible, the minimum risk strategy may include the AS handing control over to a human operator.

\begin{example}
A satellite may have to rely on extended periods of loss of communications with ground stations, e.g. during eclipses. In these periods, it must be able to make decisions about situations and failures that occur on-board, from loss of power to loss of orientation. These may require reverting to ‘safe modes’ where the satellite has a series of basic self-preservation modes that protect functions (e.g. orientate solar panels to the sun and point antenna at ground and listen for commands). Analogous behaviours need to be defined for many autonomous systems, e.g. robotic underwater vehicles where they must also not endanger other vessels / humans while in safe mode. A particular issue is that of autonomous road vehicles where there are occupants and 3rd parties to protect, and the vehicle may have to cope with failures as well as unexpected operating conditions.
\end{example}

\begin{note}
Although in many situations handover of control to a human operator can be an appropriate strategy to adopt, there are many challenges that must be considered to ensure this is done safely.

\begin{itemize}
    \item \textbf{Handover time} - It is crucial to assess whether a human operator can react quickly enough when they are expected to take over in order to do so safely. Experiments have shown that it can take quite a long time for a human operator to take control of an AS when required. For example it has been seen to take around 35 seconds for a human driver to stabilise the lateral control of an autonomous car (see also \cite{gold2013take} and \cite{merat2014transition}). The consideration here is not just the reaction time of the operator, but how long it takes until the operator is in a position to be able to operate the system safely. This includes the need for the operator to gain sufficient situational and contextual awareness.
    
    \item \textbf{Monitoring} - Successful handover may require the human to monitor the state of the AS. Monitoring is not a stimulating task and it is therefore something that humans are generally not good at. Consideration should therefore be given to whether it is reasonable to expect a human to effectively monitor the behaviour of an AS over long periods of time. This is particularly the case if the human operator is rarely asked to do anything more engaging. Even when a human is monitoring an AS, it is likely that they may often miss the requirement to intervene. There can also be problems with ensuring that all of the required information is available to the human monitor in a timely manner for them to safely intervene. This is even more challenging if the operator is remote from the AS.
    
    \item \textbf{Competency} - Since handovers to human operators may be quite rare, human operators will often have less opportunity to practice, and can become de‐skilled as a result. It is unrealistic to expect humans in many situations with autonomous functions to be able to maintain up to date skills, with associated formal certification and competence (like commercial pilots have to do today). The effects of this are exacerbated since the requirement to handover will generally occur during difficult or unusual situations for which higher competency may be required.
    
    \item \textbf{Unsafe Override} - The AS may be behaving in a safe manner, but if an operator monitoring the system does not understand what the system is doing and why, then the human operator may incorrectly decide to override and take control. Because the operator did not correctly understand the situation, the override may actually put the AS into a less safe state. It may be inherently difficult for a human operator to interpret the decisions that the AS is taking at any point in time and why those decisions are being taken. 
    
    \item \textbf{Authority} - Safety issues can arise when there is ambiguity over who has explicit authority or when authority is transitioned between the human operator and the AS or vice versa. The ambiguity can lead to unexpected actions which could be unsafe, or lead to mode confusion since it becomes unclear which agent is in control of the AS at any point in time. 
\end{itemize}

\end{note}

The selected Minimum Risk Strategies (\ArtAN)and a justification for their sufficiency shall be documented (\ArtAM).

\subsection*{Artefact \ArtAL: Stakeholder Risk Acceptance Definition}\label{art:LL}

Stakeholder Risk Acceptance Definition describes which risks the different stakeholders involved are prepared to hold. The risks each stakeholder are prepared to hold and manage are important as they can significantly change the risks which have to be mitigated as part of the AS functionality, and hence the mitigation strategy.

\begin{note}
It is often the case that the stakeholders that hold the risk are different from the stakeholders that make the decision on risk acceptability. The developers are therefore making this judgment of behalf of those who hold the risk. For instance for an autonomous car, the risk is held by the the vehicle occupants and other road users. The people who make the judgment of whether the risk associated with the autonomy is acceptable however are the automotive engineers who design and build the vehicle.  It is therefore important that the basis for risk acceptance is documented and communicated.
\end{note}

\Activity{Demonstrate Risk Strategy is Satisfied Outside ODM \label{art:OO} \ArtAO}{act:ASoutStratDemo}

It shall be demonstrated that the minimum risk strategy defined for outside the ODM (\ArtAM) will be satisfied by the AS during operation. This requires that evidence is generated which shows that the requirements of the strategy are implemented by the AS. This shall be achieved by performing verification activities, which could take a number of forms. The verification process for AS is discussed in detail in Stage~\stageref{8}. The results of verification activities shall be documented in the outside ODM verification report (\ArtAO).

\Activity{Instantiate Out of Context Operation Assurance Case Pattern \label{art:QQ} \ArtAQ}{act:ASoutPattern}

This activity requires as input the AS hazardous scenarios assurance argument pattern (\ArtAP), as well as the artefacts from Activities~\actref{act:assOut2}{22} to~\actref{art:OO}{26}. The activity uses these artefacts to create an instantiated out of context operation assurance argument  which demonstrates that the hazardous scenarios have been correctly identified.

\subsection*{Artefact \ArtAP: AS Hazardous Scenarios Assurance Case Pattern \label{art:PP}}

\begin{figure}[h]
    \centering
    \includegraphics[width=1.1\linewidth]{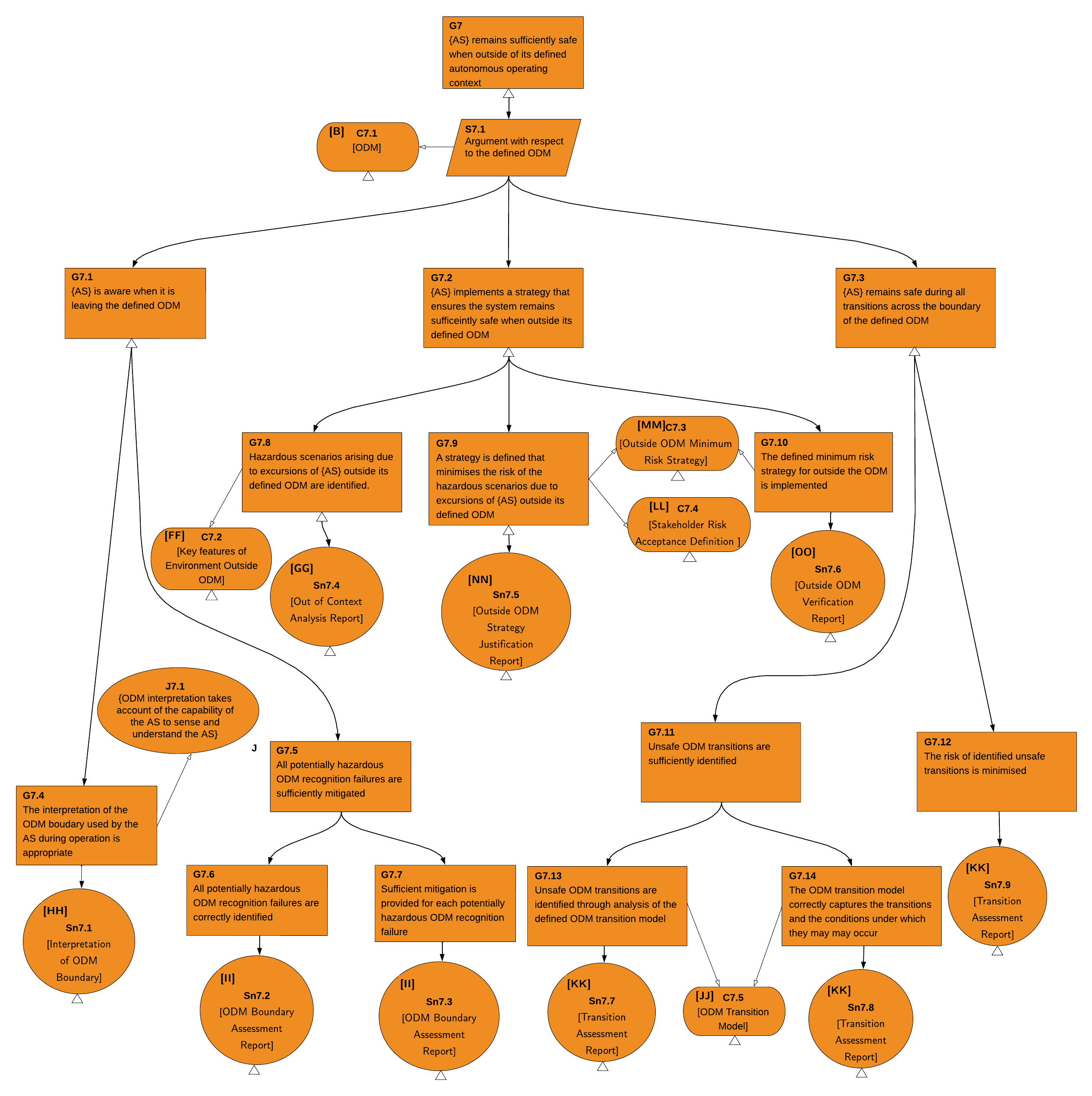}
    \caption{\ArtAP~:~Argument Pattern for Out of Context Operation Assurance} 
    \label{fig:outArg}
\end{figure}

The argument pattern relating to this stage is shown in Figure~\ref{fig:outArg} and key elements from the pattern are described in the following sections.

\subsection*{G7}

The top claim in this argument pattern is that the AS will remain sufficiently safe even if it were to move outside of the defined autonomous operating context. This is achieved through consideration of the defined ODM (\ArtB). It is firstly demonstrated that the AS is aware of when it is leaving the ODM (G7.1). Secondly it is demonstrated that if the AS does leave the ODM, a strategy is implemented that ensures the AS remains sufficiently safe (G7.2). Lastly it must also be demonstrated that the AS remains safe as it transitions across the ODM boundary.

\subsection*{G7.1}

To demonstrate that the AS is aware when it's leaving the defined ODM, it must be shown that the way in which the AS interprets the boundary to the ODM (defined in \ArtAH) is appropriate. An explicit justification for this must be provided as part of the argument (J7.1) that should take account of the capability that the AS has to sense and understand the operating environment. It must be shown that any failures in recognising the ODM boundary that may be hazardous have been identified, and that sufficient mitigation is provided for those failures. The ODM boundary assessment report (\ArtI) provides evidence for this.

\subsection*{G7.2}

To demonstrate that the AS remains safe when operating outside of the defined ODM, the results of the out of context analysis (\ArtAG) provide evidence that the hazardous scenarios arising from such excursions have been identified. It is then shown using the outside ODM strategy justification report (\ArtAN) that the chosen strategy minimises the risk associated with those hazardous scenarios to an acceptable level. The outside ODM verification report (\ArtAO) is used to demonstrate that the minimum risk strategy has been correctly implemented.

\subsection*{G7.3}

To demonstrate that the AS will remain safe as it transitions across the ODM boundary, it must be shown that unsafe transitions are identified (G7.11) and the risk of those transitions is minimised (G7.12). The transition assessment report (\ArtAK) is used as evidence that the unsafe transitions are identified correctly through analysis of the transition model. It is also used to demonstrate that the transition model itself is correct in the way it models the transitions and the conditions under which those transitions occur.
\clearpage
\stage{AS Verification Assurance}

\subsection*{Objectives}
\begin{enumerate}
\item  Generate evidence to demonstrate that the safety requirements specified for the AS are satisfied
\item Justify the sufficiency of the verification activities
\item Create the AS Verification Argument
\end{enumerate}

\subsection*{Inputs to the Stage}

\begin{itemize}
\item[\ArtB]: Operational Domain Model
\item[\ArtD]: Autonomous Capabilities Definition
\item[\ArtE]: Operating Scenarios Definition
\item[\ArtL]: SOC
\item[\ArtQ]: Safety Requirements for tier n
\item[\ArtAU]: AS Verification Argument Pattern
\end{itemize}

\subsection*{Outputs of the Stage}

\begin{itemize}
\item[\ArtAR]: Verification Strategy
\item[\ArtAZ]: Verification Plan
\item[\ArtAS]: AS Verification Log
\item[\ArtAT]: Verification Results
\item[\ArtAV]: AS Verification Argument
\end{itemize}

\subsection*{Description of the Stage}

\begin{figure}[h]
    \centering
    \includegraphics[width=1\linewidth]{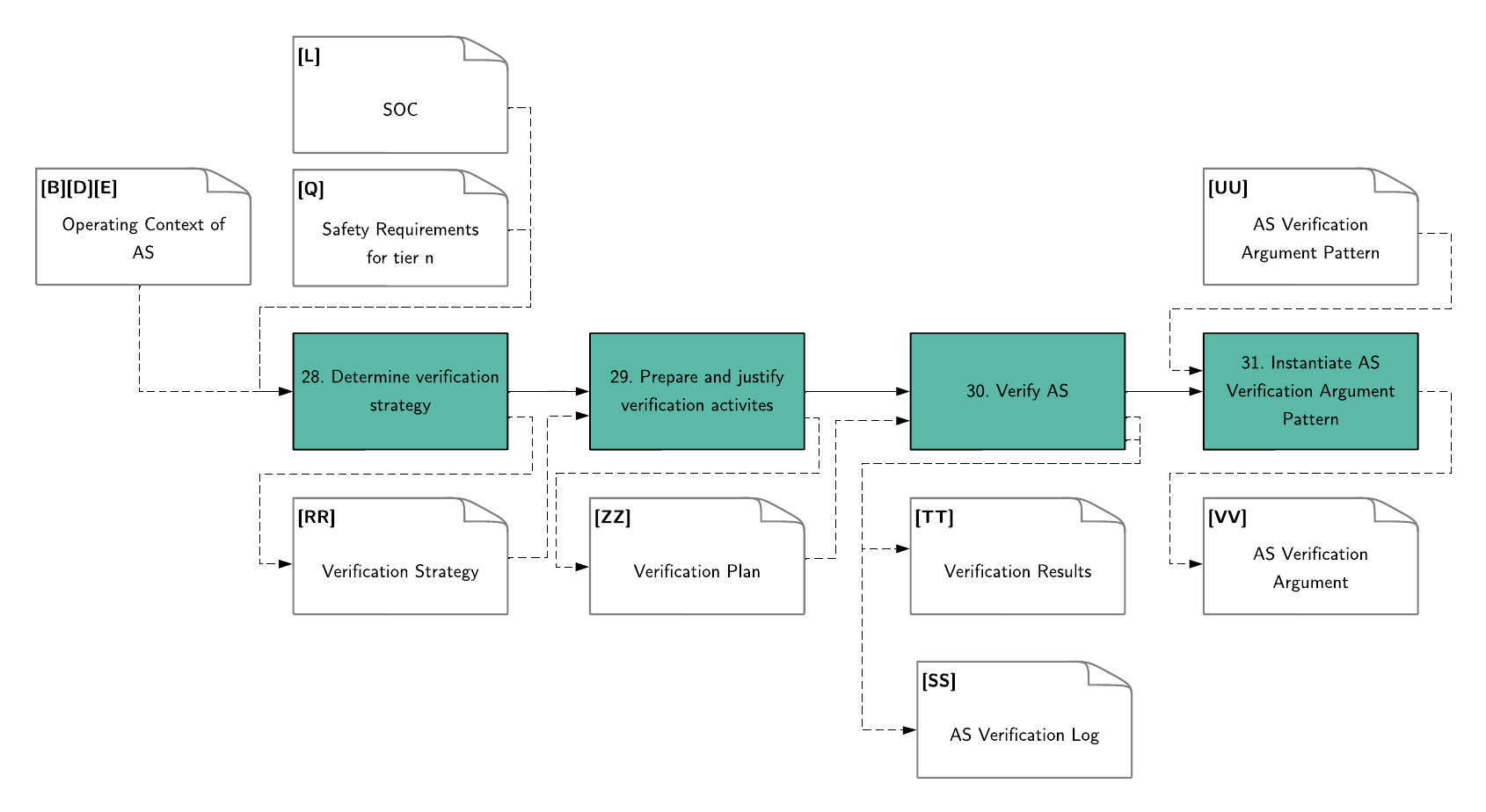}
    \caption{AS Verification Process}
    \label{fig:VerifProcess}
\end{figure}

The purpose of verification is to generate evidence to demonstrate that the safety requirements specified for the AS are satisfied within the defined operating context. Verification may be carried out for different tiers of the system decomposition. For example, as illustrated in Figure \ref{fig:VerifProcess}, this may involve performing verification of the requirements on individual components of the AS and on the sub-systems created through integrating components together, as well as verification of the AS platform as a whole. At each tier, the verification is performed with respect to the safety requirements defined for that particular tier.

\begin{note}
In many regards, the approaches to verification of AS are no different from those used for traditional systems. However, as the behaviour of AS is often less bounded than for traditional systems, this can give rise to particular challenges, particularly when verifying AS that operate in complex environments. In \cite{song2021concepts} the following non-exhaustive list of challenges to testing AS are identified:
\begin{enumerate}
    \item  Unpredictable Environment: This unpredictability adds uncertainties to testing as the AS can run into environmental variables that may have been unknown during design.
    \item System and Scenario Complexity: It is unclear how to define scenarios that include all features involved in the operational environment and the system itself and how to check whether the scenarios used for testing reflect the actual situations encountered.
    \item Data Accessibility: To be able to test the systems, good quality data is required. This data can be difficult and costly to collect, interpret and validate for AS.
    \item Missing Standards and Guidelines: No accepted standards and guidelines are established for testing of AS. This makes determining the sufficiency of the testing activities used more difficult, and harder to justify.
\end{enumerate}
\end{note} 

\begin{note}
For low-level tiers, such as individual components that use particular technology, particular approaches may be required for which further, more detailed guidance is required (such as that provided for components that use machine learning in \cite{AAIP2021a}).
\end{note}

\Activity{Determine Verification Strategy \label{art:RR}  \ArtAR}{act:verifStrat}

An appropriate verification strategy shall be adopted. This shall describe how, in general, the project approached verification, and justify how the various verification activities performed work together to give adequate overall verification.

At the highest level this will require a decision on whether verification shall be undertaken using testing, formal verification or a combination of both approaches. The advantages and disadvantages of using either approach for the verification of AS are discussed below.

\textbf{Testing}

Testing involves exposing the system to a defined set of inputs and checking that the outputs or behaviour of the system is as specified and safe. The main advantage of using testing is that evidence may be generated against the real system operating in the target environment. This means that demonstrating the relevance of the verification evidence to the AS operation is comparatively straightforward. Some testing of AS may however be performed in a controlled environment, such as a test site or a laboratory setting, requiring that the similarity to the actual operating environment is demonstrated. The main disadvantage is that testing cannot be carried out exhaustively, so achieving an acceptable level of coverage of the operating domain can be very challenging, particularly for AS in complex operating environments that represent a very large state space. Providing justification for the acceptability of the level of coverage that is achieved is also a major challenge.

\textbf{\emph{Simulation}}

Testing does not always need to be performed using the target AS in its intended operating environment. Particularly for AS, simulation may also be used for the purposes of testing. Simulation may be used for a number of reasons: it can allow verification activities to begin earlier in the development lifecycle, since a fully developed system is not required; it can also enable tests to be carried out that would otherwise be difficult to create (due to their rarity in the real world) or dangerous to perform (since they would require exposing people to unnecessary risk); it can also be used as a way of increasing the coverage that can be achieved through testing, by enabling test cases to be created quickly and cheaply. The disadvantage of using simulation is that all simulations are models of the real world and justifying the representativeness of the simulation models can be challenging.

``In-the-loop'' simulation is an approach that can be used as part of the verification of an AS in which real system components are used to provide inputs to the simulation. Using an in-the-loop approach can help to ensure that the data used by the simulation during verification is representative of the data to which the AS will be exposed during operation.

\begin{example}
There are many examples that illustrate the use of in-the-loop simulations for AS in many different applications, such as autonomous driving (\cite{deng2008hardware}) and unmanned aerial vehicles (\cite{lepej2017flexible} and \cite{mueller2007hardware}).
\end{example}

\textbf{Formal Verification}

Formal verification involves using mathematical techniques to prove that a formally specified model of (some aspect of) the AS will always satisfy a set of formally specified properties. The main advantage of using formal verification techniques is that the properties are proved to be true for the entire scope of the model of the AS, meaning that the challenge of demonstrating sufficient coverage that existed for testing is avoided. The disadvantage of using formal analysis is being able to demonstrate that the analysis results are a true reflection of the real system in operation. Firstly, in order to support verification, the formally specified properties that are defined must sufficiently capture the intent of the safety requirements being verified. Secondly, the formal model, upon which the properties are proven to hold, must be sufficiently representative of the system itself, and any elements of the operating environment included in the model. All models require abstractions and assumptions to be made. For complicated AS operating in complex environments demonstrating the acceptability of the abstractions and assumptions made is challenging. Due to these challenges, formal methods should be used with caution for AS, and should be done so in combination with testing. An overview providing further details of different applications of formal methods to AS is available at \cite{luckcuck2021using}.

The chosen verification strategy shall be documented (\ArtAR).

\Activity{Prepare and justify verification activities}{act:verifAct}

Having decided upon a verification strategy, the activities required to be undertaken to carry out that strategy shall be defined. The sufficiency of those planned verification activities shall be justified to explain how they provide sufficient evidence with respect to the safety requirements. The verification activities and their justification shall be documented as part of a verification plan (\ArtAZ).

\subsection*{Artefact \ArtAZ: Verification Plan}\label{art:ZZ}

The verification plan shall explicitly detail the verification activities as well as documenting the rationale for those activities. The plan is a more concrete implementation of the strategy, defining what should be done (e.g. specific tests) and how (e.g. environment, tools). Some of the key considerations that should be included in the verification log are given below.

\textbf{Testing}

Details of the test cases shall be provided along with the required result for that test case. A justification for the sufficiency of the coverage that the test cases provide shall be included in the verification plan. For AS the coverage should in particular consider:

\begin{itemize}
    \item Safety requirements - Are there test cases relating to each of the safety requirements defined for the relevant tier?
    \item ODM - Are there a sufficient range of test cases for each of the relevant ODM features? Are a sufficient range of combinations of ODM features represented by the test cases?
    \item Operating Scenarios - Do the test cases provide sufficient coverage of the possible operating scenarios of the AS? \cite{tahir2020coverage} provides a survey of the coverage criteria that have been proposed for verification of autonomous vehicles and possible  techniques for maximizing coverage.
    \item Edge cases - Do the test cases include sufficient examples of edge cases (low probability, hard to predict events or situations with high potential safety impact) to which the AS may be exposed during operation? 
\end{itemize}

Details of the test environment shall be included in the verification plan. A justification shall also be provided that the test environment is sufficiently representative of the real operating domain of the AS. This justification is required in order to provide confidence that the results obtained for the test cases for the AS in operation would match the results observed in the test environment. This is particularly important for simulation-based testing, where justification must be provided that the simulation tools and models are sufficiently valid to support the claims made as a result of the tests.

\textbf{Formal Verification}

The formal properties shall be specified in the verification plan along with the rationale for the specification and a justification that the specified properties are equivalent to the relevant safety requirements. The formal models that are used for verification shall be documented in the verification plan along with a justification for all assumptions and abstractions made in the model, both with respect to the AS itself, and with respect to the operating environment.

\Activity{Verify AS \label{art:TT}  \ArtAT}{act:verify}

The verification activities described in the AS verification plan (\ArtAZ) shall be carried out and the results evaluated with respect to the defined safety requirements. Explanations for how the results that are observed indicate that the safety requirements have been met shall be provided where necessary. The verification results shall be explicitly documented (\ArtAT).

\subsection*{Artefact \ArtAS: AS Verification Log}\label{art:SS}
All verification activities shall be recorded in the AS Verification Log, including what is actually done, when and by whom, as well as any problems that occurred during verification and how they were fixed. Decisions that are implemented during testing shall be recorded. 

\Activity{Instantiate AS Verification Argument Pattern \label{art:VV} \ArtAV}{act:VerifPattern}

This activity requires as input the AS Verification argument pattern (\ArtAU), as well as the artefacts from Activities~\actref{act:verifStrat}{28}, ~\actref{act:verifAct}{29}  and ~\actref{act:verify}{30} (\ArtAR, \ArtAZ, \ArtAS and \ArtAT). The activity uses these artefacts to create an instantiated AS verification assurance argument (\ArtAV) which demonstrates how the verification activities show that the defined safety requirements are sufficiently satisfied by the AS.

\subsection*{Artefact \ArtAU: AS Verification Argument Pattern}\label{art:UU}

\begin{figure}[h]
    \centering
    \includegraphics[width=1\linewidth]{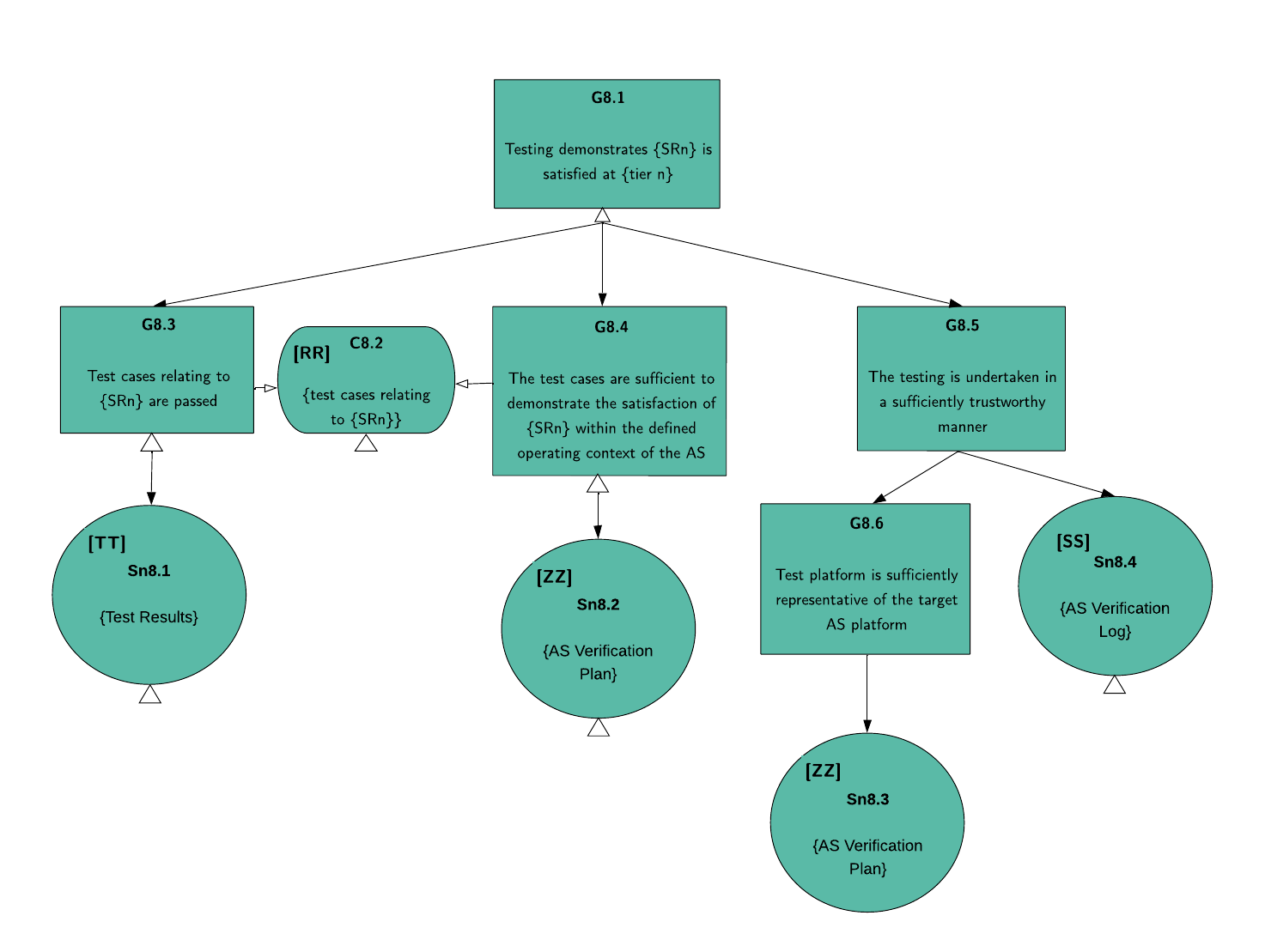}
    \caption{\ArtAU:~Argument Pattern for testing as part of AS Verification} 
    \label{fig:VerifArg}
\end{figure}

The argument pattern relating to this stage is shown in Figure~\ref{fig:VerifArg} and key elements from the pattern are described in the following sections.

\begin{figure}[h]
    \centering
    \includegraphics[width=1\linewidth]{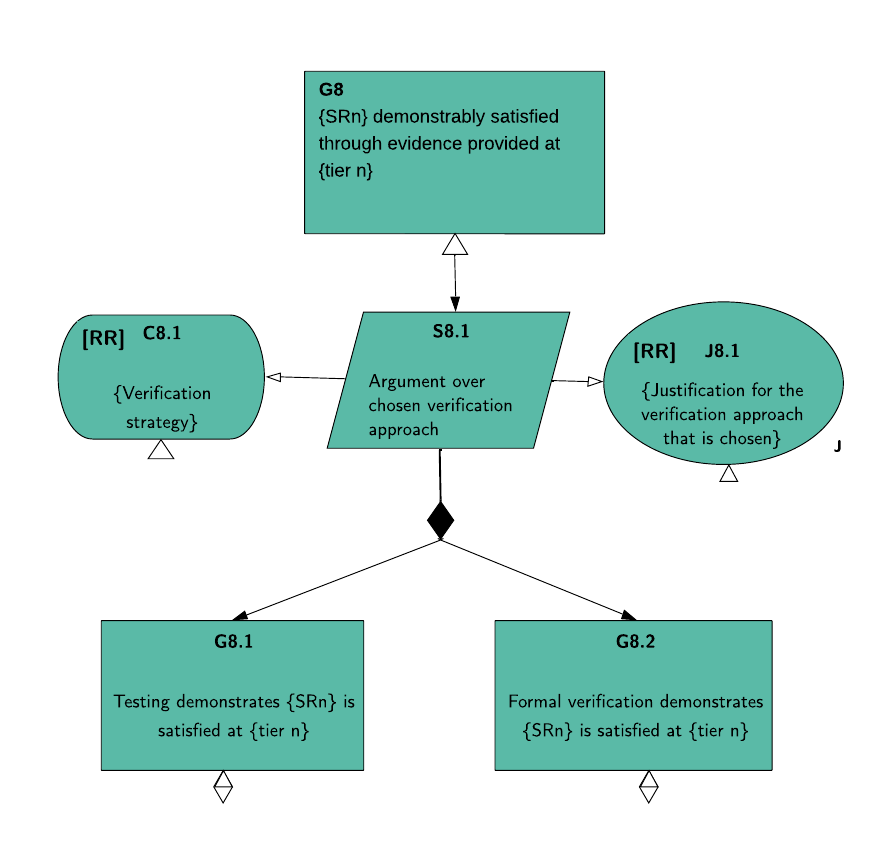}
    \caption{\ArtAU:~Argument Pattern for AS Verification} 
    \label{fig:VerifArgOver}
\end{figure}

\begin{figure}[h]
    \centering
    \includegraphics[width=1\linewidth]{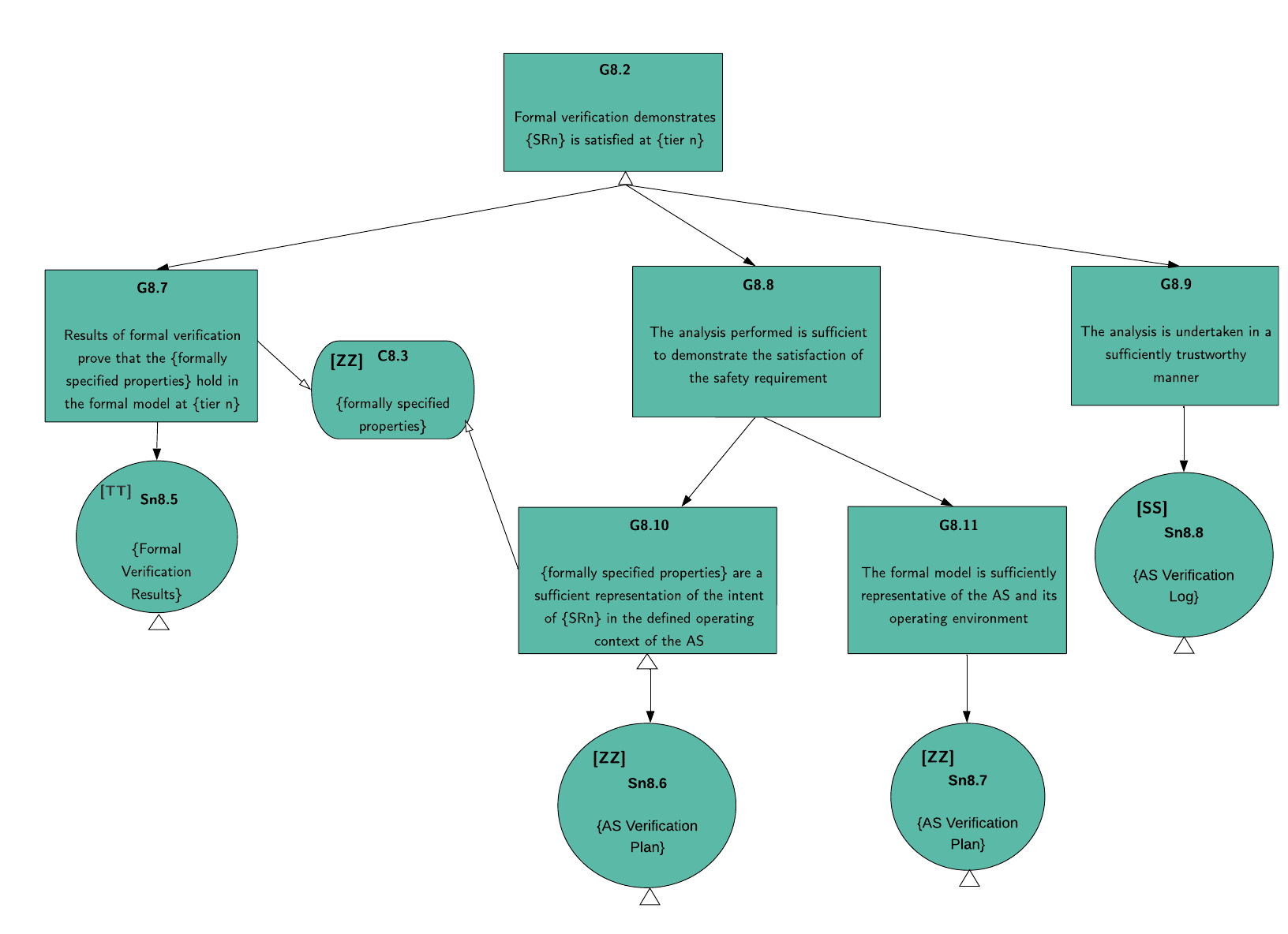}
    \caption{\ArtAU:~Argument Pattern for formal verification as part of AS Verification} 
    \label{fig:VerifArgFormal}
\end{figure}

\subsection*{G8} This argument is created for each tier, to demonstrate that the verification evidence that is provided for that tier is sufficient to show that the safety requirements defined at that tier are satisfied.

\subsection*{S8.1} The claim is supported by making an argument over the chosen verification strategy that is defined in \ArtAR. A justification for why the defined strategy has been chosen must also be provided (J8.1). The strategy may involve testing, formal verification, or some combination of these approaches. Where testing is used, claim G8.2 must be instantiated. Where formal verification is used, claim G8.6 must be instantiated.

\subsection*{G8.1} Where testing is being used, for each of the defined safety requirements it must be demonstrated that the testing undertaken demonstrates that the requirement is satisfied. To support this claim it is necessary to show that the tests that were performed have been passed (G8.3), that sufficient tests have been performed for each requirement (G8.4), and that the tests were undertaken in a manner such that the results obtained are trustworthy (G8.5). 

\subsection*{G8.5} In order to show that the testing has been performed in a trustworthy manner, the rigour of the testing process is considered as well as sufficiency of the test platform. The verification log (\ArtAS) can be used as evidence of the process that was followed, as well as the suitability of the people and tools that were used to implement the testing. The verification plan (\ArtAZ) provides evidence that the test platform that was used to perform the tests is sufficiently representative of the target system. 

\subsection*{G8.2} Where formal verification is being used, for each of the defined safety requirements it must be demonstrated that the verification undertaken demonstrates that the requirement is satisfied. To support this claim it is necessary to show that the formally specified properties have been proven to be satisfied (G8.7), that the analysis that is performed is sufficient to demonstrate each requirement is met (G8.8), and that the analysis is undertaken in a manner such that the results obtained are trustworthy (G8.9).

\subsection*{G8.8} To demonstrate that the analysis performed is sufficient, it is necessary to demonstrate firstly that the properties that have been specified for the formal analysis are sufficient to capture the intent of the safety requirement that needs to be demonstrated (G8.10). The verification plan (\ArtAZ) should provide this justification. Secondly, it imust be demonstrated that the formal model that is used to undertake the analysis is a sufficiently accurate reflection of the real AS in its operating environment. This claim is made at G8.11 and is again supported by evidence from the verification plan (\ArtAZ).

\clearpage
\section{Afterword}

It would not have been possible to produce this document without the numerous insightful interactions of the authors with a wide range of experts across industry and academia. We cannot acknowledge them all personally here, but their contributions are very much appreciated. In particular we would like to thank the following AAIP Visiting Fellows and colleagues who kindly reviewed and provided feedback on an initial draft of this document:

\begin{itemize}
    \item Alec Banks (DSTL)
   \item Lydia Gauerhof (Robert Bosch GmbH)
   \item Roger Rivett (Jaguar Land Rover (retired))
   \item Simon Smith (CACI)
    \item Shakir Laher (NHS Digital)
    \item Rasmus Adler (Fraunhofer IESE)
\end{itemize}

We would very much value feedback on the guidance provided in this document. We would in particular encourage the reader where appropriate to apply this guidance to the development of systems and share those experiences with the authors.

This work has been funded by Lloyds Register Foundation and the University of York through the Centre for Assuring Autonomy https://www.york.ac.uk/assuring-autonomy.

\clearpage
\bibliographystyle{plainurl}
\bibliography{main}

\end{document}